\pdfoutput=1
\documentclass[traditabstract]{aa} 
\usepackage{graphicx}
\usepackage{txfonts}
\usepackage[round]{natbib}
\bibpunct{(}{)}{;}{a}{}{,}
\begin{document}
\title{Molecular gas content in strongly-lensed $z\sim 1.5-3$ star-forming galaxies with low IR luminosities
\thanks{Based on observations carried out with the IRAM Plateau de Bure
Interferometer and the IRAM 30~m telescope. IRAM is supported by CNRS/INSU
(France), the MPG (Germany) and the IGN (Spain).}}


\author{M. Dessauges-Zavadsky\inst{1},
        M. Zamojski\inst{1},
        D. Schaerer\inst{1,2},
        F. Combes\inst{3},
        E. Egami\inst{4},
        A.~M. Swinbank\inst{5},\\
        J. Richard\inst{6}, 
        P. Sklias\inst{1},
        T.~D. Rawle\inst{7},
        M. Rex\inst{4},
        J.-P. Kneib\inst{8,9},
        F. Boone\inst{2},
        \and
        A. Blain\inst{10}
	}

\offprints{miroslava.dessauges@unige.ch}

\institute{Observatoire de Gen\`eve, Universit\'e de Gen\`eve, 
           51 Ch. des Maillettes, 1290 Versoix, Switzerland
	   \and 
	       CNRS, IRAP, 14 Avenue E. Belin, 31400 Toulouse, France
	   \and
	       Observatoire de Paris, LERMA, 61 Avenue de l'Observatoire, 75014 Paris, France
	   \and 
	       Steward Observatory, University of Arizona, 933 North Cherry Avenue, Tucson, AZ 85721, USA
	   \and
	       Institute for Computational Cosmology, Durham University, South Road, Durham DH1 3LE, UK
	   \and
	       CRAL, Observatoire de Lyon, Universit\'e Lyon 1, 9 Avenue Ch. Andr\'e, 69561 Saint Genis Laval Cedex, France
	   \and
	       ESAC, ESA, PO Box 78, Villanueva de la Canada, 28691 Madrid, Spain
	   \and
	       Laboratoire d'Astrophysique, Ecole Polytechnique F\'ed\'erale de Lausanne (EPFL), Observatoire de Sauverny, 1290 Versoix, Switzerland
	   \and
	       Aix Marseille Universit\'e, CNRS, LAM, UMR 7326, 13388, Marseille, France    
	   \and
	       Department of Physics \& Astronomy, University of Leicester, University Road, Leicester LE1 7RH, UK  
           }

\date{}

\authorrunning{Dessauges-Zavadsky et~al.}

\titlerunning{Molecular gas content in strongly-lensed $z\sim 1.5-3$ star-forming galaxies}

 
\abstract
{To extend the molecular gas measurements to more typical star-forming galaxies (SFGs) with ${\rm SFR} < 40~\rm M_{\sun}~yr^{-1}$ and $M_* < 2.5\times 10^{10}~\rm M_{\sun}$ at $z\sim 1.5-3$, we have observed CO emission with the IRAM Plateau de Bure Interferometer and 30~m telescope for five strongly-lensed galaxies, selected from the {\it Herschel} Lensing Survey. These observations are combined with a compilation of CO measurements from the literature. From this, we infer the CO luminosity correction factors $r_{2,1} = 0.81\pm 0.20$ and $r_{3,1} = 0.57\pm 0.15$ for the $J=2$ and $J=3$ CO transitions, respectively, valid for SFGs at $z>1$. The combined sample of CO-detected SFGs at $z>1$ shows a large spread in star formation efficiency (SFE) with a dispersion of 0.33~dex, such that the SFE extend well beyond the low values of local spirals and overlap the distribution of $z>1$ sub-mm galaxies. We find that the spread in SFE (or equivalently in molecular gas depletion timescale) is due to variations of several physical parameters, primarily the specific star formation rate, but also stellar mass and redshift. Correlations of the SFE with the offset from the main-sequence and the compactness of the starburst are less clear. The possible increase of the molecular gas depletion timescale with $M_*$ now revealed by low stellar mass SFGs at $z>1$ and also observed at $z=0$ is in contrast to the constant molecular gas depletion timescale generally admitted and refutes the linearity of the Kennicutt-Schmidt relation. A net rise of the molecular gas fraction ($f_{\rm gas}$) is observed from $z\sim 0.2$ to $z\sim 1.2$, followed by a very mild increase toward higher redshifts, as found in earlier studies. At each redshift the molecular gas fraction shows a large dispersion, mainly due to the dependence of $f_{\rm gas}$ on stellar mass, producing a gradient of increasing $f_{\rm gas}$ with decreasing $M_*$. We provide the first measurement of the molecular gas fraction of $z>1$ SFGs at the low-$M_*$ end between $10^{9.4} < M_*/{\rm M}_{\sun} < 10^{9.9}$, reaching a mean $\langle f_{\rm gas}\rangle = 0.69\pm 0.18$ which shows a clear $f_{\rm gas}$ upturn at these lower stellar masses. Finally, we find evidence for a non-universal dust-to-gas ratio among high-redshift SFGs and sub-mm galaxies, local spirals and ultra-luminous infrared galaxies with near-solar metallicities, as inferred from a homogeneous analysis of their rest-frame 850~$\mu$m luminosity per unit gas mass. $z>1$ SFGs show a trend for a lower $L_{\nu}(850\,\mu {\rm m})/M_{\rm gas}$ mean by 0.33~dex compared to the other galaxy populations.}


\maketitle
%

\section{Introduction}
\label{sect:introduction}

Gas, stars, dust, and metals are the basic galaxy constituents. They determine 
the majority of observable properties in galaxies at all wavelengths. 
Therefore, the characterization of these observable properties from present up 
to high-redshifts provides stringent tests 
and anchors for the physical processes that regulate the galaxy evolution. 

Several empirical relationships between gas, stars, dust, and metals have been
observationally identified. The first relationship is the correlation between 
the star formation rate (SFR) and stellar mass ($M_*$). It determines the 
so-called ``main-sequence'', 
representing the locus where local and high-redshift star-forming galaxies 
(SFGs) lie in the SFR--$M_*$ plane. The correlation is slightly sublinear and 
evolves with redshift, such that high-redshift galaxies form more stars per 
unit time than the low-redshift ones with the same stellar mass 
\citep[e.g.,][]{noeske07,daddi07,elbaz07,rodighiero10,salmi12}. This implies an 
increase of the specific star formation rate, defined as the ratio of the star 
formation rate over the stellar mass, with redshift. The second relationship is 
the correlation between the gas-phase metallicity and stellar mass 
\citep[e.g.,][]{tremonti04,savaglio05,erb06,maiolino08}. It shows a redshift 
evolution toward lower metallicities at higher redshifts. More recently,
\citet{mannucci10,mannucci11} found the ``fundamental'' metallicity relation 
(FMR) that connects at the same time metallicity, stellar mass, and star 
formation rate, as revealed by local and $z<2.5$ galaxies. 
The third important relationship is the Kennicutt-Schmidt relation, or the 
star-formation relation, that relates the star formation rate surface density 
to the total gas ($\rm H\,I + H_2$) surface density through a power law 
\citep{kennicutt98a}. Mostly constrained by local galaxies so far 
\citep{leroy08,bigiel08,bigiel11}, the tightest correlation is observed between 
the star formation rate surface density and the H$_2$ gas surface density well 
parametrized by a linear relation, 
meaning that the molecular gas is being consumed at a constant rate within a 
molecular gas depletion timescale of about 1.5~Gyr. However, both the COLD GASS 
survey of local massive galaxies ($\log (M_*/{\rm M_{\sun}}) > 10$) from 
\citet{saintonge11} and the recent ALLSMOG survey of local low-mass galaxies 
($9 < \log (M_*/{\rm M_{\sun}}) < 10$) from \citet{bothwell14}
show an increase of the molecular gas depletion timescale with stellar mass 
and thus bring evidence against a linear Kennicutt-Schmidt relation. 
A linear Kennicutt-Schmidt relation for the H$_2$ gas 
surface densities seems to hold for high-redshift SFGs, 
but with a much shorter molecular gas depletion timescale of about 0.7~Gyr 
\citep{tacconi13}, even though they are more gas-rich 
\citep[e.g.,][]{daddi10a,tacconi10,genzel10}. Finally, there is the 
relationship between the dust-to-gas ratio and metallicity seen in nearby 
galaxies \citep[e.g.,][]{issa90,dwek98,edmunds01,inoue03,draine07,leroy11}. 
Attempts to measure the dust-to-gas ratios of high-redshift galaxies support a 
similar correlation, as well as a trend toward smaller dust-to-gas ratios at 
higher redshifts \citep{saintonge13,chen13}.

Most of these empirical correlations are both qualitatively and quantitatively 
consistent with the so-called ``bathtub'' model, which gives a good analytical 
representation of the gross features of the star-forming galaxy evolution 
\citep[e.g.,][]{bouche10,lagos12,lilly13,dekel14}. 
Solely based on the equation of conservation of gas mass in a galaxy, this 
model assumes that galaxies lie in a quasi-steady state equilibrium, where 
their ability to form stars is regulated by the availability of gas replenished 
through the accretion rate dictated by cosmology and the amount of material 
they return into the intergalactic medium through gas outflows. The predicted 
redshift and mass dependences are the following (although slightly differing 
from authors to authors): the average gas accretion rate varies as 
$M_{\rm halo}^{1.1} (1+z)^{2.2}$, the specific star formation rate as 
$(1+z)^{2.2}$, the gas depletion timescale as $(1+z)^{-1.5}$, and the molecular 
gas fraction steadily increases with redshift and decreases with stellar mass. 
In agreement with cosmological hydrodynamic simulations from 
\citet{dave11,dave12}, the latter, in addition, predict a decrease of the gas 
depletion timescale with the stellar mass, instead of a constant gas depletion 
timescale.

Improvements in the sensitivity of the IRAM Plateau de Bure Interferometer have 
made it possible over the last years to start getting a census of the molecular 
gas content in star-forming galaxies near the peak of the cosmic star formation activity.
However, the samples of CO-detected objects at $z=1-1.5$ and $z=2-3$ are still 
small and confined to the high-SFR and high-$M_*$ end of main-sequence SFGs 
\citep{daddi10a,genzel10,tacconi10,tacconi13}. 
In this paper, we extend the dynamical range of star formation rates and 
stellar masses of star-forming galaxies with observationally constrained 
molecular gas contents below $\rm SFR < 40~M_{\sun}~yr^{-1}$ and $M_* < 
2.5\times 10^{10}~\rm M_{\sun}$, and thus reach the $L^{\star}$ to 
sub-$L^{\star}$ domain of $z\sim 1.5-3$ galaxies \citep{gruppioni13}. 
This domain of physical parameters is accessible to CO emission measurements 
only with the help of gravitational lensing, a technique that proved to be 
efficient by a few objects already\citep{baker04,coppin07,saintonge13}. 
Therefore, our five galaxies selected for CO follow-up observations come from 
the {\it Herschel} Lensing Survey of massive galaxy cluster fields 
\citep[HLS;][]{egami10}, designed to detect lensed, high-redshift background 
galaxies and probe more typical, intrinsically fainter galaxies than those 
identified in large-area, blank-field surveys. 
The gaseous, stellar, and dust properties inferred for these (sub-)$L^{\star}$ 
main-sequence star-forming galaxies \citep[see also][]{sklias14} are put 
face-to-face with a large comparison sample of local and high-redshift galaxies 
with CO measurements reported in the literature, which together provide new 
tests and anchors for galaxy evolution models.

In Sect.~\ref{sect:sample} we describe the target selection and their physical properties, 
and present the comparison sample of CO-detected galaxies from the literature. 
In Sect.~\ref{sect:obs+reduction} we report on CO observations performed with 
the IRAM Plateau de Bure Interferometer and 30~m telescope and discuss the CO 
results. In Sect.~\ref{sect:luminosity-corrections} we infer the CO luminosity 
correction factors for the $J=2$ and $J=3$ CO rotational transitions. 
The gaseous and stellar properties of our strongly-lensed galaxies are placed 
in the general context of galaxies with CO measurements in 
Sect.~\ref{sect:discussion}, where the main objective is to understand what 
drives the large spread in star formation efficiency
observed in high-redshift SFGs. In Sect.~\ref{sect:fgas} we explore the 
redshift evolution and the stellar mass dependence of their molecular gas 
fractions. In Sect.~\ref{sect:dust-to-gas} we discuss the universality of the 
dust-to-gas ratio inferred from a homogeneous analysis of the rest-frame 
850~$\mu$m continuum.
Summary and conclusions are given in Sect.~\ref{sect:summary}. Individual CO 
properties and inferred kinematics of our selected strongly-lensed galaxies are 
described in Appendix~\ref{sect:appendix}.

Throughout the paper, we adopt the initial mass function (IMF) of 
\citet{chabrier03} and scale the values from the literature by the factor of 
1.7 when the \citet{salpeter55} IMF is used. The designation to ``gas'' always 
refers to the molecular gas (H$_2$) only and not the total gas 
($\rm H\,I + H_2$).  All the molecular gas masses are derived from the observed 
CO emission via the ``Galactic'' CO(1--0)--H$_2$ conversion factor 
$X_{\rm CO} = 2\times 10^{20}~\rm cm^{-2}/(K~km~s^{-1})$, or $\alpha = 
4.36~\rm M_{\sun}/(K~km~s^{-1}~pc^2)$ which includes the correction factor of 
1.36 for helium. We use the cosmology with $\rm H_0 = 70~km~s^{-1}~Mpc^{-1}$, 
$\Omega_{\rm M} = 0.29$, and $\Omega_{\Lambda} = 0.71$.

%

\begin{table*}
\caption{Physical properties derived from SED fitting}             
\label{tab:SED-properties}      
\centering          
\begin{tabular}{l c c c c c c c c }     
\hline\hline       
Source & $z_{\rm H\alpha}$ & $\mu$\tablefootmark{a} & $L_{\rm IR}$\tablefootmark{b} & $M_*$\tablefootmark{c} & SFR\tablefootmark{c} & $M_{\rm dust}$\tablefootmark{d} & $T_{\rm dust}$\tablefootmark{d} & $L_{\nu}(850\,\mu{\rm m})$\tablefootmark{d} \\
 & & & ($\rm 10^{11}~L_{\sun}$) & ($\rm 10^{10}~M_{\sun}$) & ($\rm M_{\sun}~yr^{-1}$) & ($\rm 10^7~M_{\sun}$) & (K) & ($\rm 10^{29}~erg~s^{-1}~Hz^{-1}$) \\
\hline                    
A68-C0        & 1.5864                & 30 & $1.18\pm 0.07$                  & $2.0\pm 0.65$  & $9^{+3.3}_{-3.9}$  & 4.7  & 34.5 & 5.0 \\
A68-HLS115    & 1.5869                & 15 & $3.42\pm 0.16$                  & $0.81^{+0.17}_{-0.20}$ & $25^{+2.6}_{-6}$ & 3.0  & 37.5 & 7.7 \\
MACS0451-arc  & 2.013                 & 49 & $1.36\pm 0.13$\tablefootmark{e} & $0.25\pm 0.02$ & $19^{+0.6}_{-0.2}$ & 0.6  & 47.4 & 3.4 \\
A2218-Mult    & 3.104                 & 14 & $2.26\pm 0.15$                  &      & & & \\
A68-h7	      & 2.15\tablefootmark{f} & 3  & $18.3^{+0.9}_{-1.2}$                  & $15.4^{+0.9}_{-3.8}$ & $76\pm 3.4$ & 18.2 & 43.3 & 27.6 \\
MS\,1512-cB58 & 2.729                 & 30 & $3.04\pm 0.07$                  & $0.44\pm 0.03$ & $37^{+2.4}_{-2.9}$ & 1.2  & 50.1 & 2.0 \\
Cosmic Eye    & 3.0733                & 28 & $3.41^{+0.16}_{-0.23}$                  & $2.4\pm 0.06$  & $33\pm 0.6$ & 2.2  & 46.3 & 3.3 \\
\hline                  
\end{tabular}
\tablefoot{
All values correspond to intrinsic values corrected from magnification factors 
and come from \citet{sklias14} except for A2218-Mult. 
\tablefoottext{a}{Magnification factors derived from robust lens modeling 
\citep{richard07,richard11}.}
\tablefoottext{b}{Infrared luminosities integrated over the [8,1000]~$\mu$m 
interval.}
\tablefoottext{c}{Stellar masses and star formation rates as derived from the
best energy conserving SED fits, obtained under the hypothesis of an 
extinction, $A_{\rm V}$, fixed at the observed ratio of $L_{\rm IR}$ over 
$L_{\rm UV}$ following the prescriptions of \citet{schaerer13}.}
\tablefoottext{d}{Dust masses, dust temperatures, and rest-frame 850~$\mu$m 
luminosities derived from a modified black-body fit applied to the 
{\it Herschel} photometry with the $\beta$-slope fixed to 1.5 for 
$T_{\rm dust}$ and to 1.8 for $M_{\rm dust}$ and $L_{\nu}(850\,\mu{\rm m})$.}
\tablefoottext{e}{IR luminosity corrected from the AGN contribution identified
by the excess of the flux in the {\it Herschel}/PACS\,100\,$\mu$m band observed 
in the southern part of the MACS0451-arc, indicating very hot dust. A 
decomposition of the IR emission of the southern part between the AGN and 
stellar emission indicates a starburst-component contributing to about 45\% of 
the total $L_{\rm IR}(\rm south)$ (for a detailed analysis, see Zamojski 
et~al.\ in prep).}
\tablefoottext{f}{Redshift determined from weak absorption lines in the 
rest-frame UV instead of the H$\alpha$ line.}
}
\end{table*}
%

\begin{figure}
\centering
\includegraphics[width=9cm,clip]{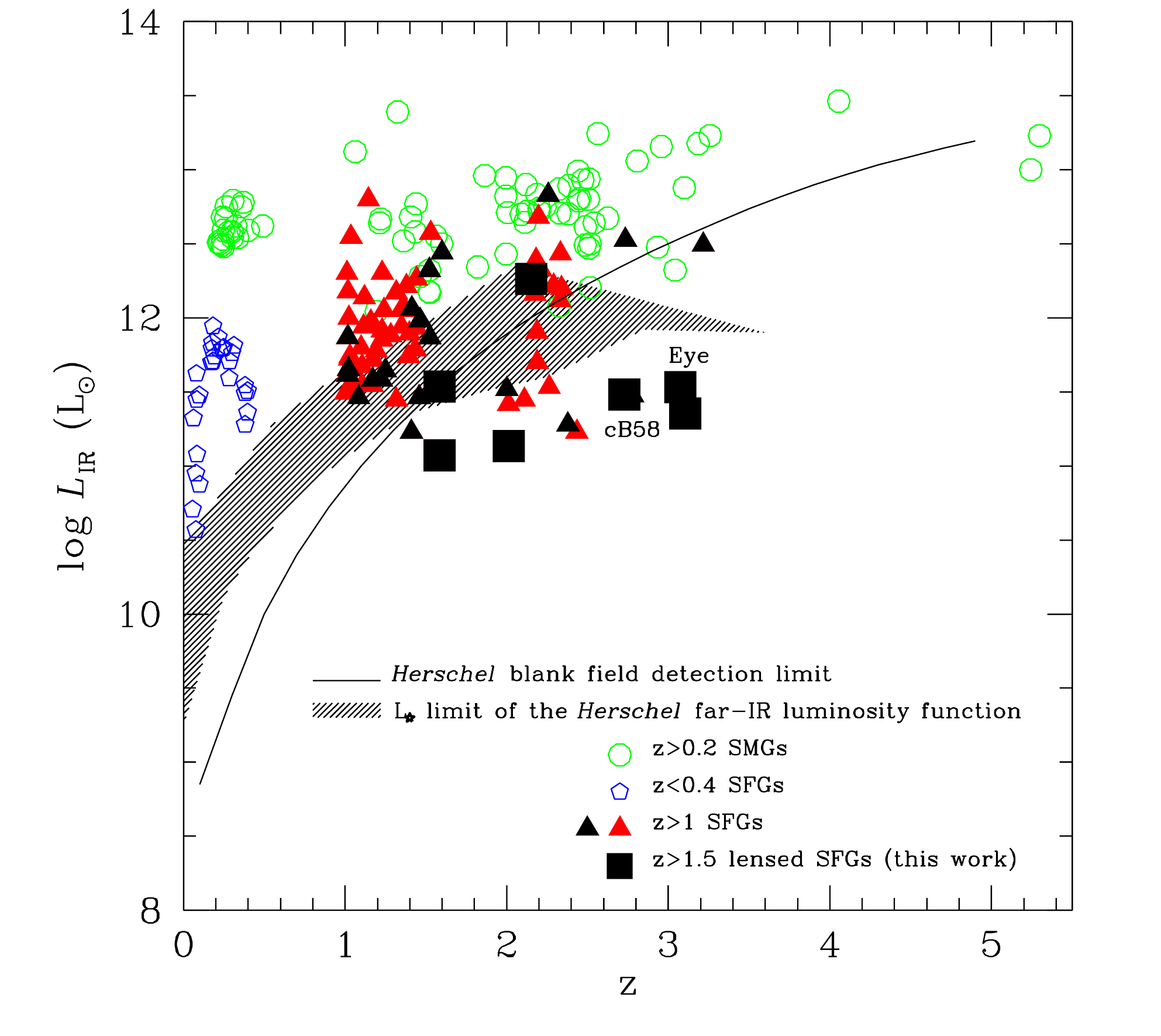} 
\caption{IR luminosities as a function of redshift of our low-$L_{\rm IR}$ 
selected SFGs (squares) compared to our compilation of galaxies with CO 
measurements from the literature (see Sect.~\ref{sect:literature}). Symbols in 
black refer to $L_{\rm IR}$ as derived from the {\it Herschel} far-IR 
photometry. 
Our SFGs populate the regime with the lower $L_{\rm IR}$ reported at $z>1.5$: 
they reach the $L^{\star}$ and below domain as delimited by the hatched area 
showing the $L^{\star}$ limit of the {\it Herschel} far-IR luminosity function 
as determined by \citet[][upper band]{magnelli13} and \citet[][lower band]
{gruppioni13} and extend the blank field studies to fainter luminosities 
as it is shown by the solid line which defines the minimal $L_{\rm IR}$ at 
each redshift that produces a flux $\gtrsim 2~\rm mJy$ in the 
{\it Herschel}/PACS\,160\,$\mu$m band \citep[$\sim 3\,\sigma$ detection limit 
in GOODS-N,][]{elbaz11}.}
\label{fig:LIR-z}
\end{figure}
%

\begin{figure}
\centering
\includegraphics[width=9cm,clip]{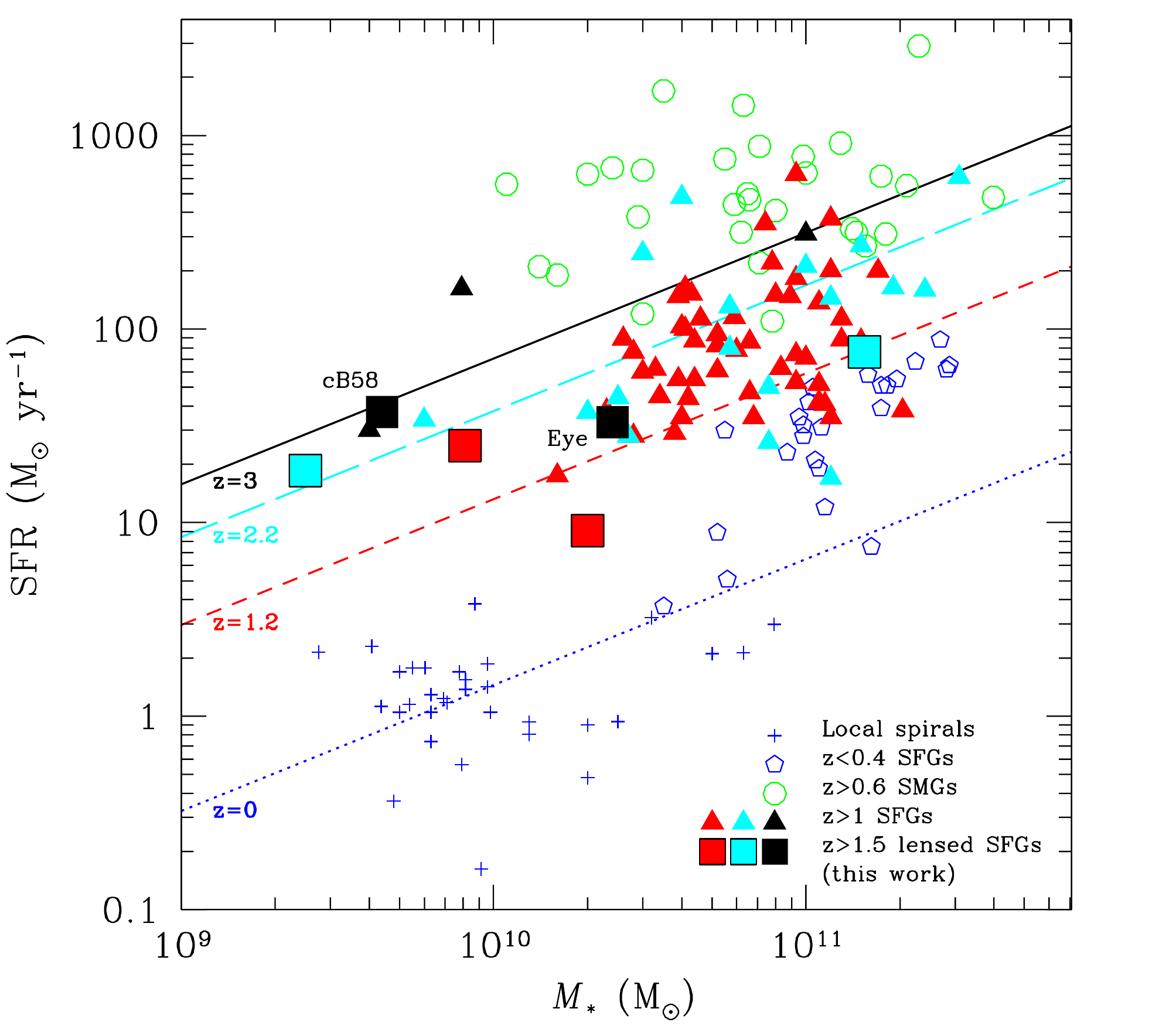} 
\caption{Location of our low-$L_{\rm IR}$ selected SFGs (squares) in the 
star-formation--stellar-mass plane compared to our compilation of galaxies with 
CO measurements from the literature (see Sect.~\ref{sect:literature}). Our 
(sub-)$L^{\star}$ galaxies probe one order of magnitude smaller stellar masses 
than the bulk of $z>1$ SFGs 
from the literature. The blue dotted, red short-dashed, cyan long-dashed, and 
black solid lines are the best-fits of SFR--$M_*$ MS at $z\simeq 0$, 1.2, 2.2, 
and 3, respectively, parametrized by Eq.~(\ref{eq:MS}). The color-coding of 
SFGs (our sample plus the $z>1$ SFGs from the literature) refers to three 
redshift intervals: $\langle z_{1.2}\rangle = [1,1.6]$ (red), 
$\langle z_{2.2}\rangle = [2,2.5]$ (cyan), and $\langle z_{3.0}\rangle = 
[2.7,3.2]$ (black) in line with the MS fits.}
\label{fig:SFR-Mstars}
\end{figure}
%

\section{High-redshift samples of CO-detected galaxies}
\label{sect:sample}

\subsection{Target selection}
\label{sect:target-selection}

The HLS is providing us with unique targets ideal for CO follow-up studies. We 
used this dataset to select 
galaxies at $z\sim 1.5-3$ with low intrinsic (delensed) IR luminosities 
$L_{\rm IR} < 4\times 10^{11}~\rm L_{\sun}$, as derived from the 
{\it Herschel}/PACS (100 and 160~$\mu$m) and SPIRE (250, 350, and 500~$\mu$m) 
SED modelling. These galaxies are of particular interest, because they allow 
probing the molecular gas content at high redshift in a regime of star 
formation rates $\rm SFR < 40~M_{\sun}~yr^{-1}$ still very poorly explored, 
while being typical of ``normal'' star-forming galaxies. In addition, we 
required that our targets (1)~are strongly lensed with well-known magnification 
factors derived from robust lens modelling \citep{richard07,richard11} in order 
to make the CO emission of these intrinsically faint $L_{\rm IR}$ objects 
accessible to current millimeter instruments, (2)~have known spectroscopic 
redshifts from our optical and near-IR observing campaigns 
\citep{richard07,richard11}, (3)~have well-sampled global SEDs from optical, 
near-IR to IR obtained with ground-based telescopes, the {\it Hubble} Space 
Telescope (HST), and the {\it Spitzer} satellite in order to have tight 
estimates of their stellar masses and star formation rates, and (4)~have 
high-resolution HST images useful to get information on their morphology.

The four galaxies---A68-C0, A68-HLS115, MACS0451-arc, and A2218-Mult---selected 
for CO observations with the IRAM interferometer at the Plateau de Bure, 
France, do satisfy all the above specifications. The fifth object---A68-h7---
observed with the IRAM 30~m telescope at Pico Veleta, Spain, has a higher 
intrinsic IR luminosity and hence has properties resembling more the 
high-redshift galaxies with CO measurements already available. 
A description of these selected targets at $z>1.5$, as well as their 
detailed multi-wavelength SED analysis from optical to far-IR/sub-mm can be 
found in \citet[][except for A2218-Mult]{sklias14}. In 
Table~\ref{tab:SED-properties} we summarize their physical properties:
spectroscopic redshifts ($z_{\rm H\alpha}$), magnification factors ($\mu$), IR 
luminosities ($L_{\rm IR}$), stellar masses ($M_*$), star formation rates 
(SFR), dust masses ($M_{\rm dust}$), dust temperatures ($T_{\rm dust}$), and 
rest-frame 850~$\mu$m luminosities ($L_{\nu}(850\,\mu{\rm m})$). We add to this 
sample, MS\,1512-cB58 \citep[][hereafter `cB58']{yee96} and the Cosmic Eye 
\citep[][hereafter `Eye']{smail07}, two well-known strongly-lensed galaxies 
with similar characteristics, for which we provide in \citet{sklias14} revised 
physical properties obtained from updated SED fitting and {\it Herschel} 
photometry.

In Fig.~\ref{fig:LIR-z} we show the intrinsic IR luminosities of the 
low-$L_{\rm IR}$ selected SFGs as a function of redshift and compare them to 
the compilation of galaxies with CO measurements from the literature 
(Sect.~\ref{sect:literature}). Our sample of SFGs populates the regime with the 
lower $L_{\rm IR}$ reported at $z>1.5$ as expected from the selection criterion 
and reach the $L^{\star}$ to sub-$L^{\star}$ domain according to the 
{\it Herschel} far-IR luminosity function of galaxies at $z\sim 1.5-3$ with 
$\log (L^{\star}/{\rm L}_{\sun})$ varying between 11.4 and 12.3 
(see the hatched area in Fig.~\ref{fig:LIR-z} as delimited from 
\citet{gruppioni13} and \citet{magnelli13}). These $L_{\rm IR}$, derived from 
{\it Herschel} far-IR photometry, also fall below the {\it Herschel} detection 
limit and are only accessible with gravitational lensing. All the data points 
from the literature below the {\it Herschel} detection limit refer either to 
other lensed galaxies with {\it Herschel} $L_{\rm IR}$ measurements or to 
galaxies with $L_{\rm IR}$ measurements determined from their star formation 
rates via the \citet{kennicutt98b} relation, $L_{\rm IR}~\rm (L_{\sun}) = 
10^{10}\times SFR~(M_{\sun}~yr^{-1})$, scaled to the \citet{chabrier03} IMF. 

Our sample of strongly-lensed galaxies also probes very small stellar masses 
$M_* < 2.5\times 10^{10}~\rm M_{\sun}$ (see Table~\ref{tab:SED-properties}), on 
average one order of magnitude smaller compared to the stellar masses of the 
bulk of $z>1$ SFGs with CO measurements reported in the literature; making our 
sample of particular interest for the study of the molecular gas content in a 
new regime of smaller stellar masses in addition to the lower star formation \
rates. This is in line with the well-defined empirical SFR--$M_*$ relation. 
Our galaxies indeed nicely follow the main-sequence (MS) of $z\simeq 1.2$, 2.2, 
and 3 galaxies, respectively, 
as shown in Fig.~\ref{fig:SFR-Mstars}. We consider here the best-fit 
parametrization of the SFR--$M_*$ MS given by \citet{tacconi13} based on the 
samples of \citet{bouche10}, \citet{noeske07}, \citet{daddi07}, 
\citet{rodighiero10}, and \citet{salmi12}:
\begin{eqnarray}\label{eq:MS}
{\rm SFR_{MS}}(z,M_*)~{\rm (M_{\sun}~yr^{-1})} & = & \nonumber \\
45 (M_*/6.6\times 10^{10}~({\rm M_{\sun}}))^{0.65} ((1+z)/2.2)^{2.8}\,, & & 
\end{eqnarray}
which we compute at three redshifts, the redshift medians of our comparison 
sample of CO-detected $z>1$ SFGs from the literature 
(Sect.~\ref{sect:literature}), plus our sample of low-$L_{\rm IR}$ selected 
galaxies, within three redshift intervals: $\langle z_{1.2}\rangle = [1,1.6]$, 
$\langle z_{2.2}\rangle = [2,2.5]$, and $\langle z_{3.0}\rangle = [2.7,3.2]$. 
No excess in the specific star formation rate is observed among our objects, 
they all lie within $0.3 < {\rm sSFR/sSFR_{MS}} \equiv 
\frac{{\rm SFR}/M_*}{{\rm SFR_{MS}}/M_*} < 3$, the admitted thickness of the 
main-sequence \citep[e.g.,][]{magdis12b}. On the other hand one 
galaxy---the Eye---is located below the MS at its respective redshift with 
$\rm sSFR/sSFR_{MS} = 0.27$. 

%

\subsection{Comparison sample}
\label{sect:literature}

In order to have a comparison sample to put face to face with our sample of 
low-$L_{\rm IR}$ selected galaxies, we built up a large compilation of local 
and high-redshift galaxies with CO measurements reported in the literature. We 
refer to the same comparison sample throughout the paper. This compilation is 
exhaustive for 
star-forming galaxies that have the highest priority in the comparative work to 
our sample of galaxies, very complete for high-redshift sub-mm galaxies (SMGs), 
but remains incomplete for all local galaxies including the spiral galaxies, 
luminous IR galaxies (LIRGs), and ultra-luminous IR galaxies (ULIRGs). All 
these CO-detected galaxies are classified along the following sub-categories 
and are designated by the same symbols (but not by the same colors) in all 
figures.

\begin{itemize}
\item {\it Local spirals} with $L_{\rm IR} < 10^{11}~\rm L_{\sun}$ (crosses):\\
\citet{helfer03}, \citet{gao04}, \citet{leroy08,leroy09}, 
\citet{papadopoulos12}, \citet{wilson12}, and \citet{bothwell14}.
\\

\item {\it Local LIRGs} with $10^{11} < L_{\rm IR} < 10^{12}~\rm L_{\sun}$ (filled diamonds):\\
\citet{solomon97}, \citet{gao04}, \citet{wilson08}, \citet{leech10}, and 
\citet{papadopoulos12}.
\\

\item {\it Local ULIRGs} with $L_{\rm IR} > 10^{12}~\rm L_{\sun}$ (small open circles):\\
\citet{solomon97} and \citet{papadopoulos12}.
\\

\item {\it High-redshift SMGs/ULIRGs} at $z>0.2$ with $L_{\rm IR} > 10^{12}~\rm L_{\sun}$ (big open circles):\\
\citet{greve05}, \citet{kneib05}, \citet{iono06}, \citet{tacconi06}, 
\citet{daddi09a,daddi09b}, \citet{knudsen09}, \citet{weiss09},
\citet{bothwell10,bothwell13}, \citet{carilli10}, \citet{harris10}, 
\citet{ivison10,ivison11}, \citet{riechers10,riechers11}, \citet{swinbank10}, 
\citet{yan10}, \citet{braun11}, \citet{casey11}, 
\citet{combes11,combes12,combes13}, \citet{danielson11}, \citet{fu12}, 
\citet{sharon13}, \citet{rawle14}, and \citet{magdis14}.
\\

\item {\it SFGs} at $z<0.4$ (open pentagons) and at $z>1$ (filled triangles) including field and lensed galaxies: \\
\citet{dannerbauer09}, \citet{geach09}, \citet{geach11}, \citet{knudsen09}, 
\citet{aravena10,aravena12,aravena14}, \citet{genzel10},
\citet{tacconi10,tacconi13}, \citet{daddi10a,daddi14}, \citet{johansson12}, 
\citet{magdis12a,magdis12b}, \citet{magnelli12}, \citet{bauermeister13}, 
\citet{saintonge13}\footnote{For the physical properties of the 8~o'clock arc (stellar mass and star formation rate), we use the values derived in \citet{dessauges11}.}, \citet{tan13}, and \citet{magdis14}.
\end{itemize}

%

\begin{table*}
\caption{Observation summary}             
\label{tab:observations-log}      
\centering          
\begin{tabular}{l l l c c c l l c c}     
\hline\hline       
Source & \multicolumn{2}{c}{Coordinates} & $T_{\rm on-source}$ & CO   & $\nu_{\rm obs}$\tablefootmark{a} & \multicolumn{2}{c}{Synthesized beam\tablefootmark{b}} & Bandwidth\tablefootmark{c} & rms\tablefootmark{d} \\ 
       & RA (J2000) & DEC (J2000)        & (hours)             & line & (GHz)                            & Size ($\arcsec$) & PA ($\degr$)                       & (MHz)                      & (mJy) \\
\hline                    
A68-C0                  & 00:37:07.404 & +09:09:26.57 &  5.2 & 2--1 &  89.131 & $6.41\times 5.53$ & 174 & 15 & 0.72 \\
A68-HLS115              & 00:37:09.503 & +09:09:03.80 &  4.6 & 2--1 &  89.128 & $5.94\times 5.79$ & 111 & 15 & 0.71 \\
MACS0451-arc            & 04:51:57.093 & +00:06:10.44 & 30.2 & 3--2 & 114.768 & $5.45\times 4.12$ & 163 & 19 & 0.84 \\
A2218-Mult              & 16:35:48.919 & +66:12:13.81 &  9.2 & 3--2 &  84.258 & $6.35\times 5.90$ & 90\ & 15 & 0.77 \\
A68-h7\tablefootmark{e} & 00:37:01.41  & +09:10:22.31 &  5.5 & 3--2 & 109.777 &                   &	& 24 & 0.94 \\
\hline                  
\end{tabular}
\tablefoot{
\tablefoottext{a}{Observed line frequency used for tuning.}
\tablefoottext{b}{Beam resulting after combining all available imaging with 
natural weighting. These are the beam sizes and position angles displayed in 
Fig.~\ref{fig:COmaps}.}
\tablefoottext{c}{Bandwidth achieved after resampling the PdBI data to a
resolution of 50~km~s$^{-1}$ and the 30~m telescope data to a resolution of 
65~km~s$^{-1}$. Such resampling perfectly suits observed CO lines with full 
widths at half maximum $> 200~\rm km~s^{-1}$.}
\tablefoottext{d}{Noise per beam obtained for the specified bandwidth and 
averaged over the full 3.6~GHz spectral range after excluding channels where CO 
emission is detected.}
\tablefoottext{e}{Source observed at the 30~m single dish telescope, while 
all the other sources were observed with the PdBI.}
}
\end{table*}
%

\begin{table*}
\caption{CO emission properties}             
\label{tab:COresults}      
\centering          
\begin{tabular}{l c c c c c c c c}     
\hline\hline       
Source & $z_{\rm CO}$ & CO   & $\rm S/N_{det}$\tablefootmark{a} & $\rm FWHM_{CO}$\tablefootmark{b} & $F_{\rm CO}$\tablefootmark{b} & $L'_{\rm CO(1-0)}$\tablefootmark{c} & $M_{\rm gas}$\tablefootmark{d} & $f_{\rm gas}$\tablefootmark{d} \\ 
       &              & line & 				        & (km~s$^{-1}$)			            & (Jy~km~s$^{-1}$)	            & ($10^9~\rm K~km~s^{-1}~pc^2$)	  & ($10^{10}~\rm M_{\sun}$)	   & \\
\hline                    
A68-C0        & 1.5854 & 2--1 & 11         & $334\pm 76$ & $1.89\pm 0.28$                    & $2.6\pm 0.4$    & 1.2    & 0.38 \\
A68-HLS115    & 1.5859 & 2--1 & 14         & $267\pm 18$ & $2.00\pm 0.30$                    & $5.5\pm 0.9$    & 2.4    & 0.75 \\
MACS0451-arc  & 2.0118 & 3--2 & 4--5       & $261\pm 41$ & $1.27\pm 0.32$                    & $1.0\pm 0.3$    & 0.4    & 0.62 \\
A2218-Mult    & 3.104  & 3--2 & undetected & --          & $<0.62$\tablefootmark{e}          & $<3.8$ & $<1.7$ &  \\
A68-h7        & 2.1529 & 3--2 & 3--4       & $282\pm 61$ & $1.12\pm 0.22$                    & $16.9\pm 3.3$   & 7.4    & 0.33 \\
MS\,1512-cB58 & 2.727  & 1--0 &            & 174         & $0.052\pm 0.013$\tablefootmark{f} & $0.6\pm 0.15$ & 0.3 & 0.41 \\
Cosmic Eye    & 3.074  & 1--0 &            & 200         & $0.077\pm 0.013$\tablefootmark{f} & $1.2\pm 0.2$ & 0.5 & 0.18 \\
\hline                  
\end{tabular}
\tablefoot{
\tablefoottext{a}{Signal-to-noise ratio of the CO emission detection.}
\tablefoottext{b}{Full widths half maximum and `observed' CO(2--1) or CO(3--2) 
line integrated fluxes in Jy~km~s$^{-1}$ and their $1\,\sigma$ errors, as 
derived from fitting Gaussian function(s) to the observed CO profile.}
\tablefoottext{c}{Lensing-corrected CO(1--0) integrated line luminosities 
obtained with the following prescriptions: 
(1)~the \citet{solomon97} formula 
$L'_{\rm CO(2-1),CO(3-2)}~({\rm K~km~s^{-1}~pc^2}) = 
3.25\times 10^7 F_{\rm CO(2-1),CO(3-2)} \nu_{\rm obs}^{-2} D_{\rm L}^2 
(1+z)^{-3}$, where $\nu_{\rm obs}$ is the observed line frequency in GHz and 
$D_{\rm L}$ is the luminosity distance in Mpc;
(2)~the correction factors r$_{2,1} = L'_{\rm CO(2-1)}/L'_{\rm CO(1-0)} = 0.81$ 
and r$_{3,1} = L'_{\rm CO(3-2)}/L'_{\rm CO(1-0)} = 0.57$ adopted for SFGs to 
account for the CO(2--1) and CO(3--2) transitions being slightly sub-thermally 
excited (see Sect.~\ref{sect:luminosity-corrections});
and (3)~the lensing magnification correction, $\mu$, given in 
Table~\ref{tab:SED-properties}.}
\tablefoottext{d}{Molecular gas masses, $M_{\rm gas} = 
\alpha L'_{\rm CO(1-0)}$, obtained by assuming the ``Galactic'' CO--H$_2$ 
conversion factor $X_{\rm CO} = 2\times 10^{20}~\rm cm^{-2}/(K~km~s^{-1})$, or 
$\alpha = 4.36~\rm M_{\sun}/(K~km~s^{-1}~pc^2)$ including the correction factor 
of 1.36 for helium. The molecular gas fraction is expressed as $f_{\rm gas} = 
M_{\rm gas}/(M_{\rm gas}+M_*)$. Both quantities correspond to intrinsic values 
corrected from magnification factors.}
\tablefoottext{e}{$4\,\sigma$ upper limit estimated by assuming a full width 
half maximum $\rm FWHM_{CO} = 200~\rm km~s^{-1}$ for the undetected CO(3--2) 
line.}
\tablefoottext{f}{`Observed' CO(1--0) integrated line fluxes and their 
$1\,\sigma$ errors from \citet{riechers10}.}
}
\end{table*}
%

\section{Observations, data reduction, and results}
\label{sect:obs+reduction}


\subsection{Plateau de Bure data}

Table~\ref{tab:observations-log} summarizes the IRAM Plateau de Bure 
Interferometer (PdBI) observations of the galaxies---A68-C0, A68-HLS115, 
MACS0451-arc, and A2218-Mult---selected from the HLS with intrinsic 
$L_{\rm IR} < 4\times 10^{11}~\rm L_{\sun}$. The observations were conducted 
under typical summer conditions between June and September 2011 using 5 
antennas in the compact D-configuration. The compact D-configuration provides 
the highest sensitivity with the smallest spatial resolution ($\sim 5.5''$ at 
our typical tuned frequencies), which is ideal for detection projects. The 
frequencies were tuned to the expected redshifted frequency of the CO(2--1) or 
CO(3--2) line chosen according to the redshift of the targets in order to 
perform all the observations in the 3~mm band. On-source integration times were 
varying between 5 hours and 30 hours in the extreme case of MACS0451-arc. We 
used the WideX correlator that provides a continuous frequency coverage of 
3.6~GHz in dual-polarization with a fixed channel spacing of 1.95~MHz 
resolution. 

Standard data reduction was performed using the IRAM GILDAS software packages 
CLIC and MAP, where bandpass calibration was done using observations of 
calibrators that were best adapted to each of our targets individually. All 
data were mapped with the CLEAN procedure using the ``clark'' deconvolution
algorithm and combined with ``natural'' weighting, resulting in synthesized 
beams listed in Table~\ref{tab:observations-log}. The final noise per beam in 
all our cleaned, weighted images reaches an rms between 0.71 and 0.84~mJy over 
a resolution of 15~MHz ($50~\rm km~s^{-1}$). The velocity-integrated maps of 
the CO emission are obtained by averaging the cleaned, weighted images over the 
spectral channels where emission is detected, and the corresponding spectra are 
obtained by spatially integrating the cleaned, weighted images over the 
$1\,\sigma$ CO detection contours.

%

\subsection{30~m telescope data}

Table~\ref{tab:observations-log} also summarizes the IRAM 30~m telescope 
observations of the HLS source---A68-h7---undertaken on September 3-5, 2011 
under good summer conditions. We used the four single pixel heterodyne EMIR 
receivers, two centred on the E0 band (3~mm) and two on the E1 band (2~mm) 
tuned, respectively, to the redshifted frequencies of the CO(3--2) and CO(4--3) 
lines. The data were recorded using the WILMA autocorrelator providing a 
spectral resolution of 2~MHz. The observations were conducted in 
wobbler-switching mode, with a switching frequency of 0.5~Hz and a symmetrical 
azimuthal wobbler throw of $50''$ to maximize the baseline stability. Series of 
12 ON/OFF subscans of 30 seconds each were performed and calibrations were 
repeated every 6 minutes. A total on-source integration time of 5.5 hours was 
obtained on this target.

The data reduction was completed with the IRAM GILDAS software package CLASS.
The scans obtained with the two receivers tuned on the CO(3--2) line 
were averaged 
using the temporal scan length as weight. The resulting 3~mm spectrum was then 
Hanning smoothed to a resolution of 24~MHz ($65~\rm km~s^{-1}$) 
and reaches an rms noise level of 0.94~mJy.

%

\begin{figure*}
\centering
\raisebox{7pt}{\includegraphics[origin=0 0,angle=-90,height=6.8cm,clip]{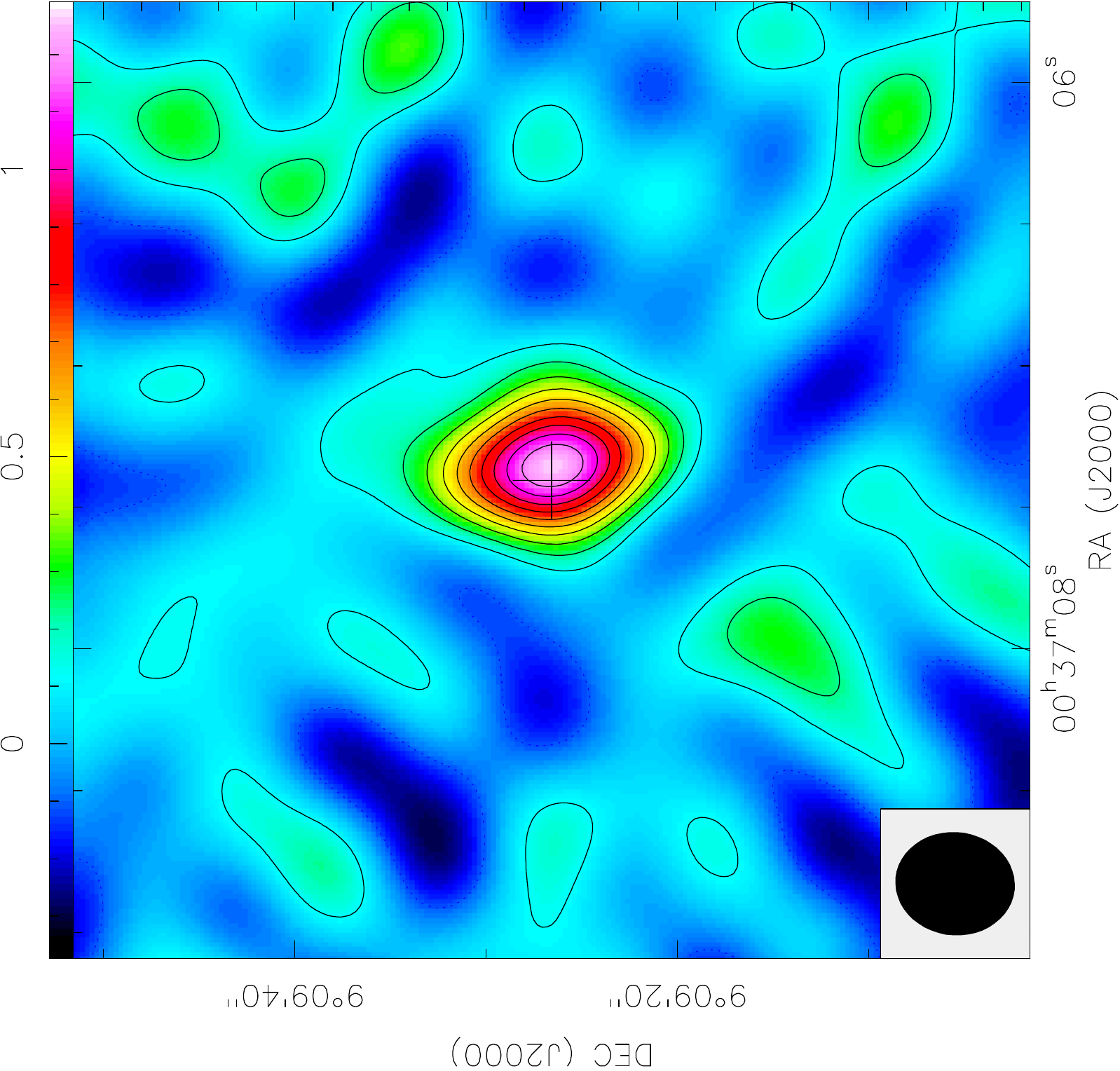}}     \hspace{1cm} \includegraphics[height=7.15cm,clip]{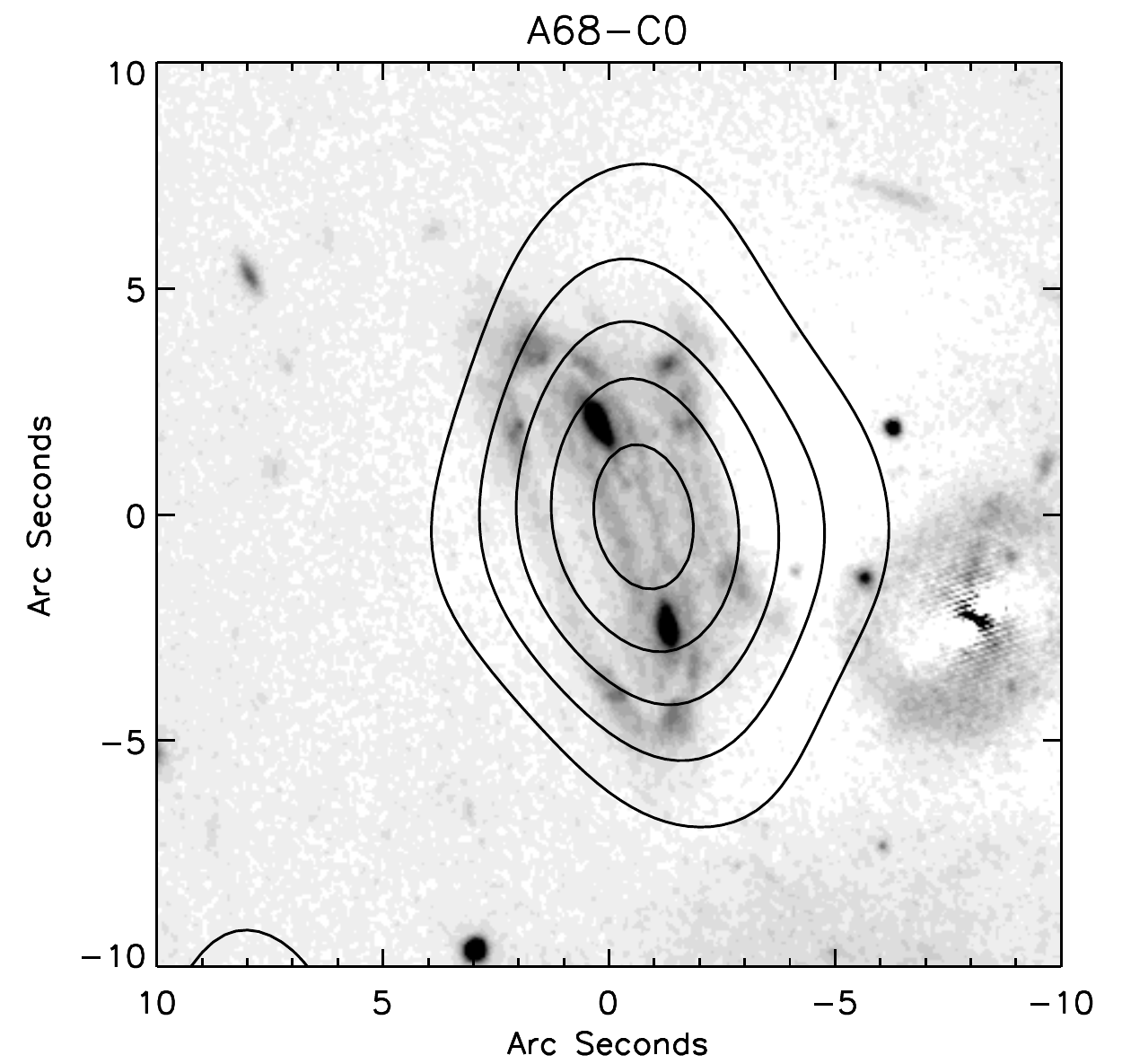}
\raisebox{6pt}{\includegraphics[origin=0 0,angle=-90,height=6.8cm,clip]{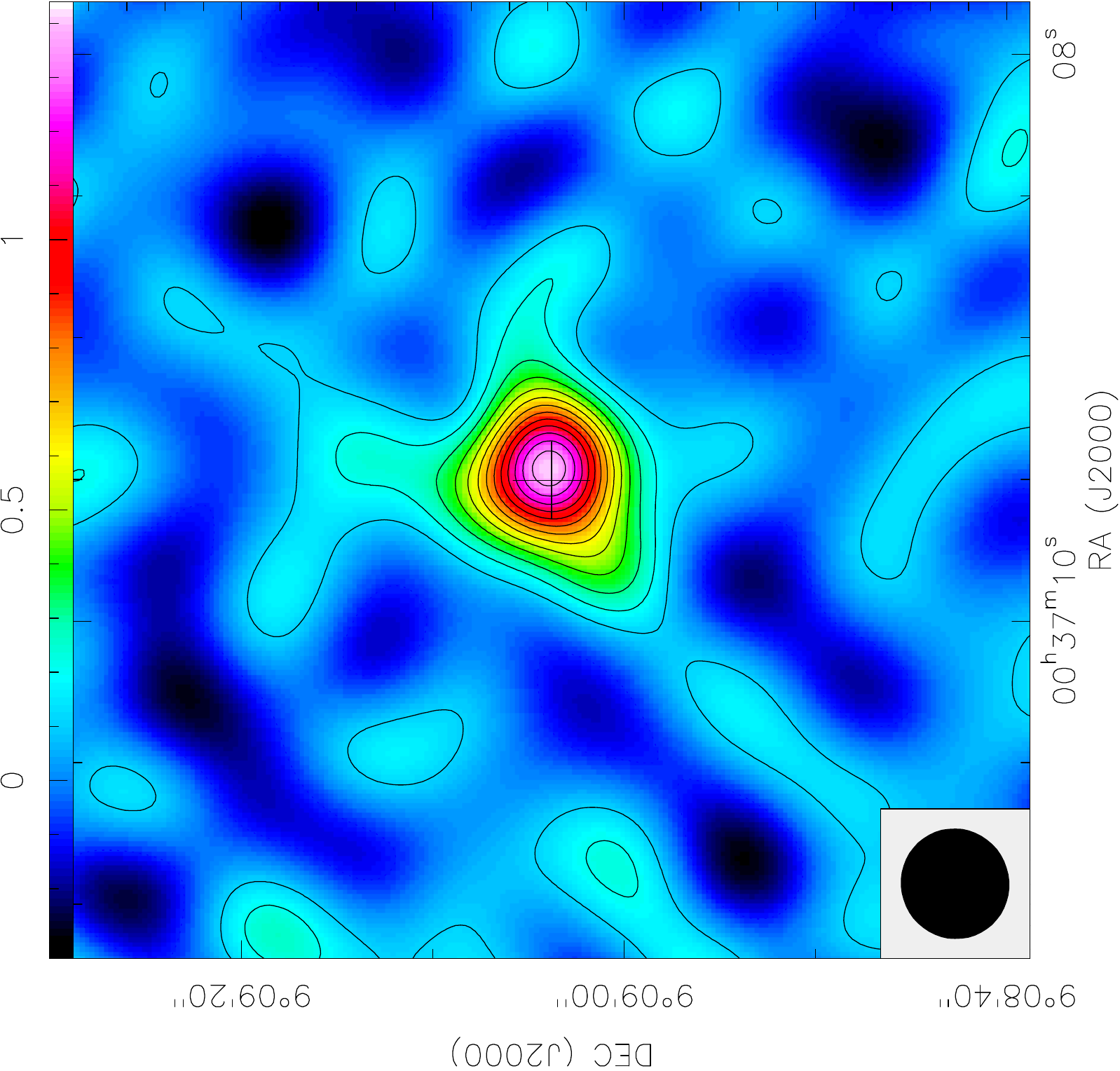}} \hspace{1cm} \includegraphics[height=7.1cm,clip]{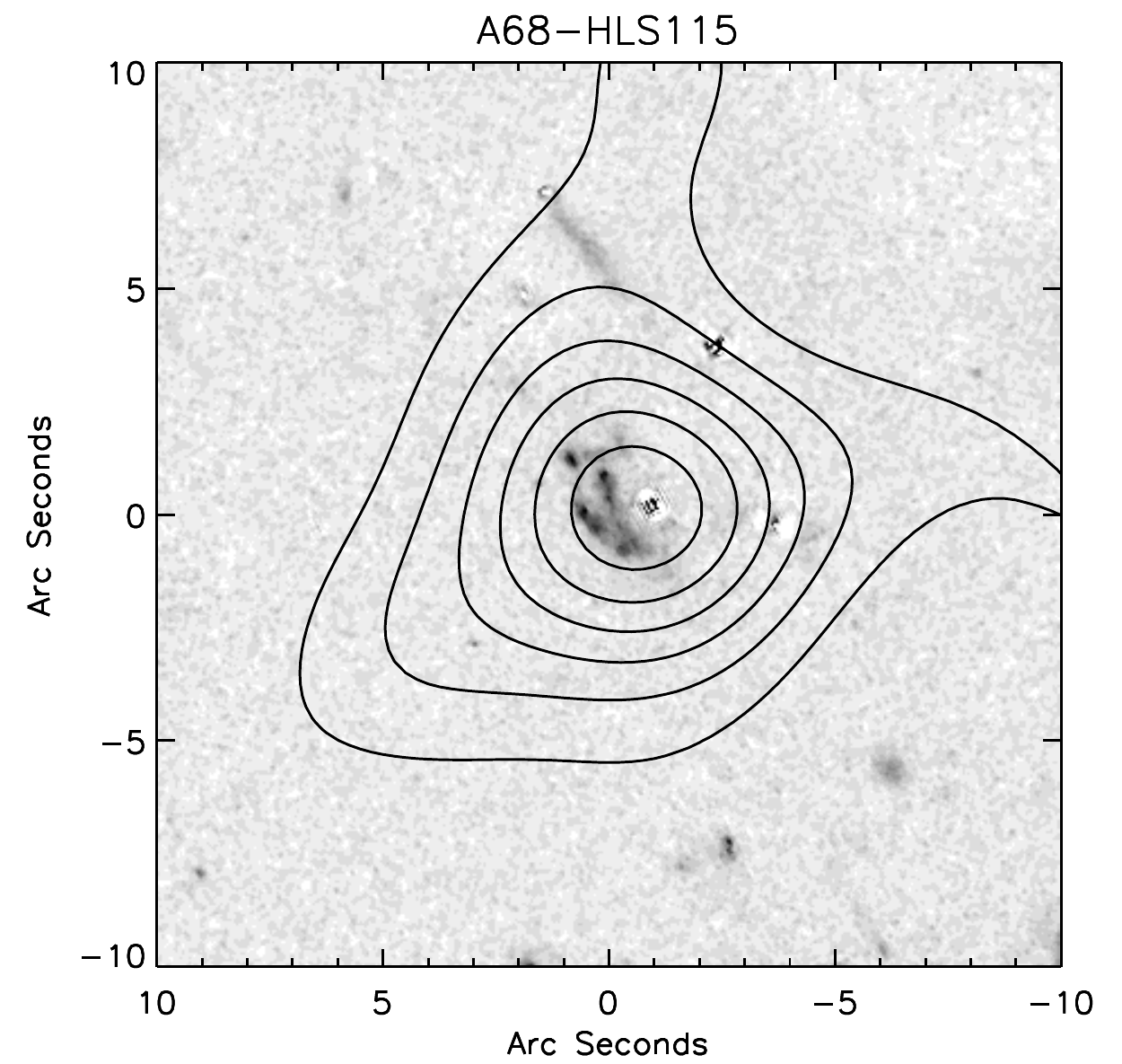}
\raisebox{11pt}{\includegraphics[origin=0 0,angle=-90,height=6.8cm,clip]{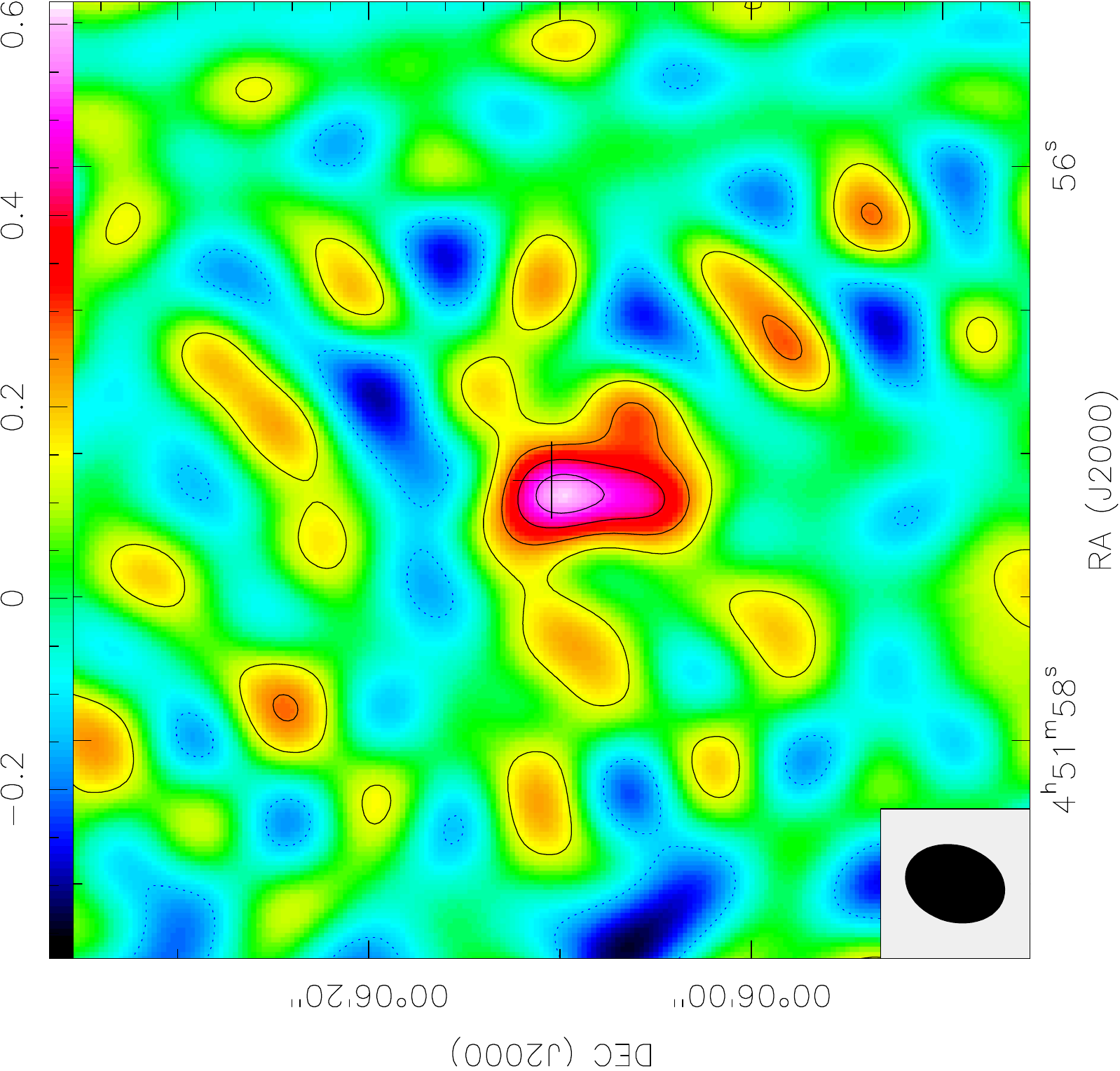}}   \hspace{1cm} \includegraphics[height=7.5cm,clip]{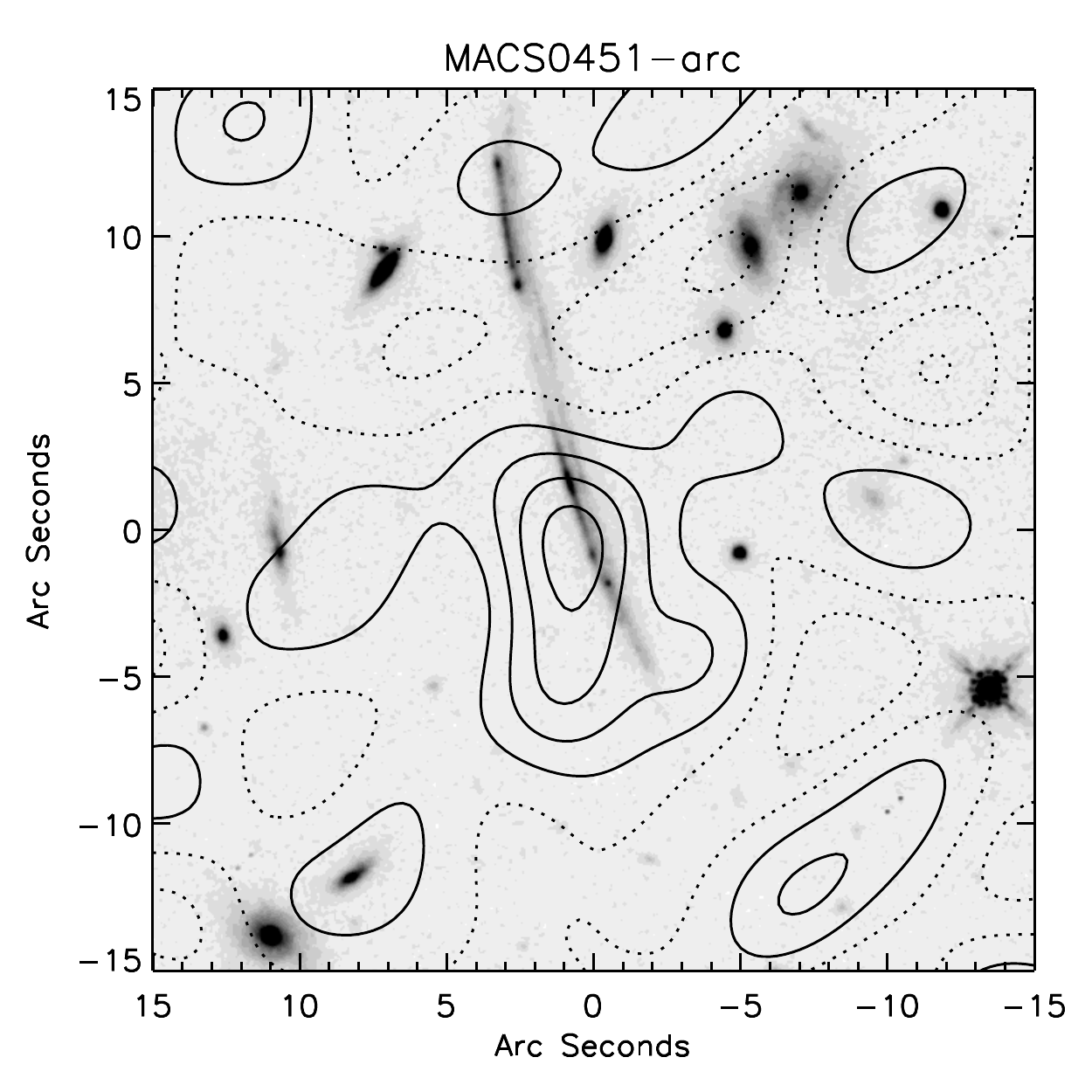}
\caption{From top to bottom, we plot A68-C0, A68-HLS115 and MACS0451-arc. 
{\it Left panels:} Velocity-integrated, cleaned maps of the CO(2--1) or 
CO(3--2) emission detected in three of our low-$L_{\rm IR}$ selected galaxies 
observed with the PdBI. The maps are integrated over the cyan-shaded channels 
shown in Fig.~\ref{fig:COspectra}. Contour levels start at $\pm 1\,\sigma$ and 
are in steps of $1\,\sigma$. The size and orientation of the beam is indicated 
by the black ellipse in the bottom left corner. None of these three CO 
detections is spatially resolved. The cross in each pannel corresponds to the 
coordinates of the optical HST continuum position as listed in 
Table~\ref{tab:observations-log} and is $\pm 2\arcsec$ ($\pm 17~\rm kpc$ at
$z\simeq2$) in size. The coding of the color bar is in units of integrated
flux $\rm Jy~km~s^{-1}$. 
{\it Right panels:} CO(2--1) or CO(3--2) contours overlaid on the HST images in 
the F110W band for A68-C0, the F702W band for A68-HLS115, and the F140W band 
for MACS0451-arc. Contour levels start at $\pm 2\,\sigma$ and are in steps of 
$2\,\sigma$, except for MACS0451-arc where they start at $\pm 1\,\sigma$ and 
are in steps of $1\,\sigma$.}
\label{fig:COmaps}
\end{figure*}
%

\begin{figure*}
\centering
\includegraphics[width=15cm,clip]{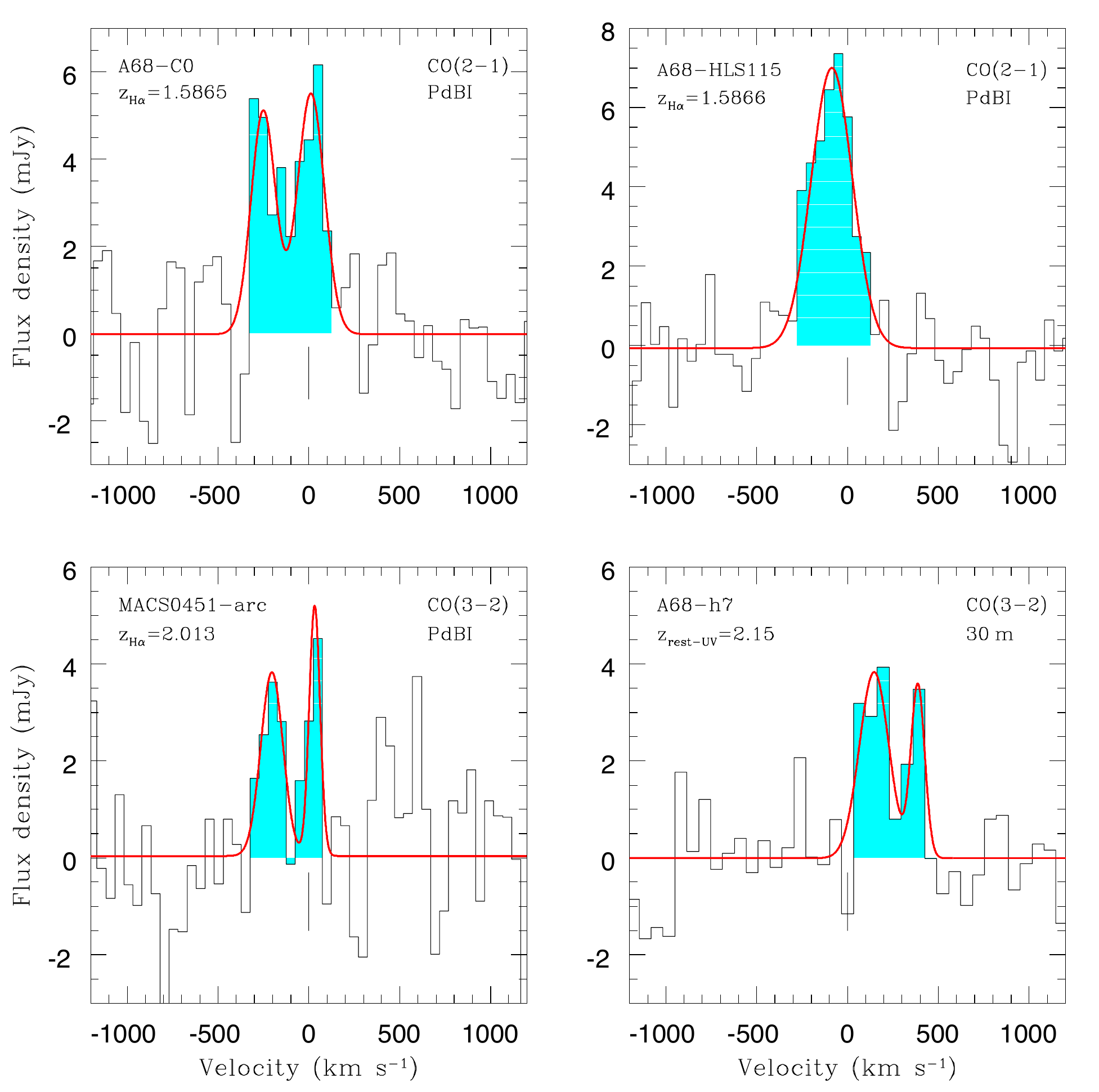} 
\caption{Spectra of the CO(2--1) or CO(3--2) emission line detected in our 
low-$L_{\rm IR}$ selected galaxies, binned in steps of $50~\rm km~s^{-1}$ 
except for A68-h7 where bins have a resolution of $65~\rm km~s^{-1}$. The 
cyan-shaded regions indicate channels where positive emission is detected. 
These channels have been used to derive total CO integrated fluxes and 
velocity-integrated maps shown in Fig.~\ref{fig:COmaps}. The solid red lines 
are the best-fitting single or double Gaussian profiles to the observed CO line 
profiles. 
The zero velocity marked by the vertical bar corresponds to the tuning 
frequency of our targets listed in Table~\ref{tab:observations-log} and
derived from the H$\alpha$ line or rest-frame UV absorption lines redshifts.}
\label{fig:COspectra}
\end{figure*}
%

\subsection{CO results}
\label{sect:results}

The CO emission has been successfully detected at the expected frequency for 
four out of the five low-$L_{\rm IR}$ selected galaxies, it remains undetected 
in A2218-Mult.
CO velocity-integrated maps are shown in the left panels of 
Fig.~\ref{fig:COmaps} for the three positive detections obtained with PdBI. In 
the right panels of Fig.~\ref{fig:COmaps} we show the CO contours overlaid on 
the HST images of the corresponding galaxies. None of these three CO detections 
is spatially resolved, the observed spatial extension of the CO emission in our 
objects is similar to the PdBI beam size. 

The resulting spectra of the CO detections can be found in 
Fig.~\ref{fig:COspectra}. There is no evidence for continuum in any of our 
targets. The properties of the CO emission---velocity centroid ($z_{\rm CO}$), 
the line full width half maximum ($\rm FWHM_{CO}$), and the observed integrated 
line flux ($F_{\rm CO}$)---are evaluated by applying a single or double Gaussian 
fitting procedure on the observed line profiles based on the nonlinear $\chi^2$ 
minimization and the Levenberg-Marquardt algorithm. Errors on the values of 
$z_{\rm CO}$, $\rm FWHM_{CO}$, and $F_{\rm CO}$ are estimated using a Monte 
Carlo approach, whereby the observed spectrum is perturbed with a random 
realization of the error spectrum and refitted. The process is repeated 1000 
times and the error in each quantity is taken to be the standard deviation of 
the values generated by the 1000 Monte Carlo runs. The derived best-fitting 
results are shown in Fig.~\ref{fig:COspectra}. In addition to the measure of CO 
line integrated fluxes from spectra, we have obtained independent integrated 
flux measurements for A68-C0, A68-HLS115 and MACS0451-arc by fitting either a 
circular or elliptical Gaussian model to the two-dimensional CO emission 
observed in velocity-integrated maps. The two respective methods lead to 
consistent integrated CO line fluxes within $1\,\sigma$ errors. 

Table~\ref{tab:COresults} summarizes the measured CO emission properties. To 
convert the measured CO(2--1) and CO(3--2) luminosities to the fundamental 
CO(1--0) luminosity, which at the end gives a measure of the total H$_2$ mass, 
we apply the luminosity correction factors r$_{2,1} = 
L'_{\rm CO(2-1)}/L'_{\rm CO(1-0)} = 0.81\pm 0.20$ and r$_{3,1} = 
L'_{\rm CO(3-2)}/L'_{\rm CO(1-0)} = 0.57\pm 0.15$ to account for the lower 
Rayleigh-Jeans brightness temperature of the 2--1 and 3--2 transitions relative 
to 1--0, as determined in Sect.~\ref{sect:luminosity-corrections}. To estimate 
the uncertainties on the luminosities, we simply propagate the uncertainties on 
the CO line integrated fluxes. The molecular gas masses, $M_{\rm gas}$, and gas 
fractions, $f_{\rm gas} = M_{\rm gas}/(M_{\rm gas}+M_*)$, correspond to values 
computed with the ``Galactic'' CO--H$_2$ conversion factor $X_{\rm CO} = 
2\times 10^{20}~\rm cm^{-2}/(K~km~s^{-1})$, or $\alpha = 
4.36~\rm M_{\sun}/(K~km~s^{-1}~pc^2)$ which includes the correction factor of 
1.36 for helium. In the case of the CO(3--2) non-detection in A2218-Mult, we 
provide the $4\,\sigma$ upper limits computed from the rms noise achieved in 
the PdBI observations of this galaxy and assume a typical full width half 
maximum $\rm FWHM_{CO} = 200~\rm km~s^{-1}$. The CO properties and inferred 
kinematics of the individual low-$L_{\rm IR}$ selected targets are described in 
Appendix~\ref{sect:appendix}.

%

\section{CO luminosity correction factors}
\label{sect:luminosity-corrections}

The measure of the molecular gas mass requires the luminosity of the 
fundamental CO(1--0) line, $L'_{\rm CO(1-0)}$. Nevertheless, as soon as we are 
interested in objects at $z>0.4$, accessing the fundamental CO(1--0) line at 
115.27~GHz becomes impossible with mm/sub-mm receivers, and CO(1--0) must be 
replaced by rotationally excited $J>1$ CO transitions. Consequently, 
corrections for the ratio of the intrinsic Rayleigh-Jeans brightness 
temperatures in the 1--0 line to that in the rotationally excited line become
necessary and need to be determined to access the fundamental 
$L'_{\rm CO(1-0)}$. The corresponding CO luminosity correction factors vary 
with the $J$-transition of the CO line considered, the optical thickness, the 
thermal excitation and gas density of the medium, and hence the galactic type 
and redshift \citep{papadopoulos12,lagos12,narayanan14}.

\begin{figure}
\centering
\includegraphics[width=9cm,clip]{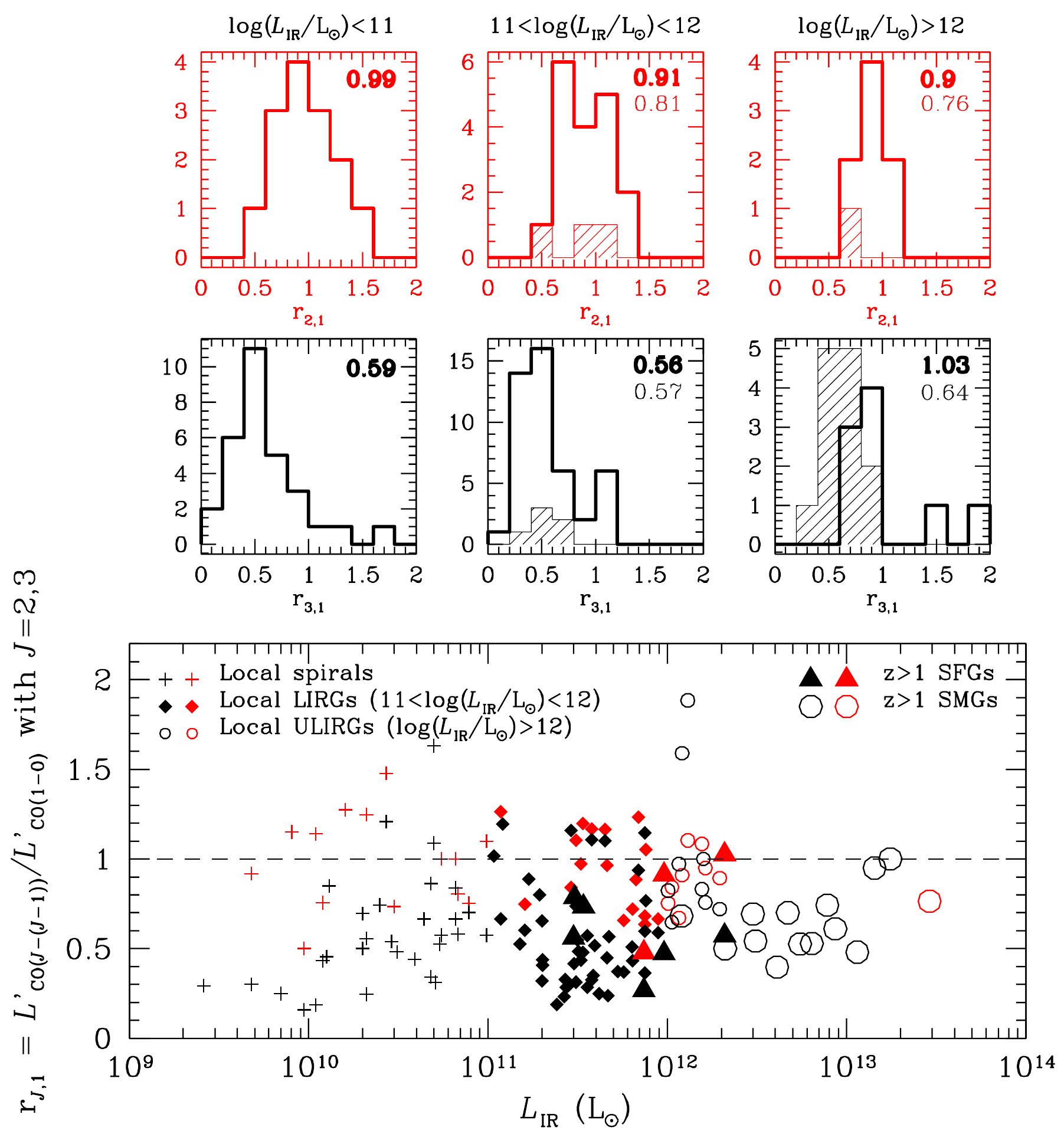} 
\caption{{\it Bottom panel:} Luminosity correction factors, r$_{2,1} = 
L'_{\rm CO(2-1)}/L'_{\rm CO(1-0)}$ (in red) and r$_{3,1} = 
L'_{\rm CO(3-2)}/L'_{\rm CO(1-0)}$ (in black), plotted as a function of the IR
luminosity for galaxies from our comparison sample (see 
Sect.~\ref{sect:literature}). 
{\it Upper, middle panels:} r$_{2,1}$ and r$_{3,1}$ distributions, 
respectively, of $z=0$ galaxies (open histograms) and $z>1$ galaxies (hatched 
histograms) plotted for three $L_{\rm IR}$ intervals: from left to right 
$L_{\rm IR} < 10^{11}~\rm L_{\sun}$ (the `spiral' regime), 
$10^{11}~{\rm L}_{\sun} < L_{\rm IR} < 10^{12}~{\rm L}_{\sun}$ (the `LIRG' 
regime), and $L_{\rm IR} > 10^{12}~\rm L_{\sun}$ (the `ULIRG' regime). The 
numbers in the upper right corner of each panel are the mean values of the 
r$_{J,1}$ distributions (``thick'' numbers refer to $z=0$ galaxies and ``thin'' 
numbers to $z>1$ galaxies).}
\label{fig:luminosity-corrections}
\end{figure}

Large efforts are done to get measurements of CO(1--0) and high-$J$ CO lines 
within the same objects to determine the CO spectral line energy distribution 
(SLED) of local and high-redshift galaxies \citep[e.g.,][]{weiss07,
dannerbauer09,papadopoulos10,daddi14}. The comparison sample of CO-detected 
galaxies from the literature (Sect.~\ref{sect:literature}), particularly 
exhaustive at high redshift, offers a unique opportunity to obtain a 
comprehensive view on the CO luminosity correction factors, defined as 
r$_{J,1} = L'_{{\rm CO}(J-(J-1))}/L'_{{\rm CO}(1-0)}$, for the $J=2$ and $J=3$ 
CO rotational transitions. In Fig.~\ref{fig:luminosity-corrections} we plot the 
r$_{2,1}$ and r$_{3,1}$ correction factors as a function of the IR luminosity 
for galaxies from the comparison sample. The histograms show, respectively, 
r$_{2,1}$ and r$_{3,1}$ separately for $z=0$ galaxies (open histograms) and 
$z>1$ galaxies (hatched histograms), and for three $L_{\rm IR}$ intervals: 
$L_{\rm IR} < 10^{11}~\rm L_{\sun}$ (the `spiral' regime), 
$10^{11}~{\rm L}_{\sun} < L_{\rm IR} < 10^{12}~{\rm L}_{\sun}$ (the `LIRG' 
regime), and $L_{\rm IR} > 10^{12}~\rm L_{\sun}$ (the `ULIRG' regime). 

The mean $\langle \rm r_{2,1}\rangle > 0.9$ and $\langle \rm r_{3,1}\rangle 
\simeq 0.55-0.6$ values of $z=0$ galaxies within the `spiral' and `LIRG' 
regimes reveal relatively well-excited CO(2--1) and CO(3--2) lines. The 
observed luminosity correction factors are in line with the $1\gtrsim 
\rm r_{2,1} \gtrsim r_{3,1} \gtrsim r_{4,1} \gtrsim ...$ sequence which is 
expected if the average state is dominated by one warm optically thick and 
thermally excited phase \citep{papadopoulos12}. On the other hand, $z=0$ 
galaxies within the `ULIRG' regime show a predominance of a high-excitation 
phase with $\langle \rm r_{3,1}\rangle \gtrsim \langle r_{2,1}\rangle$ and a 
mean $\langle \rm r_{3,1}\rangle$ close to unity. This suggests highly-excited 
media and optically thin CO SLEDs in the higher $L_{\rm IR}$ $z=0$ galaxies. 
This view is supported by the models of \citet{lagos12} that predict flatter CO 
SLEDs for the brightest IR galaxies than for the fainter IR counterparts at 
$z=0$.

For $z>1$ galaxies the available r$_{2,1}$ and r$_{3,1}$ statistics is still 
small, except for r$_{3,1}$ within the `ULIRG' regime. Both in the `LIRG' and 
`ULIRG' regimes, the mean $\langle \rm r_{2,1}\rangle \simeq 0.8$ and 
$\langle \rm r_{3,1}\rangle \simeq 0.6$ values point toward a slightly 
lower-excitation medium in high-redshift galaxies compared to $z=0$ galaxies, 
where the higher-$J$ CO lines are fainter as the subthermal excitation sets in 
more rapidly. The r$_{3,1}$ correction factors within the `ULIRG' regime also 
show no significant sign of evolution toward higher excitation in the higher 
$L_{\rm IR}$ $z>1$ galaxies, unlike what is observed in $z=0$ galaxies. These 
results again are in a nice agreement with the models of \citet{lagos12} that 
predict (1)~shallower CO SLEDs for high-redshift galaxies compared to their 
$z=0$ counterparts at a fixed IR luminosity, and (2)~smaller differences in the 
CO SLEDs of faint- and bright-IR galaxies at $z>1$ than for $z=0$ galaxies. 
This is due to the increasing average gas kinetic temperature in molecular 
clouds with redshift. Given the fact that the observed high-$J$ CO transitions 
of our sample of $z>1$ galaxies are on average not highly excited, we may 
consider the ``Galactic'' CO--H$_2$ conversion factor as a sensible assumption 
for both SFGs and SMGs at $z>1$. 

%

%

Throughout the paper, we adopt the following luminosity correction factors 
for SFGs (including our low-$L_{\rm IR}$ selected galaxies) at $z>1$: 
r$_{2,1} = L'_{{\rm CO}(2-1)}/L'_{{\rm CO}(1-0)} = 0.81\pm 0.20$ and
r$_{3,1} = L'_{{\rm CO}(3-2)}/L'_{{\rm CO}(1-0)} = 0.57\pm 0.15$. 
These are the means of observed luminosity correction factors as measured at 
high redshift, instead of extrapolations from partial CO SLEDs or CO luminosity 
correction factors derived for local galaxies. Compared to the canonical value 
r$_{3,1} = 0.5$ assumed in the literature for $z>1$ SFGs 
\citep{tacconi13,saintonge13}, our r$_{3,1}$ value is 14\% higher, but well 
within $1\,\sigma$. The difference between our r$_{2,1}$ value and the values 
found in the literature for $z>1$ SFGs is smaller: r$_{2,1} = 0.84$ in 
\citet{daddi10a} and r$_{2,1} = 0.75$ in \citet{magnelli12}. The 
luminosity correction factors we derived for both high-redshift SFGs and SMGs 
also well agree with the values compiled by \citet{carilli13}, although based 
on other prescriptions for the $z>1$ SFGs.

%

\begin{figure}
\centering
\includegraphics[width=9cm,clip]{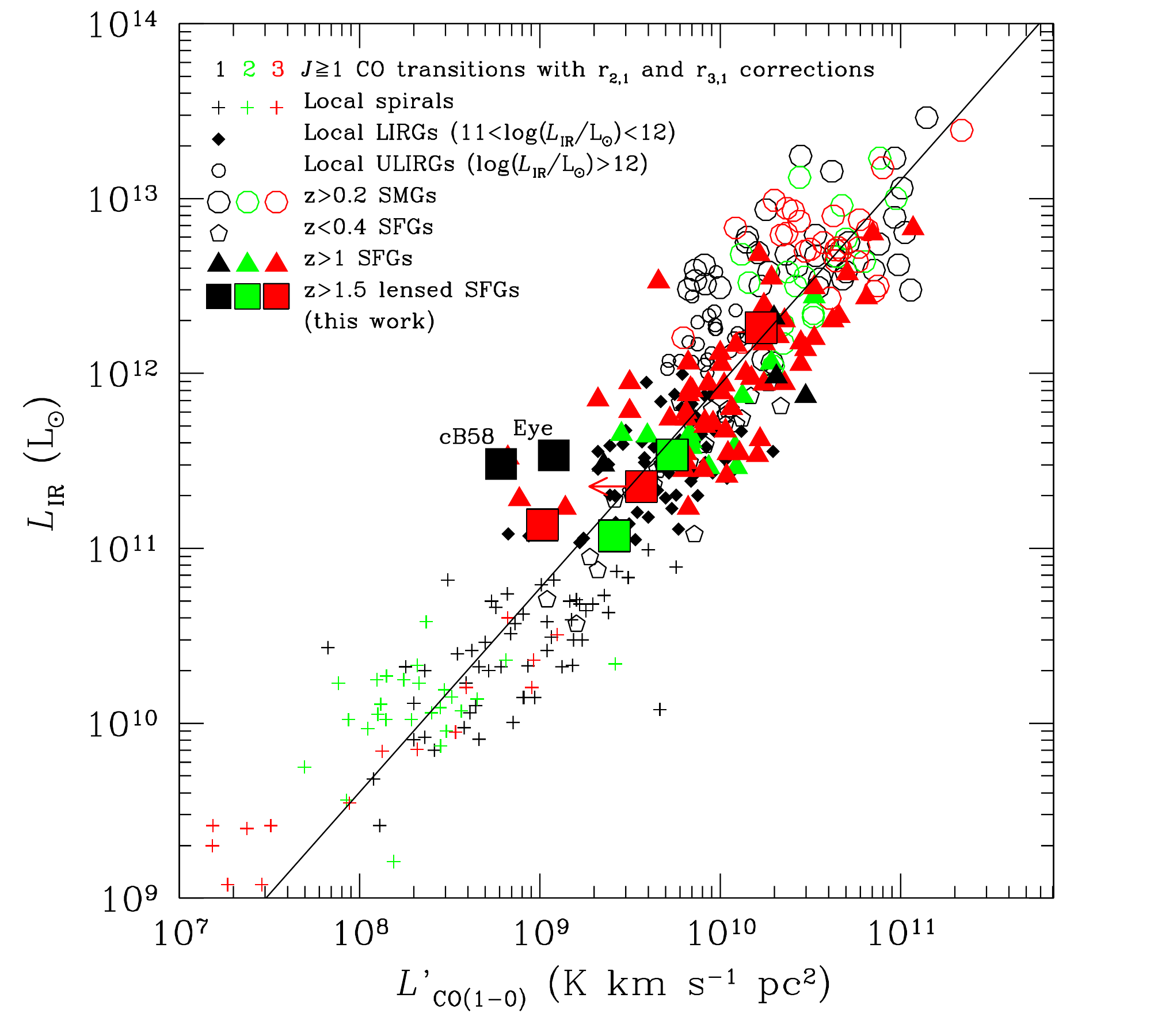}    
\caption{IR luminosities as a function of CO(1--0) luminosities of our 
low-$L_{\rm IR}$ selected SFGs (squares) compared to our compilation of
galaxies with CO measurements from the literature (see 
Sect.~\ref{sect:literature}). The color-coding of the various symbols refers 
to the $J=1$, 2, and 3 CO transitions used to infer $L'_{\rm CO(1-0)}$ by 
applying the CO luminosity correction factors determined in 
Sect.~\ref{sect:luminosity-corrections}. The best-fitting bisector linear 
relation to the entire sample of galaxies, $\log (L_{\rm IR}) = (1.17\pm 0.03) 
\log (L'_{\rm CO(1-0)}) + (0.28\pm 0.23)$, is plotted as a solid line. The $z>1$ SFGs have on average a scatter in $L_{\rm IR}$ as large as 1~dex at a 
given value of $L'_{\rm CO(1-0)}$ and a $1\,\sigma$ dispersion of 0.38~dex in 
the $y$-direction about the best-fitting $L_{\rm IR}$--$L'_{\rm CO(1-0)}$ 
relation.}
\label{fig:LIR-LCO}
\end{figure}
%

\section{Low-$L_{\rm IR}$ selected galaxies in the general context of galaxies with CO measurements}
\label{sect:discussion}

We discuss the overall physical properties derived from the new CO measurements 
achieved in our low-$L_{\rm IR}$ selected galaxies. We combine and compare 
their CO luminosities, IR luminosities, star formation efficiencies, and 
molecular gas depletion timescales
to our compilation of CO-detected galaxies from the literature 
(Sect.~\ref{sect:literature}). The goal is to investigate whether 
the extended dynamical range toward lower star formation rates $\rm SFR < 
40~M_{\sun}~yr^{-1}$ and smaller stellar masses $M_* < 2.5\times 
10^{10}~\rm M_{\sun}$ shows evidence for new trends and correlations.

%

\begin{figure}
\centering
\includegraphics[width=9cm,clip]{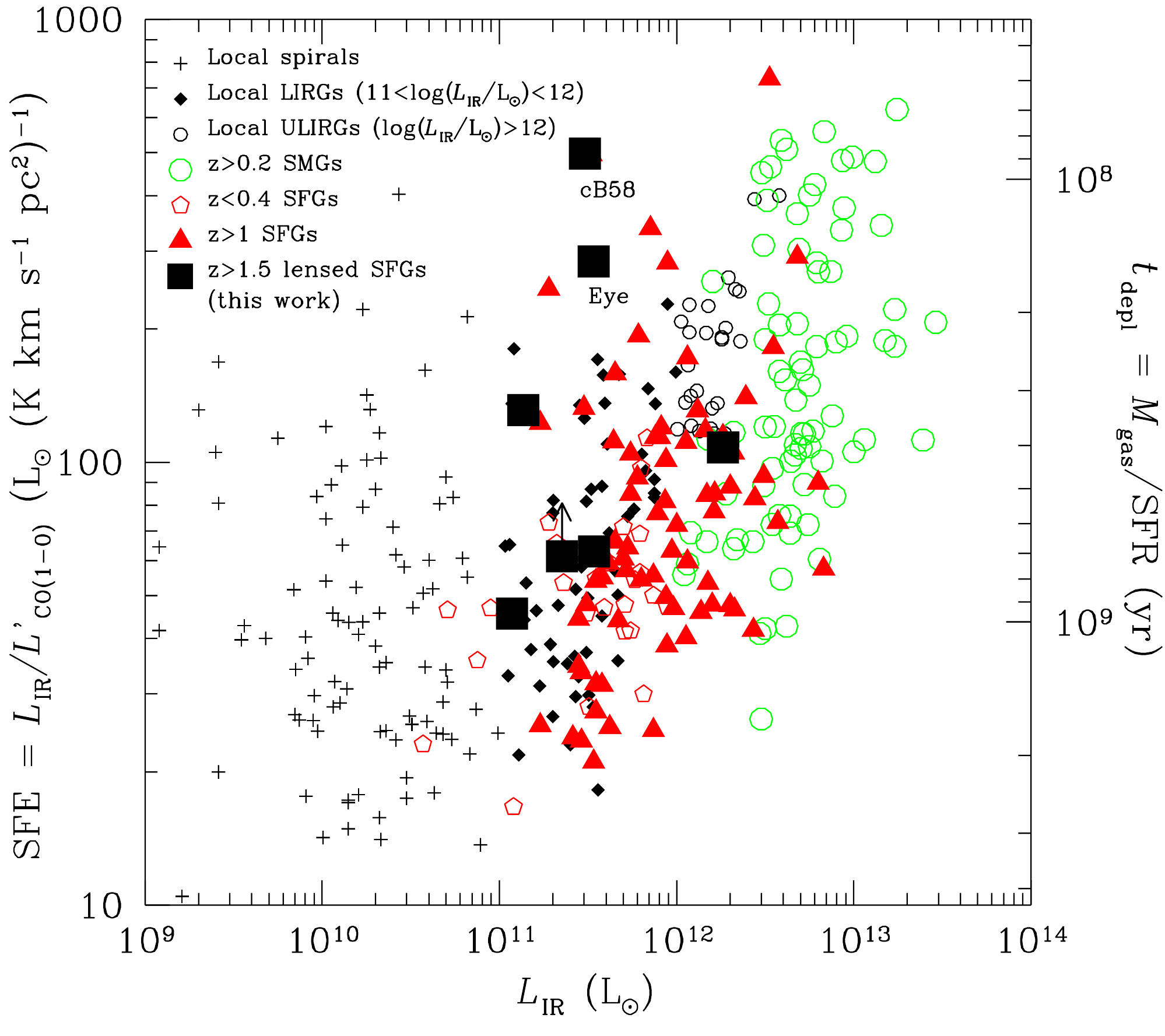}
\caption{Star formation efficiencies, ${\rm SFE} = 
L_{\rm IR}/L'_{\rm CO(1-0)}$ (left y-axis), or molecular gas depletion 
timescales, $t_{\rm depl} = M_{\rm gas}/{\rm SFR}$, assuming a ``Galactic'' 
CO--H$_2$ conversion factor for all galaxies (right y-axis), as a function of 
IR luminosities of our low-$L_{\rm IR}$ selected SFGs (squares) compared to our 
compilation of galaxies with CO measurements from the literature (see 
Sect.~\ref{sect:literature}). This illustrates in another way the large 
dispersion about the $L_{\rm IR}$--$L'_{\rm CO(1-0)}$ relation observed in 
Fig.~\ref{fig:LIR-LCO}.} 
\label{fig:SFE-LIR}
\end{figure}
%

\subsection{CO luminosity to IR luminosity relation}

It is now well established that there is a 
relation between $L_{\rm IR}$ and $L'_{\rm CO(1-0)}$ 
\citep[see the review by][]{carilli13} and this despite the uncertainties 
on $L'_{\rm CO(1-0)}$ mostly inferred from higher $J>1$ CO rotational 
transitions and on $L_{\rm IR}$ often derived from various star formation 
tracers (UV, H$\alpha$, mid-IR, sub-mm, or radio) instead of direct far-IR 
photometry. The first high-redshift datasets showed evidence for a `bimodal' 
$L_{\rm IR}$--$L'_{\rm CO(1-0)}$ behaviour between the so-called `sequence of 
disks' or `MS star-forming galaxies' which includes local spirals plus 
high-redshift SFGs, and the `sequence of starbursts' or `mergers' which 
includes local ULIRGs plus high-redshift SMGs/ULIRGs \citep[e.g.,][]
{daddi10b,genzel10,sargent14}. 

The results for the most recent compilation of CO-detected galaxies from the 
literature (Sect.~\ref{sect:literature}), including our sample of 
low-$L_{\rm IR}$ selected SFGs, are shown in Fig.~\ref{fig:LIR-LCO}. The 
color-coding allows to track the original $J=1$, 2, or 3 CO transition used to 
compute the CO(1--0) luminosity. Our new sources populate a new regime of 
low IR luminosities, $L_{\rm IR} < 4\times 10^{11}~\rm L_{\sun}$, and low 
CO(1--0) luminosities.
They overlap with the domain of local LIRGs, their $L_{\rm IR}$ counterparts at 
$z=0$, and perfectly extend the observed $L_{\rm IR}$ versus $L'_{\rm CO(1-0)}$ 
distribution of previously studied $z>1$ SFGs with higher IR luminosities. They 
show an increased scatter toward lower $L'_{\rm CO(1-0)}$ at given values of 
$L_{\rm IR}$. This is a general trend which is highlighted by the new 
compilation of CO-detected SFGs at $z>1$, in contrast to what was previously 
observed, where all the high-redshift SFGs' CO luminosities were characterized 
by higher values at given $L_{\rm IR}$ compared to the local ULIRG and 
high-redshift SMG/ULIRG populations \citep{daddi10b,genzel10,carilli13}. 

As a result, a single linear relation now best reproduces the distribution 
within the $L_{\rm IR}$--$L'_{\rm CO(1-0)}$ log plane of the various 
CO-detected galaxies spanning 5 orders of magnitude in the IR luminosity, 
having redshifts between $z=0$ and 5.3, and sampling diverse galaxy types from 
main-sequence galaxies to mergers. The best-fitting bisector linear relation 
has a slope of $1.17\pm 0.03$. The `bimodal' behaviour is clearly smeared out: 
the offset between the different galaxy populations 
is embedded in the large dispersion of data points about the linear 
$L_{\rm IR}$--$L'_{\rm CO(1-0)}$ relation in log space. Indeed, the $z>1$ SFGs 
only\footnote{Those include our sample of strongly-lensed low-$L_{\rm IR}$ 
galaxies, the BzK galaxies from \citet{daddi10a}, the PHIBSS (EGS, BM/BX) 
galaxies from \citet{tacconi10,tacconi13} and \citet{genzel10}, the PEP 
galaxies from \citet{magnelli12}, and strongly-lensed SFGs from 
\citet{saintonge13} and others (see Sect.~\ref{sect:literature}).} 
show a scatter in $L_{\rm IR}$ as large as 1~dex at a given value of 
$L'_{\rm CO(1-0)}$ and a $1\,\sigma$ dispersion of 0.3~dex in the $y$-direction 
about the best-fit of their $L_{\rm IR}$--$L'_{\rm CO(1-0)}$ relation. The 
offset of 0.46--0.5~dex in the normalization between the `sequence of disks' 
and the `sequence of starbursts' reported by \citet{daddi10a} and 
\citet{sargent14} hence is within $1.5\,\sigma$ dispersion of the current 
sample of $z>1$ SFGs. 

%

\subsection{Star formation efficiency and gas depletion timescale: What determines their behaviour at high redshift\,?}
\label{sect:SFE,tdepl}

Another way to represent the CO luminosity--IR luminosity relation, that may 
help understand the dispersion in this relation, is through the star formation 
efficiency, SFE, defined as ${\rm SFE} = L_{\rm IR}/L'_{\rm CO(1-0)}$, or 
equivalently ${\rm SFE = SFR}/M_{\rm gas}$ 
by assuming for all galaxies a ``Galactic'' CO--H$_2$ conversion factor $\alpha 
= 4.36~\rm M_{\sun}/(K~km~s^{-1}~pc^2)$ which includes the correction factor of 
1.36 for helium. 
The star formation efficiency is intimately linked to the molecular gas 
depletion timescale defined as the inverse of the SFE, $t_{\rm depl} = 
M_{\rm gas}/{\rm SFR}$. This physical parameter describes how long each galaxy 
could sustain star formation at the current rate before running out of fuel, 
assuming that the gas reservoir is not replenished.

In Fig.~\ref{fig:SFE-LIR} we plot the star formation efficiency and the 
molecular gas depletion timescale as a function of the IR luminosity for our 
compilation of galaxies with CO measurements from the literature 
(Sect.~\ref{sect:literature}), including our sample of low-$L_{\rm IR}$ 
selected SFGs. This shows: (1)~High-redshift SFGs and SMGs have very comparable 
star formation efficiency distributions and spreads. Their respective mean SFE 
(very similar to their respective median SFE) at $z>1$ are 
$\langle \log ({\rm SFE})\rangle_{\rm SFGs} = 
1.89\pm 0.33~\rm L_{\sun}/(K~km~s^{-1}~pc^2)$ ($\langle t_{\rm depl}
\rangle_{\rm SFGs} = 560~\rm Myr$) and $\langle \log ({\rm SFE})
\rangle_{\rm SMGs/ULIRGs} = 2.14\pm 0.30~\rm L_{\sun}/(K~km~s^{-1}~pc^2)$ 
($\langle t_{\rm depl}\rangle_{\rm SMGs} = 315~\rm Myr$); they are the same 
within $1\,\sigma$ dispersion. As a result, $z>1$ SFGs have not their SFE 
confined to the low values of local spirals and $z>1$ SMGs/ULIRGs do not 
systematically show an excess in $L_{\rm IR}/L'_{\rm CO(1-0)}$, in contrast to
the reported `bimodality' \citep{daddi10a,daddi10b,genzel10,sargent14}. 
(2)~The IR luminosity hence appears as a weak tracer of the star formation 
efficiency, although a correlation between $L_{\rm IR}$ (or SFR) and SFE 
remains, as the slope of the $L_{\rm IR}$ versus $L'_{\rm CO(1-0)}$ relation in 
log space is not exactly equal to unity (it is equal to 1.17, see 
Fig.~\ref{fig:LIR-LCO}).

\smallskip
\noindent In what follows, we focus on $z>1$ SFGs with the aim to try to 
understand what drives their large spread in SFE and what differentiates 
galaxies with high SFE from those with low SFE. Various physical parameters may 
induce differences in the star formation efficiencies, or respectively 
molecular gas depletion timescales. 
We investigate below the link between the SFE (or $t_{\rm depl}$) and the 
following physical parameters: specific star formation rate, stellar mass, 
redshift, offset from the main-sequence, and compactness of the starburst.

%

\begin{figure}
\centering
\includegraphics[width=9cm,clip]{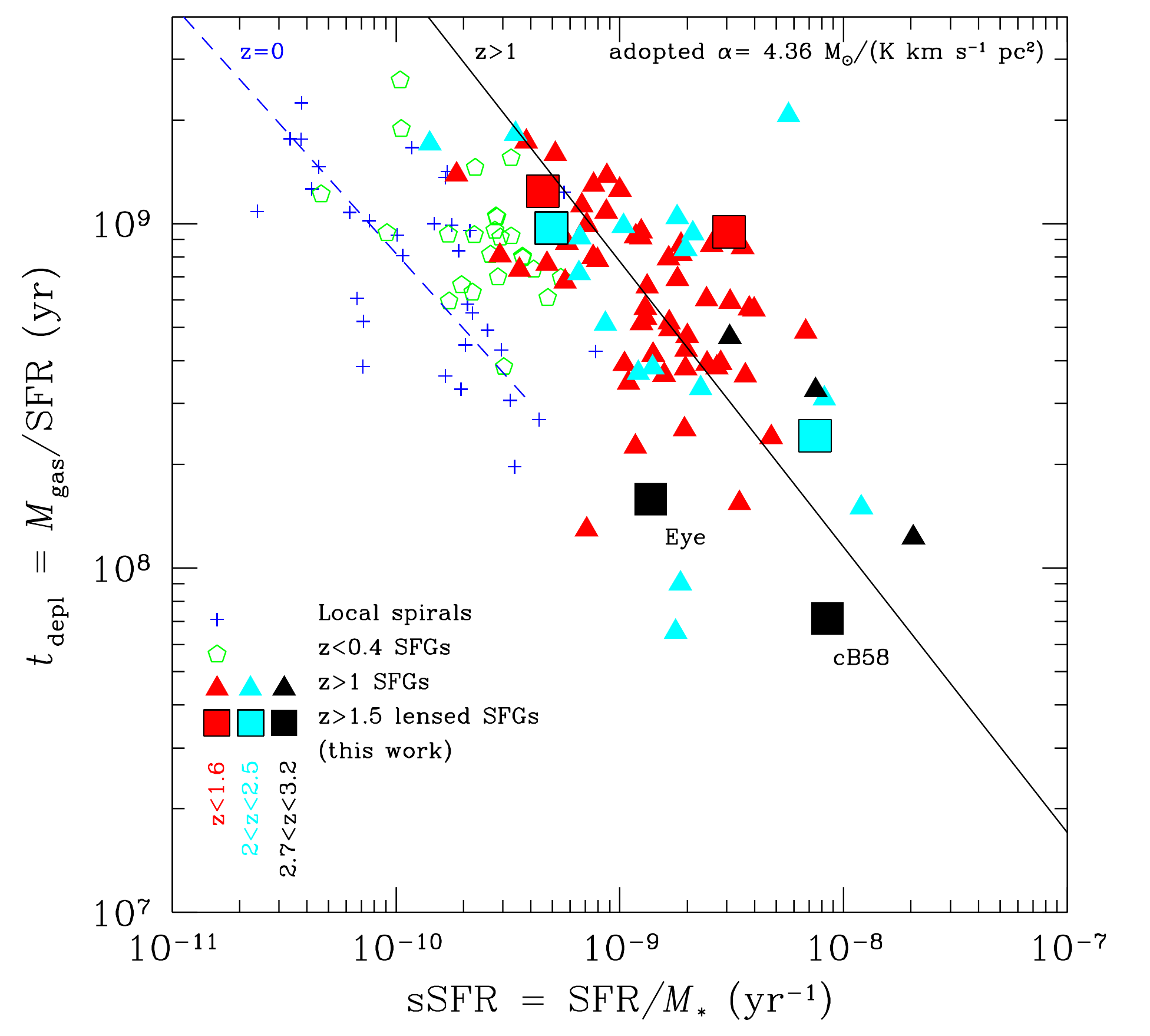} 
\caption{Molecular gas depletion timescales, $t_{\rm depl} = 
M_{\rm gas}/{\rm SFR}$, as a function of specific star formation rates, 
${\rm sSFR = SFR}/M_*$, plotted for our low-$L_{\rm IR}$ selected SFGs 
(squares) and our comparison sample of SFGs and local spirals with CO 
measurements from the literature (see Sect.~\ref{sect:literature}). The 
molecular gas depletion timescales are computed assuming a ``Galactic'' 
CO--H$_2$ conversion factor for all galaxies. $t_{\rm depl}$ 
shows a strong dependence on sSFR. The best-fitting bisector linear relation to the $z=0$ COLD GASS sample from \citet{saintonge11} is plotted as a dashed blue 
line. $z>1$ SFGs are displaced with respect to the local relation, their 
best-fitting bisector linear relation, $\log (t_{\rm depl}) = 
(-0.83\pm 0.08) \log ({\rm sSFR}) + (1.43\pm 0.70)$, is plotted as a solid 
black line. The color-coding of SFGs (our sample plus $z>1$ SFGs from the 
literature) refers to three redshift intervals: $\langle z_{1.2}\rangle = 
[1,1.6]$ (red), $\langle z_{2.2}\rangle = [2,2.5]$ (cyan), and 
$\langle z_{3.0}\rangle = [2.7,3.2]$ (black).}
\label{fig:tdepl-sSFR}
\end{figure}
%

\subsubsection{Specific star formation rate}
\label{sect:tdepl-sSFR}

A possible physical parameter which may control the SFE or $t_{\rm depl}$ of 
galaxies is the specific star formation rate, ${\rm sSFR = SFR}/M_*$, i.e.\ the 
SFR normalized by the stellar mass which gives the timescale of formation of 
all the stellar mass in a galaxy at the given present SFR. In their study of 
$z=0$ massive star-forming galaxies with $10.0 < \log (M_*/{\rm M_{\sun}}) < 
11.5$ from the COLD GASS survey, \citet{saintonge11} found the strongest 
dependence of $t_{\rm depl}$ to be precisely on the sSFR. 
Their $t_{\rm depl}$--sSFR anti-correlation in fact highlights comparable 
timescales for the gas consumption and the stellar mass formation. 
They interpret it 
as the result of the increased dynamical stirring as one proceeds to more 
strongly star-forming galaxies, since in local starburst galaxies the very 
high values of sSFR are achieved because dynamical disturbances act to compress 
the available interstellar medium atomic gas and create giant molecular clouds 
and stars.

We also find a good anti-correlation between $t_{\rm depl}$ and sSFR for $z>1$ 
SFGs as shown in Fig.~\ref{fig:tdepl-sSFR}. We overplot the best-fit relation 
obtained for the COLD GASS sample and observe that $z>1$ SFGs are generally 
distributed above the $z=0$ relation and have longer $t_{\rm depl}$ by about 
0.75~dex than local galaxies with the same sSFR (see the best-fitting relation 
obtained for $z>1$ SFGs). The observed $t_{\rm depl}$--sSFR anti-correlation 
and its shift with redshift are in agreement with the ``bathtub'' model 
predictions \citep{bouche10} and confirm the findings by \citet{saintonge11} 
and \citet{combes13}. They attribute this displacement of $z=1-3$ star-forming 
galaxies with respect to the local galaxy population 
to their significantly larger molecular gas fractions that afford longer 
molecular gas depletion times at a given value of sSFR. We do confirm in 
Sect.~\ref{sect:fgas} the increased available molecular gas reservoir as one 
proceeds toward higher redshifts, as well as, on average, as one proceeds 
toward more strongly star-forming galaxies (higher sSFR), which certainly 
triggers the $t_{\rm depl}$--sSFR anti-correlation observed at $z>1$. 
Nevertheless, while a clear evolution of the $t_{\rm depl}$-sSFR relation 
appears to be present between local and $z>1$ galaxies, no clear trend seems to 
emerge between the three redshift bins $\langle z_{1.2}\rangle = [1,1.6]$, 
$\langle z_{2.2}\rangle = [2,2.5]$, and $\langle z_{3.0}\rangle = [2.7,3.2]$. 
This agrees with the steep increase of the molecular gas fraction between $z=0$ 
and $z\sim 1$ followed by a quasi non-evolution toward higher redshifts (but 
with a large scatter) as observed in Fig.~\ref{fig:fgas-z}.

In addition to the shift of $z>1$ SFGs with respect to local galaxies 
due to their larger molecular gas fractions, 
the specific star formation rates of galaxies in the local Universe are sealed 
on low values because of the accumulation of more and more old stars in their 
bulge at $z=0$. These old stars have an important weight in the total stellar
mass budget, but none on the current SFR and thus imply low sSFR values. As a 
result, this accentuates the displacement between the respective distributions 
of local and high-redshift galaxies in the $t_{\rm depl}$--sSFR plane. 


%

\subsubsection{Stellar mass}
\label{sect:tdepl-Mstars}

Another physical parameter which may trigger differences in SFE or 
$t_{\rm depl}$ from galaxy to galaxy is the stellar mass. In their analysis of 
$z=0$ massive star-forming galaxies from the COLD GASS survey,
\citet{saintonge11} could observe an increase in $t_{\rm depl}$ by a factor of 
6 over the stellar mass range of $10^{10}~\rm M_{\sun}$ to 
$10^{11.5}~\rm M_{\sun}$, from about 0.7~Gyr to 4~Gyr. 
They assign this $t_{\rm depl}$--$M_*$ correlation to the more bursty star 
formation history of low-mass galaxies which leads to enhanced SFE, inversely 
reduced $t_{\rm depl}$, due to minor starburst events produced either by weak 
mergers associated with distant tidal encounters, variations in the 
intergalactic medium accretion rate, or secular processes within galactic 
disks. On the other hand, 
morphological quenching and feedback in high-mass galaxies prevent the 
molecular gas from forming stars, while at the same time not destroying the 
gas leads to an increased reservoir of molecular gas. 
This dependence of $t_{\rm depl}$ on $M_*$ in local SFGs has been 
recently confirmed over the full stellar mass range down to $M_* = 
10^9~\rm M_{\sun}$ by the ALLSMOG survey from \citet{bothwell14}.

\begin{figure}
\centering
\includegraphics[width=9cm,clip]{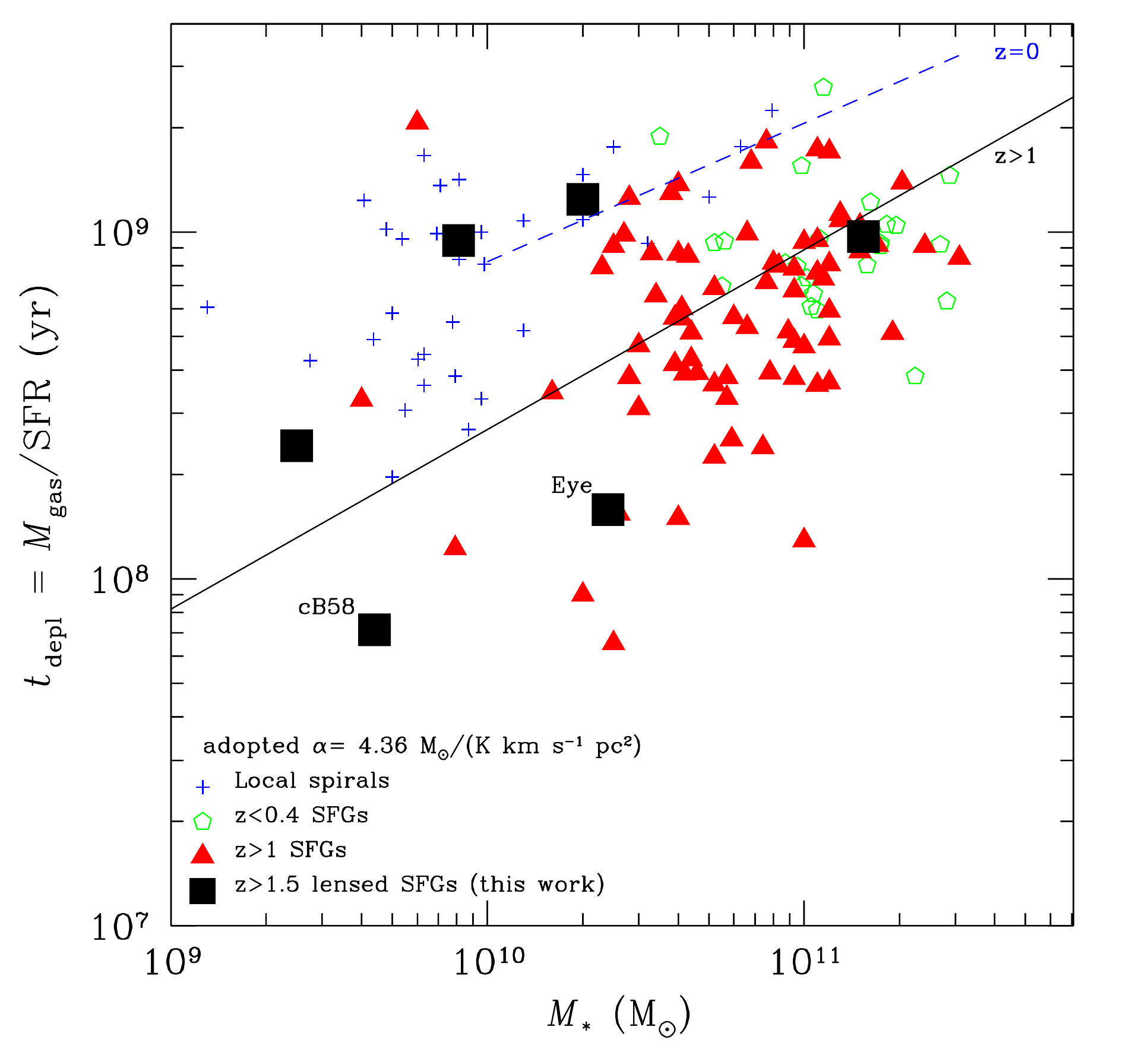} 
\caption{Molecular gas depletion timescales, $t_{\rm depl} = 
M_{\rm gas}/{\rm SFR}$, as a function of stellar masses plotted for our 
low-$L_{\rm IR}$ selected SFGs (squares) and our comparison sample of SFGs and 
local spirals with CO measurements from the literature (see 
Sect.~\ref{sect:literature}). The molecular gas depletion timescales are 
computed assuming a ``Galactic'' CO--H$_2$ conversion factor for all galaxies. 
A $t_{\rm depl}$--$M_*$ correlation emerges for $z>1$ SFGs; their best-fitting 
bisector linear relation, 
$\log (t_{\rm depl}) = (0.52\pm 0.16) \log (M_*) + (3.25\pm 1.74)$, is plotted 
as a solid black line. The best-fitting bisector linear relation to the $z=0$ 
COLD GASS sample from \citet{saintonge11} is shown by the dashed blue line.}
\label{fig:tdepl-Mstars}
\end{figure}

The increase of the molecular gas depletion timescale with the stellar mass is 
also observed in our compilation of $z>1$ SFGs with CO measurements from the 
literature, including our low-$L_{\rm IR}$ selected SFGs, as shown in 
Fig.~\ref{fig:tdepl-Mstars}.
It is confirmed with a Kendall's tau probability of 0.0017 per cent 
and a slope $M_*^{0.5}$. On average, $z>1$ SFGs have shorter $t_{\rm depl}$ 
than local galaxies with the same $M_*$.
This $t_{\rm depl}$--$M_*$ correlation in order to be compatible with the 
$t_{\rm depl}$--sSFR anti-correlation discussed above 
(Sect.~\ref{sect:tdepl-sSFR}), implies that a galaxy at $z>1$ with a given 
$M_*$ necessarily has its SFR higher than a galaxy at $z=0$ with the same 
$M_*$. This is in line with the SFR--$M_*$ MS relation (see 
Fig.~\ref{fig:SFR-Mstars}).

The increase in $t_{\rm depl}$ by a factor of 10 over the stellar mass range of 
$10^{9.4}~\rm M_{\sun}$ to $10^{11.5}~\rm M_{\sun}$, from about 0.15~Gyr to 
1.5~Gyr, we observe in our sample of $z>1$ SFGs, questions the constant 
$t_{\rm depl}$ of 0.7~Gyr found by \citet{tacconi13} in their sample of $z=1-3$ 
SFGs with stellar masses confined to $M_* > 10^{10.4}~\rm M_{\sun}$. In 
Fig.~\ref{fig:tdepl-Mstars}, we may see that the $t_{\rm depl}$--$M_*$ 
correlation at high redshift is triggered by galaxies with small stellar 
masses. This resembles the first $z=0$ surveys that were unable to highlight 
the $t_{\rm depl}$ correlation with $M_*$ because of their limited dynamical 
range in $M_*$. Indeed, in the smaller range of galaxies explored in the THINGS 
sample, \citet{bigiel08} and \citet{leroy08} initially found a constant 
molecular gas depletion timescale.

If true, such a correlation between $t_{\rm depl}$ and $M_*$ observed both in 
local and $z>1$ SFGs has two important implications. First, it is in
contradiction with cosmological hydrodynamic simulations by \citet{dave11} 
which predict the opposite, namely an anti-correlation between $M_*$ and 
$t_{\rm depl}$, where $t_{\rm depl}$ drops to larger stellar masses roughly as 
$M_*^{-0.3}$~\footnote{The dependence of $t_{\rm depl}$ on $M_*^{-0.3}$ comes 
from the star formation relation assumed ($\Sigma_{\rm SFR} \propto 
\Sigma_{\rm gas}^{1.4}$) and the empirical relation measured in \citet{dave11} 
simulations of $\Sigma_{\rm gas} \propto M_*^{3/4}$.}. Second, it refutes the 
linearity of the Kennicutt-Schmidt relation, i.e.\ the proportionality of 
SFR and molecular gas mass, such that $\Sigma_{\rm SFR} \propto 
\Sigma_{\rm gas}^N$ with $N\neq 1$. 
This is fundamental, as it contradicts one of the hypotheses of the ``bathtub'' 
model that assumes a constant molecular gas depletion timescale, and hence a 
linear Kennicutt-Schmidt relation 
\citep[e.g.,][]{bouche10,lilly13,dekel14}. Therefore, we may wonder whether the 
$t_{\rm depl}$--$M_*$ correlation really exists, or whether we simply see a 
constant $t_{\rm depl}$ embedded in a very large spread. Only getting more 
molecular gas depletion timescale measurements for $z>1$ galaxies with small 
stellar masses will bring a definitive answer. 

%

%

A way to reconcile the current empirical finding with a constant $t_{\rm depl}$ 
is to invoke the possible CO--H$_2$ conversion factor dependence on metallicity 
\citep[e.g.,][]{israel97,leroy11,feldmann12,genzel12}. Indeed, it is expected 
that galaxies with lower stellar masses have lower metallicities, according to 
the mass-metallicity relation (see Sect.~\ref{sect:introduction}),
and would hence require higher CO--H$_2$ conversion factors to be applied that 
would, in reality, lead to longer molecular gas depletion timescales in low 
stellar mass galaxies in comparison to what we get when assuming a ``Galactic'' 
CO--H$_2$ conversion factor. However, whether this is sufficient to compensate 
the $t_{\rm depl}$ dependence on $M_*^{0.6}$ needs to be further tested.

%

\subsubsection{Redshift}
\label{sect:tdepl-z}

Since the star-forming galaxies span a large interval of the Hubble time, the 
redshift could also be at the origin of their large SFE or $t_{\rm depl}$ 
dispersion. Models do predict a cosmic evolution of the molecular gas depletion 
timescale for main-sequence galaxies \citep[e.g.,][]{hopkins06,dave11,
dave12}. 
The temporal scaling of $t_{\rm depl}$ can be most easily understood when using 
the formulation of the star formation relation given by ${\rm SFR} \propto 
M_{\rm gas}/t_{\rm dyn}$, where $t_{\rm dyn}$ is the dynamical time of the star 
formation region \citep[see also][]{bouche10,genel10}. This then gives 
$t_{\rm depl} \propto t_{\rm dyn}$, which in a canonical disk model scales as 
$(1+z)^{-1.5}$ \citep{mo98}. 

\begin{figure}
\centering
\includegraphics[width=9cm,clip]{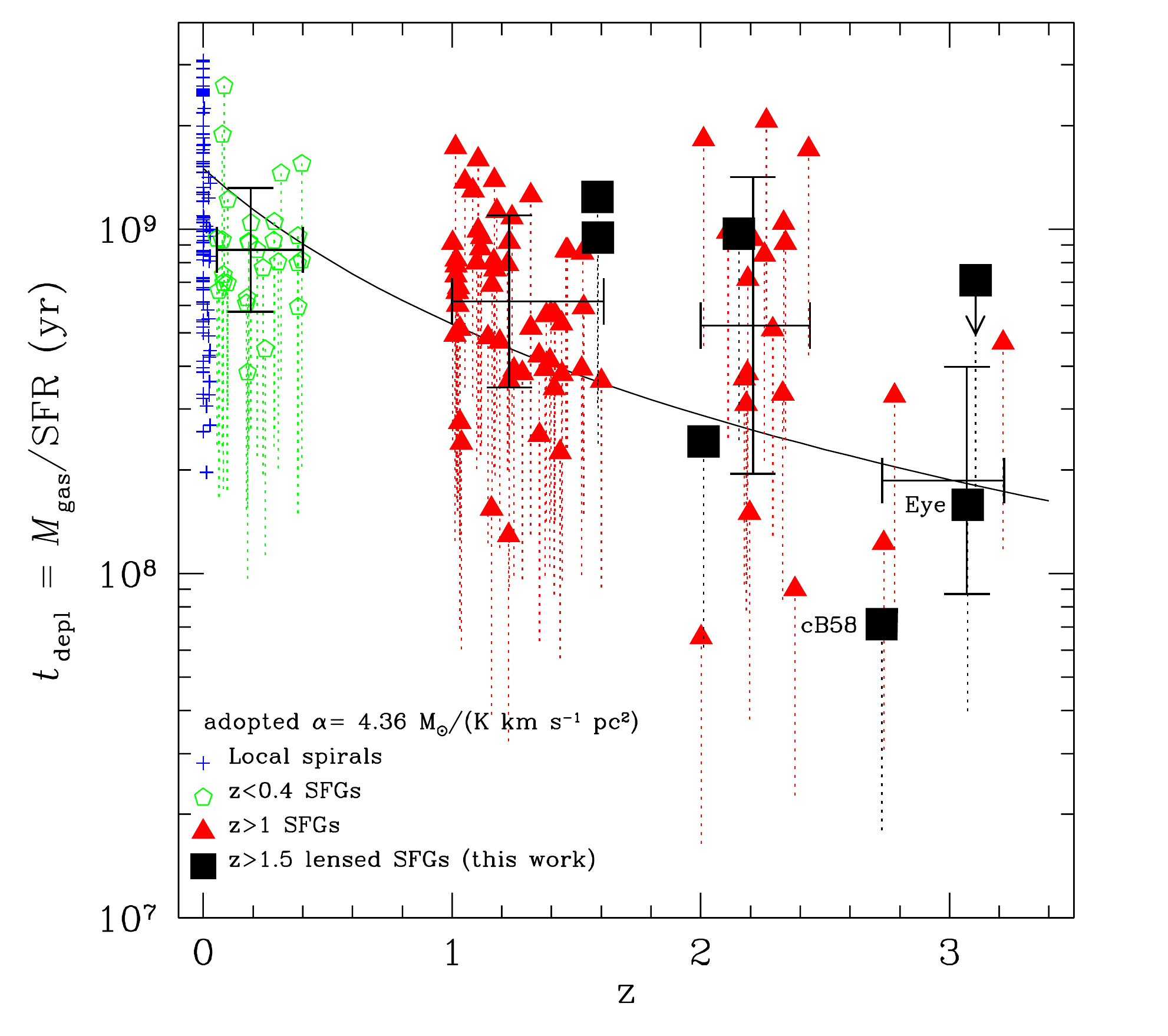} 
\caption{Molecular gas depletion timescales, $t_{\rm depl} = 
M_{\rm gas}/{\rm SFR}$, as a function of redshift plotted for our 
low-$L_{\rm IR}$ selected SFGs (squares) and our comparison sample of SFGs and 
local spirals with CO measurements from the literature (see 
Sect.~\ref{sect:literature}). The molecular gas depletion timescales are 
computed assuming a ``Galactic'' CO--H$_2$ conversion factor for all galaxies, 
but for $z<0.4$ and $z>1$ SFGs (pentagons, triangles, and squares) the 
dotted lines, in addition, show the interval of possible $t_{\rm depl}$ values 
as determined with two extreme CO--H$_2$ conversion factors $\alpha = 4.36$ 
(``Galactic'' value) and 1.1 (local ULIRG value). A redshift evolution of 
$t_{\rm depl}$ is observed in agreement with the expected 
$1.5\times (1+z)^{-1.5}$ dependence of \citet{dave12}, where the normalization 
is set to the typical depletion time of 1.5~Gyr observed in local galaxies 
(solid line). Four redshift bins are considered, $\langle z_{0.2}\rangle = 
[0.055,0.4]$, $\langle z_{1.2}\rangle = [1,1.6]$, $\langle z_{2.2}\rangle = 
[2,2.5]$, and $\langle z_{3.0}\rangle = [2.7,3.2]$, for the $t_{\rm depl}$ 
means shown by the large black crosses with their $1\,\sigma$ dispersion.}
\label{fig:tdepl-z}
\end{figure}

The decrease in $t_{\rm depl}$ with redshift is supported by observations, as 
already reported by \citet{combes13}, \citet{tacconi13}, \citet{saintonge13}, 
and \citet{santini14}. 
In Fig.~\ref{fig:tdepl-z} we plot the molecular gas depletion timescale as a 
function of redshift for our compilation of local spirals and SFGs with CO 
measurements from the literature, including our low-$L_{\rm IR}$ selected SFGs. 
Four redshift bins are considered $\langle z_{0.2}\rangle 
= [0.055,0.4]$, $\langle z_{1.2}\rangle = [1,1.6]$, $\langle z_{2.2}\rangle = 
[2,2.5]$, and $\langle z_{3.0}\rangle = [2.7,3.2]$, their respective mean 
values are 
$\langle \log (t_{\rm depl}^{\langle z_{0.2}\rangle})\rangle = 8.94\pm 0.18$ (870~Myr), 
$\langle \log (t_{\rm depl}^{\langle z_{1.2}\rangle})\rangle = 8.79\pm 0.25$ (620~Myr), 
$\langle \log (t_{\rm depl}^{\langle z_{2.2}\rangle})\rangle = 8.72\pm 0.43$ (520~Myr), and 
$\langle \log (t_{\rm depl}^{\langle z_{3.0}\rangle})\rangle = 8.27\pm 0.33$ 
(190~Myr) for a ``Galactic'' CO--H$_2$ conversion factor. This confirms a 
shallow $t_{\rm depl}$ decrease from $z=0$ to $z=3.2$, in agreement with the 
expected $(1+z)^{-1.5}$ dependence. As a result, high redshift SFGs do form 
stars with a much higher SFE, and consume the molecular gas over a much shorter 
timescale, than local SFGs. The $t_{\rm depl}$ spread and dispersion per 
redshift bin are significant, they result from the $t_{\rm depl}$--sSFR and 
$t_{\rm depl}$--$M_*$ relations discussed above (Sects.~\ref{sect:tdepl-sSFR} 
and \ref{sect:tdepl-Mstars}). Indeed, these relations induce a gradient in 
$t_{\rm depl}$ per redshift bin, such that galaxies with the higher sSFR and 
smaller $M_*$ have the shorter $t_{\rm depl}$. 


%

\subsubsection{Offset from the main-sequence}

\begin{figure}
\centering
\includegraphics[width=9cm,clip]{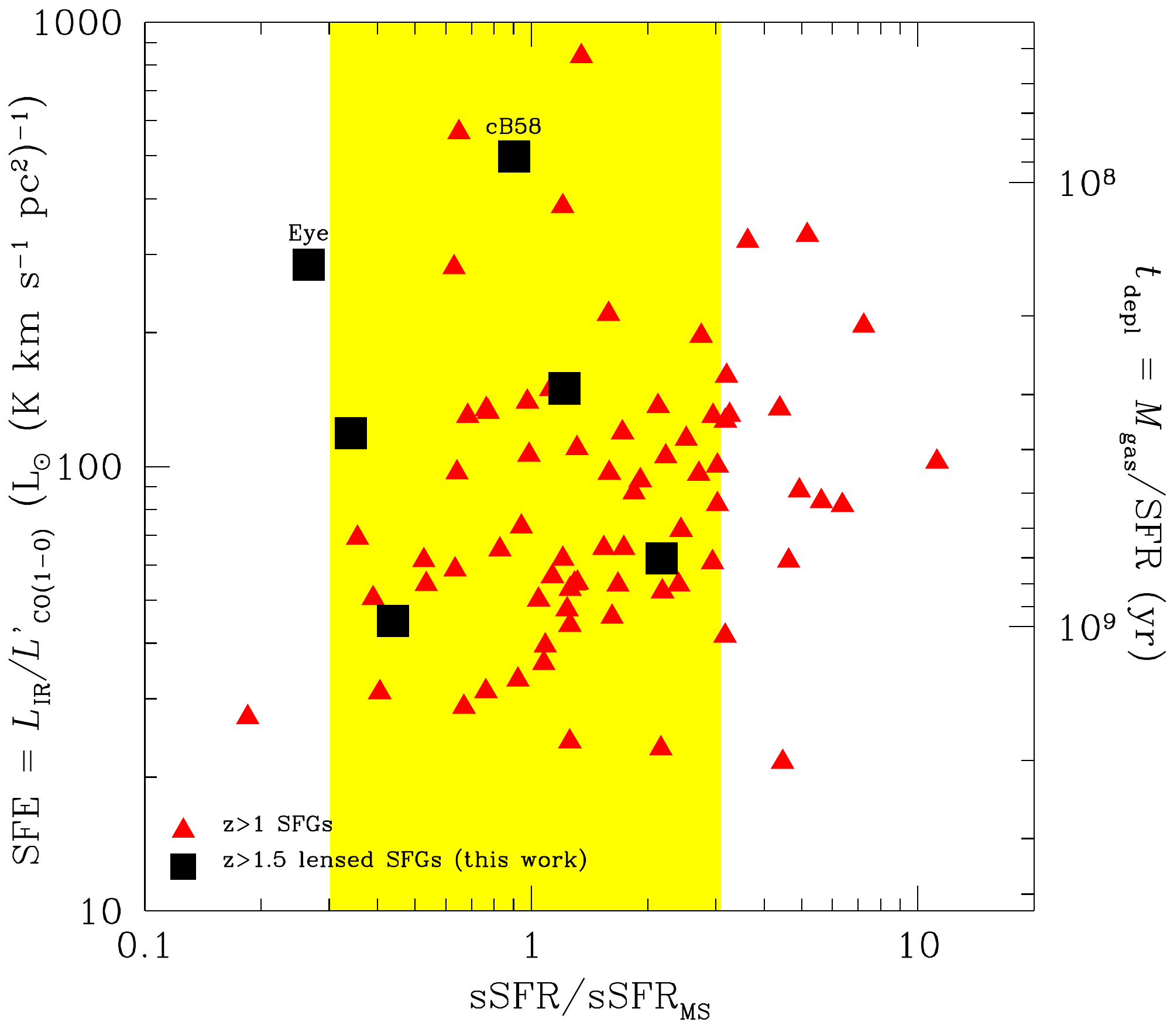} 
\caption{Star formation efficiencies, ${\rm SFE} = 
L_{\rm IR}/L'_{\rm CO(1-0)}$ (left y-axis), or molecular gas depletion 
timescales, $t_{\rm depl} = {\rm SFR}/M_{\rm gas}$, assuming a ``Galactic'' 
CO--H$_2$ conversion factor for all galaxies (right y-axis), as a function of 
galaxies' offsets from the main-sequence, $\rm sSFR/sSFR_{MS}$, plotted for our 
low-$L_{\rm IR}$ selected SFGs (squares) and our comparison sample of $z>1$ 
SFGs with CO measurements from the literature (triangles). The shaded area 
defines the accepted thickness of the MS, i.e.\ a scatter about the SFR--$M_*$ 
relation between $\rm 0.3 < sSFR/sSFR_{MS} < 3$. The enlarged sample of $z>1$ 
SFGs confirms the general trend of higher SFE for galaxies with larger offsets 
from the MS with a Kendall's tau probability of 0.025 per cent. However, 
considering MS galaxies only within the shaded area, their SFE appear roughly 
constant, 
given the Kendall's tau probability of 0.23 per cent with a large dispersion of 
0.32~dex.}
\label{fig:SFE-sSFRsSFRMS}
\end{figure}

A positive empirical correlation between SFE and the offset from the 
main-sequence, $\rm sSFR/sSFR_{MS}$, was found by \citet{magdis12b},
\citet{saintonge12}, and \citet{sargent14}, suggesting higher SFE for galaxies 
with enhanced sSFR and, in particular, for galaxies with $\rm sSFR/sSFR_{MS} 
\gg 3$, namely beyond the accepted MS thickness, and this independent of 
galaxies' redshift. In Fig.~\ref{fig:SFE-sSFRsSFRMS} we show the SFE of our 
low-$L_{\rm IR}$ selected galaxies and $z>1$ SFGs from the literature 
(Sect.~\ref{sect:literature}) as a function of their offset from the MS. The 
main-sequence, ${\rm SFR_{MS}}(z,M_*)$, is defined here by Eq.~(\ref{eq:MS}) 
and is computed at three redshifts covering three redshift intervals $\langle 
z_{1.2}\rangle = [1,1.6]$, $\langle z_{2.2}\rangle = [2,2.5]$, and $\langle 
z_{3.0}\rangle = [2.7,3.2]$ of the SFGs. While the enlarged sample of $z>1$ 
SFGs confirms the general trend of higher SFE for galaxies with larger offsets 
from the MS (Kendall's tau probability of 0.025 per cent), the MS galaxies 
themselves, with offsets from the MS restricted to $0.3 < \rm sSFR/sSFR_{MS} < 
3$ (80\% of the sample), have roughly a constant SFE given the Kendall's tau 
probability of 0.23 per cent with a large dispersion of 0.32~dex, reaching SFE 
values even higher than those of galaxies beyond the accepted MS thickness. 
Consequently, the large spread in SFE observed among MS galaxies over 1.7 
orders of magnitude cannot be attributed to the thickness of the SFR--$M_*$ 
relation, i.e.\ the relative position of galaxies with respect to the 
main-sequence. 

%

%

\begin{figure}
\centering
\includegraphics[width=9cm,clip]{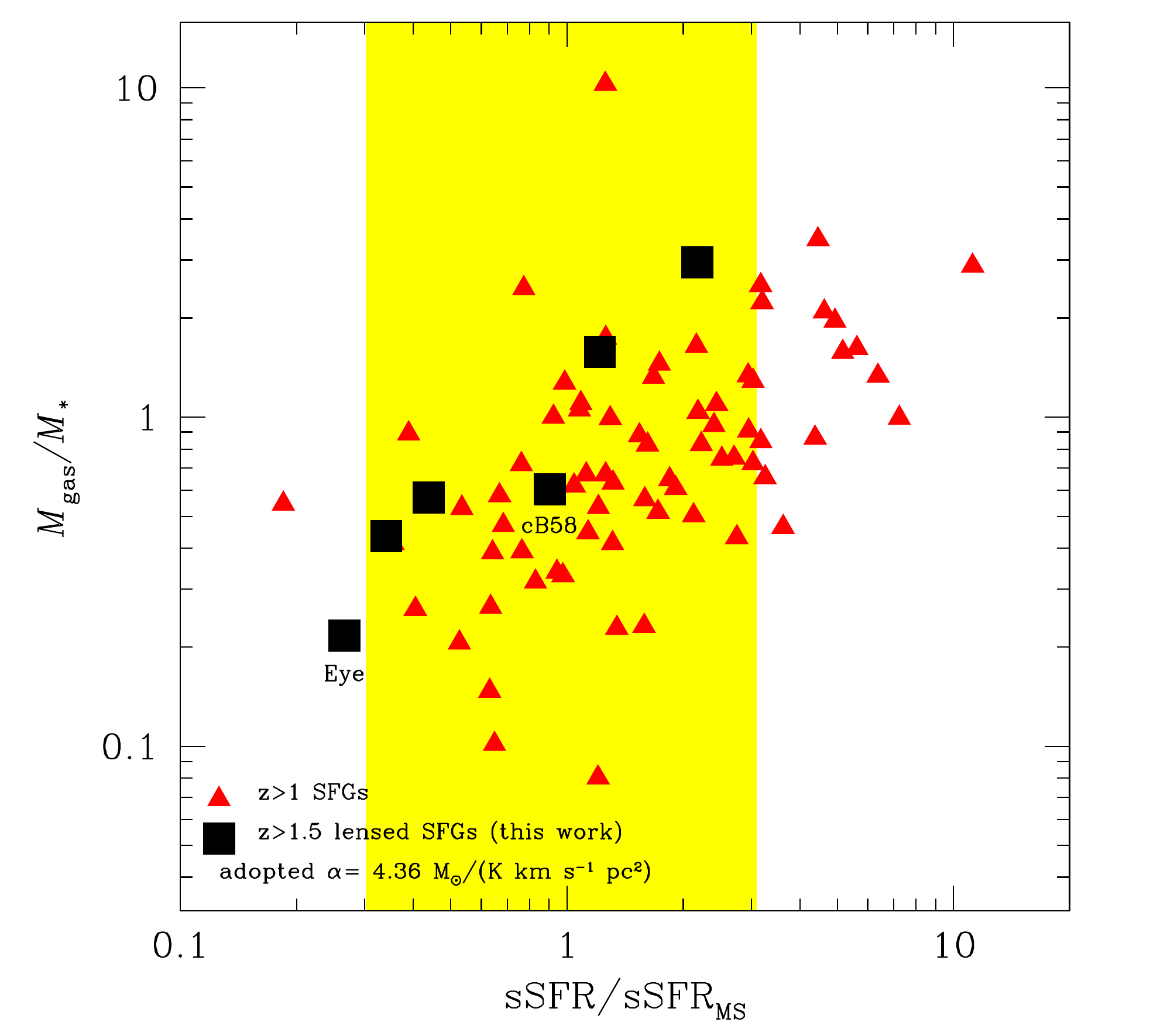} 
\caption{Molecular gas mass over stellar mass ratios, $M_{\rm gas}/M_*$, as a 
function of galaxies' offsets from the main-sequence, $\rm sSFR/sSFR_{MS}$, 
plotted for our low-$L_{\rm IR}$ selected SFGs (squares) and our comparison 
sample of $z>1$ SFGs with CO measurements from the literature (triangles). The 
shaded area defines the accepted thickness of the MS as above.
A clear correlation is observed with a Kendall's tau probability of 0 per 
cent.}
\label{fig:fgas-sSFRsSFRMS}
\end{figure}

Saying things on the other way, these results tell us that it is not the star 
formation efficiency of MS galaxies that drives the thickness of the SFR--$M_*$ 
relation, although the SFE certainly contributes to it in some way within its 
large spread. What physical parameter then dominates and triggers this 
thickness\,? In Fig.~\ref{fig:fgas-sSFRsSFRMS} we show that the strongest 
dependence of the offset from the MS is observed on $M_{\rm gas}/M_*$, the 
molecular gas mass over the stellar mass ratio, with a Kendall's tau 
probability of 0 per cent and this over two 
orders of magnitude. This confirms the findings by \citet{magdis12b} and 
\citet{sargent14}, who support that the thickness of the SFR--$M_*$ relation at 
any redshift is driven by the variations of the molecular gas fractions of MS 
galaxies rather than by variations of their star formation efficiencies. 
According to \citet{magdis12b}, this prevalence of $M_{\rm gas}/M_*$ implies 
that MS galaxies can have higher SFR (within the thickness of the MS) mainly 
because they have more raw molecular gas material to produce stars, which hence 
favours the existence of a global star-formation relation 
$L_{\rm IR}$(or SFR)--$M_{\rm gas}$
for all MS galaxies (see Fig.~\ref{fig:LIR-LCO}).

%

\subsubsection{Compactness of the starburst}

\citet{downes98} were the first to point out that the compactness of the 
nuclear starburst regions was linked to the star formation efficiency in 
extreme local ULIRGs.
\citet{combes13} found, in this sense, a weak trend for SFE to be 
anti-correlated with the half-light radii of high-redshift galaxies, $R_{1/2}$, 
or correlated with their compactness. The five SFGs with the highest SFE 
from our compilation of $z>1$ SFGs, including our low-$L_{\rm IR}$ selected 
galaxies, have $R_{1/2}$ ranging between 1.5 and 4~kpc with one object having 
$R_{1/2}$ as large as 8~kpc. These half-light radii are typical of 
high-redshift star-forming galaxies \citep{tacconi13}, and thus do not 
highlight a particular compactness among SFGs with the highest SFE. Certainly, 
we have to increase the number of available data, in parallel with the angular 
resolution of CO measurements to confirm or affirm the role of the starburst's 
compactness in the SFE (or $t_{\rm depl}$) spread.

%

%

\subsubsection{Summary}

The answer to the question ``what determines the behaviour of the star 
formation efficiency and molecular gas depletion timescale in star-forming 
galaxies at high redshift'' is not straightforward from what we have discussed 
above. The dependence of SFE (or $t_{\rm depl}$) on five physical parameters 
has been investigated. 
While the dependence appears to be the strongest on the specific star formation 
rate, there is, in addition, a trend for a correlation with the stellar mass 
and a clear evolution with redshift. The dependences on the offset from the 
main-sequence and the compactness of the starburst are less clear. 
Consequently, there is a tight interplay between the SFE (or $t_{\rm depl}$) 
and, respectively, the specific star formation rate, the stellar mass, and the 
redshift of galaxies, which implies that not only one single physical parameter 
drives the observed large spread in SFE of $z>1$ SFGs, but a combination of all 
of them.

%
%

In addition, an intrinsic spread in SFE and $t_{\rm depl}$ certainly exists due 
to other, more complex, and hence more difficult to test, effects. 
For example, some galaxies very likely are not in a quasi-steady state 
equilibrium, because the accretion rate is not constant, in contrast to what is 
assumed in the ideal ``bathtub'' model where the accretion rate and the star 
formation rate timescale are supposed to vary slowly enough. And obviously, 
environment and mergers, as well as other variability effects, like an episodic 
star formation, probably imply a not so smooth galaxy evolution.

%

%

\section{Molecular gas fraction of high-redshift galaxies}
\label{sect:fgas}

The molecular gas fraction, defined as $f_{\rm gas} = 
M_{\rm gas}/(M_{\rm gas}+M_*)$, is a very important physical quantity to know, 
but also very difficult to access, since dependent on the CO--H$_2$ conversion 
factor. 
We can also express the molecular gas fraction as:
\begin{equation}\label{eq:fgas}
f_{\rm gas} = \frac{1}{1+(t_{\rm depl} {\rm sSFR})^{-1}}\,,
\end{equation}
meaning that it is closely linked to the molecular gas depletion timescale and 
the specific star formation rate. 
Observations show a clear trend of increasing $f_{\rm gas}$ with sSFR 
\citep{tacconi13}, but with a large dispersion (larger than the systematic 
uncertainties in sSFR and $f_{\rm gas}$ measurements) induced by the different 
molecular gas depletion timescales found among $z>1$ SFGs, ranging between 
0.15~Gyr and 1.5~Gyr (see Sect.~\ref{sect:SFE,tdepl}).


The various physical processes at play in the evolution of galaxies (e.g., 
accretion, star formation, and feedback) have direct impact on the behaviour of 
their molecular gas fraction. 
Consequently, a solid way to test galaxy evolution models is to confront their 
predictions with the observed $f_{\rm gas}$ behaviour. We consider below two 
main observables, the redshift and the stellar mass, and investigate the 
respective redshift evolution and stellar mass dependence of $f_{\rm gas}$ in 
high-redshift SFGs.


%

\begin{figure}
\centering
\includegraphics[width=9cm,clip]{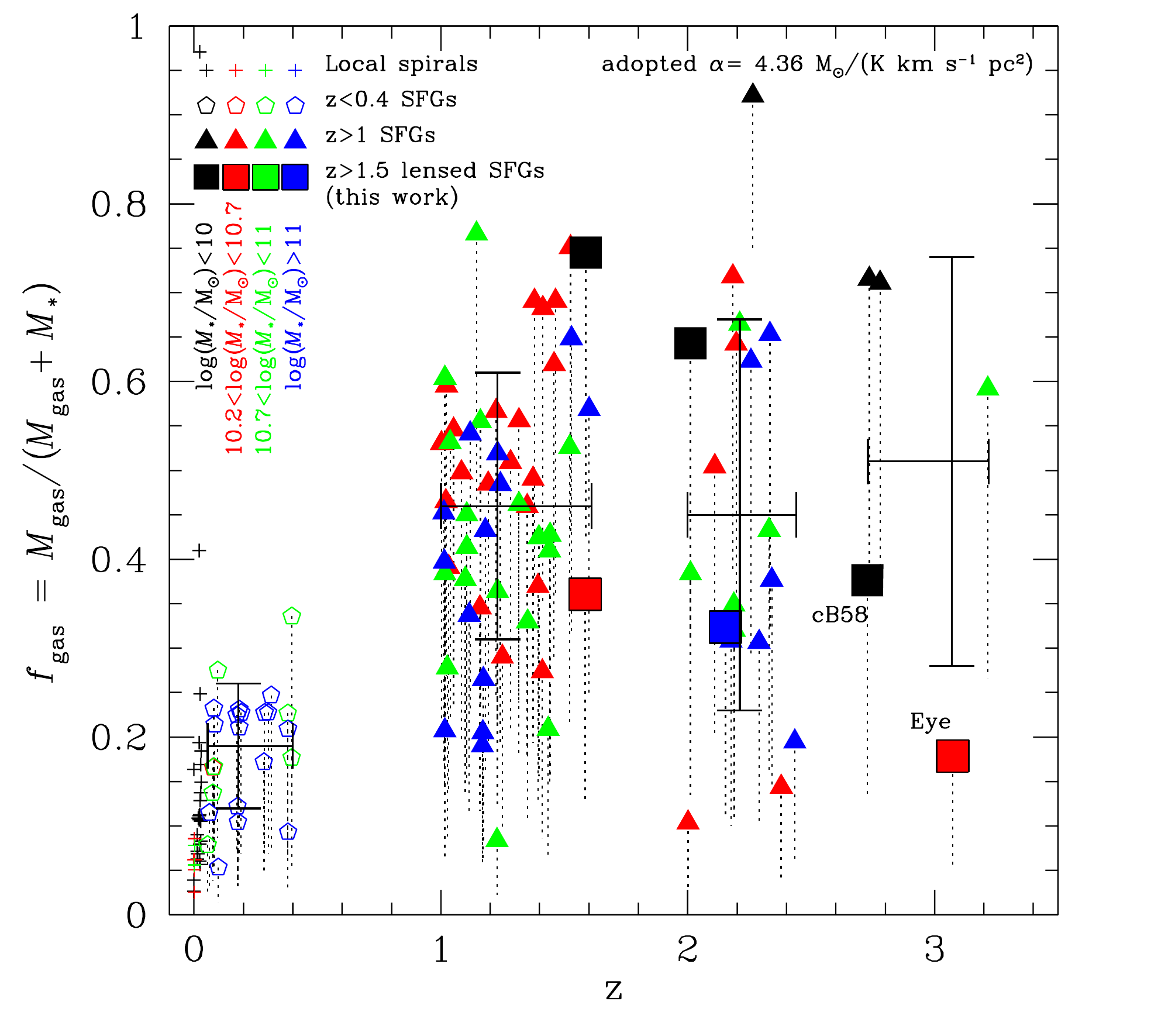} 
\caption{Molecular gas fractions, $f_{\rm gas} = 
M_{\rm gas}/(M_{\rm gas}+M_*)$, as a function of redshift plotted for our 
low-$L_{\rm IR}$ selected SFGs (squares) and our comparison sample of SFGs and 
local spirals with CO measurements from the literature (see 
Sect.~\ref{sect:literature}). The molecular gas fractions are computed assuming 
a ``Galactic'' CO--H$_2$ conversion factor for all galaxies, but for $z<0.4$ 
and $z>1$ SFGs (pentagons, triangles, and squares) the dotted lines, in 
addition, show the interval of possible $f_{\rm gas}$ values as determined 
with two extreme CO--H$_2$ conversion factors $\alpha = 4.36$ (``Galactic'' 
value) and 1.1 (local ULIRG value). Four redshift bins are considered, 
$\langle z_{0.2}\rangle = [0.055,0.4]$, $\langle z_{1.2}\rangle = [1,1.6]$, 
$\langle z_{2.2}\rangle = [2,2.5]$, and $\langle z_{3.0}\rangle = [2.7,3.2]$, 
for the $f_{\rm gas}$ means shown by the large black crosses with their 
$1\,\sigma$ dispersion. A net rise of $f_{\rm gas}$ is observed from 
$\langle z_{0.2}\rangle$ to $\langle z_{1.2}\rangle$, followed by a very mild 
increase toward higher redshifts.
The color-coding refers to four stellar mass intervals: 
$\log (M_*/{\rm M}_{\sun}) < 10$ (black), $10.2 < \log (M_*/{\rm M}_{\sun}) < 
10.7$ (red), $10.7 < \log (M_*/{\rm M}_{\sun}) < 11$ (green), and 
$\log (M_*/{\rm M}_{\sun}) > 11$ (blue), as defined in 
Sect.~\ref{sect:fgas-Mstars}.} 
\label{fig:fgas-z}
\end{figure}
%

\begin{figure}
\centering
\includegraphics[width=9cm,clip]{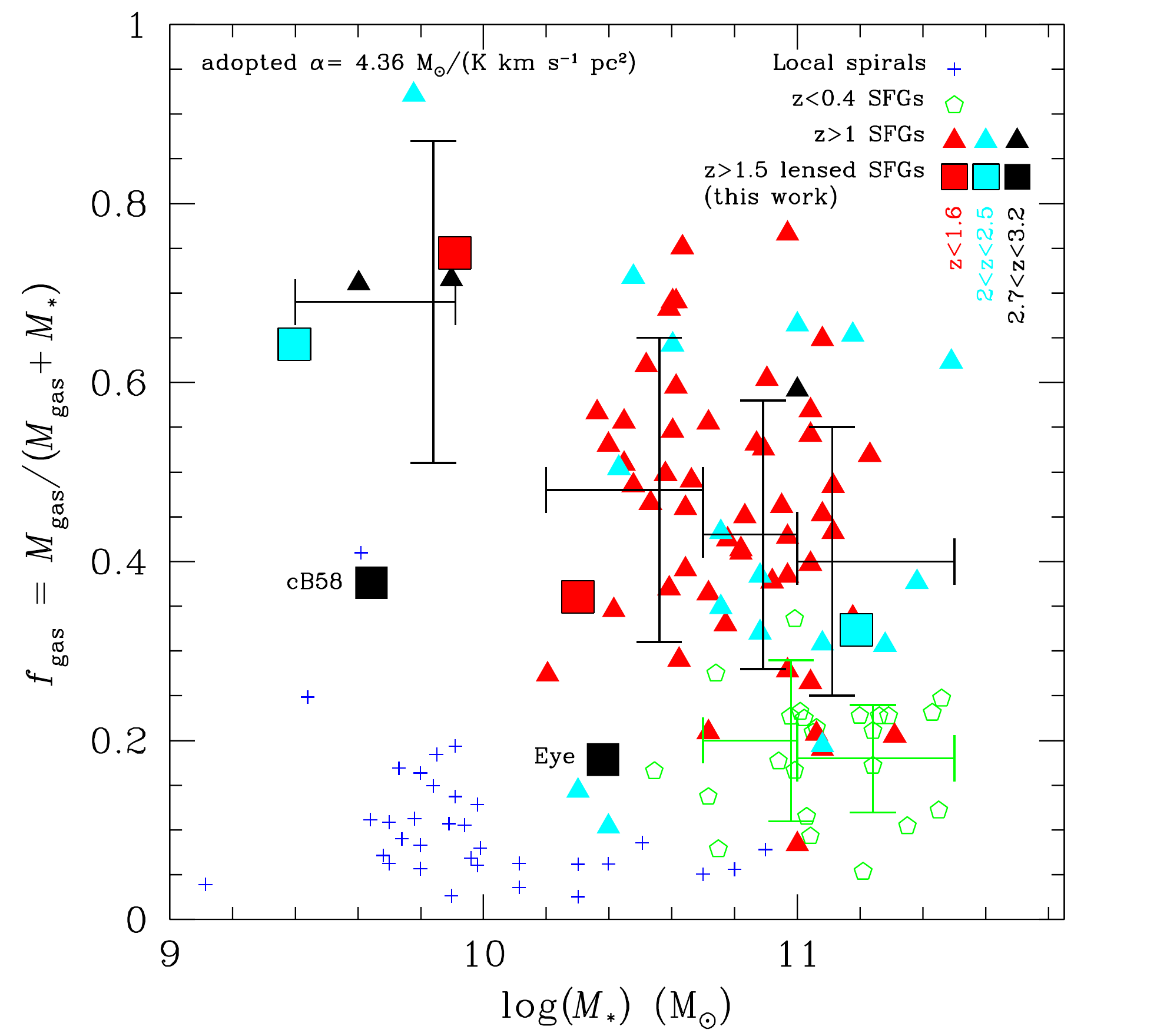} 
\caption{Molecular gas fractions, $f_{\rm gas} = 
M_{\rm gas}/(M_{\rm gas}+M_*)$, as a function of stellar masses plotted for our 
low-$L_{\rm IR}$ selected SFGs (squares) and our comparison sample of SFGs and 
local spirals with CO measurements from the literature (see 
Sect.~\ref{sect:literature}). The molecular gas fractions are computed assuming 
a ``Galactic'' CO--H$_2$ conversion factor for all galaxies. 
Four stellar mass bins are considered $9.4 < \log (M_*/{\rm M_{\sun}}) < 9.9$, 
$10.2 < \log (M_*/{\rm M_{\sun}}) < 10.7$, $10.7 < \log (M_*/{\rm M_{\sun}}) < 
11.0$, and $11.0 < \log (M_*/{\rm M_{\sun}}) < 11.5$.
The corresponding $f_{\rm gas}$ means with their $1\,\sigma$ dispersion of 
$z>1$ SFGs are shown by the large black crosses and of $z<0.4$ SFGs by the 
large green crosses. The $M_*$ dependence of $f_{\rm gas}$ sustains an upturn 
of the molecular gas fraction at the low-$M_*$ end, a mild decrease toward 
higher stellar masses of SFGs at $z>1$, and a shift toward lower $f_{\rm gas}$ 
values for SFGs at lower redshifts. 
The color-coding of SFGs (our sample plus $z>1$ SFGs from the literature) 
refers to three redshift intervals: $\langle z_{1.2}\rangle = [1,1.6]$ (red), 
$\langle z_{2.2}\rangle = [2,2.5]$ (cyan), and $\langle z_{3.0}\rangle = 
[2.7,3.2]$ (black).}
\label{fig:fgas-Mstars}
\end{figure}
%

\subsection{Redshift evolution of the molecular gas fraction}
\label{sect:fgas-z}

The cosmic evolution of the molecular gas fraction with the Hubble time 
directly results from the expansion of the Universe itself. 
New matter falling on galaxies through accretion from filaments comes from 
further out as the time progresses and, consequently, high angular momentum 
matter settles in the outer parts of galaxies. This leads to the size growth of 
galaxies with the cosmic time, such that galaxies were more compact in the 
early Universe than today. A finding first proposed by dark matter simulations 
\citep[e.g.,][]{fall80,mo98}, more recently by cosmological hydrodynamic 
simulations including cold gas accretion \citep{stewart13}, and now 
observationally confirmed \citep[e.g.,][]{bouwens04,trujillo06,buitrago08}. The 
size growth of galaxies has a direct impact on the H$_2$/H\,I ratio evolution. 
Indeed, the pressure on the midplane of galaxies strongly depends on the gas 
surface density, in the way that a smaller radius increases the gas surface 
density. 
Since the H$_2$/H\,I ratio is strongly correlated to this pressure, for the 
same gas mass the high-redshift galaxies, which are smaller (more compact) and 
hence have a higher internal pressure, will have a higher H$_2$/H\,I ratio and 
that means a higher H$_2$ content than local galaxies \citep[e.g.,][]
{obreschkow09,lagos11,lagos14}. This together with the trend for high-redshift 
galaxies to be intrinsically more gas-rich, i.e.\ have a higher total gas mass 
($\rm H\,I + H_2$), leads to even larger difference between high-redshift and 
local galaxies and hence to a net increase of the molecular gas fraction with 
redshift.

From Eq.~(\ref{eq:fgas}) we see that the evolution of the molecular gas 
fraction depends on the evolution of both $t_{\rm depl}$ and sSFR. We have 
shown in Sect.~\ref{sect:tdepl-z} that the molecular gas depletion timescale 
evolves with $(1+z)^{-1.5}$. The ``bathtub'' model, in parallel, predicts a 
continuing rise of the specific star formation rate, generally driven by the 
redshift evolution of the average accretion rate (inflow), which scales as 
$(1+z)^{\alpha}$ with $\alpha$ varying between 5/3 and 3 according to the 
prescriptions considered \citep{bouche10,dave12,lilly13}. The steeper rise of 
the specific star formation rate with respect to the decline of the molecular 
gas depletion timescale leads to the steady increase of $f_{\rm gas}$ with 
redshift. 

Observationally, the cosmic evolution of the molecular gas fraction with the 
Hubble time, in the way that higher redshift galaxies are more gas-rich, is now 
well established \citep[e.g.,][]{daddi10a,geach11,tacconi10,tacconi13,
saintonge13,sargent14,carilli13}. In Fig.~\ref{fig:fgas-z} we show 
$f_{\rm gas}$ as a function of redshift for our compilation of local spirals 
and SFGs with CO measurements from the literature, including our 
low-$L_{\rm IR}$ selected SFGs. We consider four redshift bins 
$\langle z_{0.2}\rangle = [0.055,0.4]$, $\langle z_{1.2}\rangle = [1,1.6]$, 
$\langle z_{2.2}\rangle = [2,2.5]$, and $\langle z_{3.0}\rangle = [2.7,3.2]$. 
The respective mean values $\langle f_{\rm gas}^{\langle z_{0.2}\rangle}\rangle 
= 0.19\pm 0.07$, $\langle f_{\rm gas}^{\langle z_{1.2}\rangle}\rangle = 
0.46\pm 0.15$, $\langle f_{\rm gas}^{\langle z_{2.2}\rangle}\rangle = 
0.45\pm 0.22$, and $\langle f_{\rm gas}^{\langle z_{3.0}\rangle}\rangle = 
0.51\pm 0.23$ for a ``Galactic'' CO--H$_2$ conversion factor show a net 
rise of the molecular gas fraction from $\langle z_{0.2}\rangle$ to 
$\langle z_{1.2}\rangle$, followed by a very mild increase toward higher 
redshifts, as found in earlier studies.
If further confirmed, such an $f_{\rm gas}$ redshift evolution does not really correspond to 
a steady redshift increase of $f_{\rm gas}$ as predicted by the ``bathtub'' 
model. It requires a different redshift-dependence of either $t_{\rm depl}$ or 
sSFR (see Eq.~(\ref{eq:fgas})).
The sSFR redshift evolution parametrised by \citet{lilly13} proposes
a steep sSFR increase as $(1+z)^3$ out to $z=2$ and a slow down to 
$(1+z)^{5/3}$ at $z>2$. This model is among the closest to the observations, 
since it predicts a flattening of the $f_{\rm gas}$ evolution beyond $z=2$ \citep[see also][]{saintonge13}. 

The spread and dispersion of the molecular gas fractions per redshift bin are 
significant, in particular for $z>1$ SFGs. 
As discussed in \citet{carilli13}, a large $f_{\rm gas}$ dispersion at a 
given redshift is expected: it is mainly due to the strong dependence of 
$f_{\rm gas}$ on the stellar mass (shown in Sect.~\ref{sect:fgas-Mstars}), 
which should produce an $f_{\rm gas}$ gradient per redshift bin, such that 
galaxies with the smaller $M_*$ have the larger $f_{\rm gas}$ (see 
Fig.~\ref{fig:fgas-Mstars}). To highlight the effect of this dependence, the 
color-coding of the data points in Fig.~\ref{fig:fgas-z} refers to four stellar 
mass bins. We may see that the data points within the lowest $M_*$ bin 
($\log (M_*/{\rm M}_{\sun}) < 10$) do show the larger $f_{\rm gas}$ per 
redshift bin. The expected $f_{\rm gas}$ gradient with $M_*$ per redshift bin 
for galaxies within the other $M_*$ bins is more scattered, but we still 
see its effect on the $z<0.4$ SFGs which have the smallest dispersion 
because they cover the two higher $M_*$ bins only. The stellar mass is not the 
only physical parameter inducing an $f_{\rm gas}$ dispersion per redshift bin. 
The star formation rate plays an important role too, since $f_{\rm gas}$ 
increases with SFR as a consequence of the Kennicutt-Schmidt relation (see the 
fundamental $f_{\rm gas}$--$M_*$--SFR relation proposed by \citet{santini14}), 
as well as 
environmental effects (outflows, disk hydrostatic pressure, etc.).

%

\subsection{Stellar mass dependence of the molecular gas fraction}
\label{sect:fgas-Mstars}

\citet{bouche10}, \citet{dave11}, and \citet{lagos14}, all show a drop in the 
molecular gas fraction with increasing stellar mass and an upturn of 
$f_{\rm gas}$ at the low-$M_*$ end. The location and strength of that upturn 
depends on the model adopted and, in particular, on whether any outflow is 
considered. A comparable $M_*$ dependence of the molecular gas fraction is 
predicted for both local and high-redshift galaxies, but with a shift in 
$f_{\rm gas}$ toward higher values at higher redshifts, a shift that results 
from the redshift evolution of $f_{\rm gas}$ (see Fig.~\ref{fig:fgas-z}). While 
in a closed-box model the molecular gas fraction is expected to decrease with 
the stellar mass because of the gas conversion into stars, this turns out to be 
similar in the ``bathtub'' model as the modelled galaxies approach a steady 
state where the SFR, which is proportional to the gas mass, represents the rate 
at which gas is turned into stars, follows the net gas accretion rate dictated 
by cosmology plus a gas outflow rate \citep{dekel14}.

The $M_*$ dependence of $f_{\rm gas}$ at $z>1$ has already been observationally 
constrained by \citet{tacconi13} and \citet{santini14} for galaxies at the 
intermediate/high-$M_*$ end $10.4 < \log (M_*/{\rm M}_{\sun}) < 12$ 
\citep[see also][]{magdis12b,sargent14}. We provide for the first time 
insights on $f_{\rm gas}$ of high-redshift SFGs at the low-$M_*$ end $9.4 < 
\log (M_*/{\rm M}_{\sun}) < 9.9$. Our new sample of low-$L_{\rm IR}$ selected 
SFGs 
doubles the number of galaxies at $z>1$ (3 out of a total of 6) with achieved 
$f_{\rm gas}$ measurements at the low-$M_*$ end. In Fig.~\ref{fig:fgas-Mstars} 
we show the resulting $f_{\rm gas}$ as a function of stellar mass.
We divide the sample of $z>1$ SFGs with intermediate/high-$M_*$ in three 
stellar mass bins $10.2 < \log (M_*/{\rm M_{\sun}}) < 10.7$, $10.7 < 
\log (M_*/{\rm M_{\sun}}) < 11.0$, and $11.0 < \log (M_*/{\rm M_{\sun}}) < 
11.5$, each containing a comparable number of galaxies. The respective mean 
values $\langle f_{\rm gas}^{\langle M_{*\,10.5}\rangle} \rangle = 
0.48\pm 0.17$, $\langle f_{\rm gas}^{\langle M_{*\,10.9}\rangle} \rangle = 
0.43\pm 0.15$, and $\langle f_{\rm gas}^{\langle M_{*\,11.1}\rangle} \rangle = 
0.40\pm 0.15$ for a ``Galactic'' CO--H$_2$ conversion factor show a mild 
decrease of $f_{\rm gas}$ with stellar mass within a very large dispersion. On 
the other hand, the mean $\langle f_{\rm gas}^{\langle M_{*\,{9.7}}\rangle} 
\rangle = 0.69\pm 0.18$ value of the lowest stellar mass bin $9.4 < 
\log (M_*/{\rm M}_{\sun}) < 9.9$ highlights an upturn of $f_{\rm gas}$ at the 
low-$M_*$ end of SFGs at $z>1$ as predicted.
More data in the low-$M_*$ end, tackling the $f_{\rm gas}$ upturn, may help 
disentangling specific outflow/feedback/wind recipes.

The mean values of $f_{\rm gas}$ within the two higher stellar mass bins 
defined above were also computed for SFGs at $z<0.4$ and are plotted in 
Fig.~\ref{fig:fgas-Mstars}. They nicely show the shift of the $M_*$ dependence 
toward smaller molecular gas fractions that is expected for star-forming 
galaxies at lower redshifts, because of the redshift evolution of $f_{\rm gas}$ 
(Sect.~\ref{sect:fgas-z}). This evolution is also partly responsible for the 
large dispersion observed within each $M_*$ bin of $z>1$ SFGs, as shown by the 
color-coding of the data points of $z>1$ SFGs highlighting three redshift bins.


The combination of the redshift evolution and $M_*$ dependence of the molecular 
gas fraction shows not only that the average $f_{\rm gas}$ of star-forming 
galaxies increases with redshift (Sect.~\ref{sect:fgas-z}), but that this 
increase is even more substantial for low stellar mass galaxies than for the 
high stellar mass ones given the upturn of $f_{\rm gas}$ at the low-$M_*$ end 
for $z>1$ SFGs. A behaviour judged by \citet{bouche10} and \citet{santini14} as 
being a direct result of the ``downsizing'' scenario, which claims that more 
massive galaxies formed earlier and over a shorter period of time 
\citep[e.g.,][]{cowie96,heavens04,jimenez07,thomas10}, i.e., consume their 
molecular gas more quickly.
Consequently, massive galaxies have already consumed most of their gas at high 
redshifts, while less massive galaxies still have a large fraction of molecular 
gas.


%

\section{Dust-to-gas ratio}
\label{sect:dust-to-gas}

The {\it Herschel}/PACS+SPIRE and longer wavelength data of our 
low-$L_{\rm IR}$ selected SFGs allow accessing their dust masses 
\citep[see Table~\ref{tab:SED-properties} and][]{sklias14}. When combined with 
CO emission data, we may hence investigate another key physical parameter that 
is the dust-to-gas mass ratio, $\delta_{\rm DGR} = M_{\rm dust}/M_{\rm gas}$.
This is even more important, as the dust-to-gas mass ratio and dust masses 
start to be used to estimate the CO--H$_2$ conversion factor both in local and 
high-redshift galaxies \citep[e.g.,][]{leroy11,magdis11,magdis12b,magnelli12} 
and to determine molecular gas masses when CO measurements are not available 
\citep{santini14,scoville14}. 

It is now accepted both from observations and interstellar medium evolution 
models that $\delta_{\rm DGR}$ scales linearly with the oxygen abundance 
\citep[e.g.,][]{issa90,dwek98,edmunds01,inoue03,draine07,leroy11,saintonge13,
chen13}. 
Nevertheless, the measure of the dust-to-gas mass ratio, in particular, in 
high-redshift galaxies 
remains very uncertain, because of a number of important assumptions that are 
done:
\begin{enumerate}
\item The dust mass derivation from the far-IR/sub-mm SED is tributary to
several assumptions, such as the dust model, the dust emissivity index $\beta$ 
within the modified black-body (MBB) fits, and the dust mass absorption cross section. 
The freedom on these parameters can 
lead to large variations in the $M_{\rm dust}$ estimates. 
\citet{magnelli12} already pointed out a factor of 3 difference between dust 
masses as inferred from MBB fits and the \citet{draineli07} model.
\item A CO--H$_2$ conversion factor needs to be assumed to determine the 
molecular gas mass. 
Since the CO--H$_2$ conversion factor is suspected to vary with metallicity, 
this could induce a false $\delta_{\rm DGR}$--metallicity dependence.
\item At high redshift we consider $M_{\rm H_2} \gg M_{\rm H\,I}$, or 
equivalently $M_{\rm gas} \approx M_{\rm H_2} = M_{\rm dust}/\delta_{\rm DGR}$. 
This is supported by the high molecular gas fractions measured in $z>1$ SFGs 
(see Fig.~\ref{fig:fgas-z}), which leave little room for a significant atomic 
gas component within the total mass budget 
as inferred from dynamical analysis \citep{daddi10a,tacconi10} and theoretical 
arguments like those discussed in Sect.~\ref{sect:fgas-z} 
\citep[e.g.,][]{obreschkow09,lagos11,lagos14}. However, no direct H\,I 
measurements exist in high-redshift galaxies so far.
\end{enumerate}

To alleviate the uncertainty on the above first assumption, \citet{scoville14} 
proposed to consider the long-wavelength far-IR/sub-mm continuum as a dust mass 
tracer, since the long-wavelength Rayleigh-Jeans tail of dust emission is 
generally optically thin and thus provides a direct probe of the total dust. 
They chose 850~$\mu$m as the fiducial wavelength and measured, in a homogeneous 
way, $\delta_{\rm DGR} \approx L_{\nu}(850\,\mu{\rm m})/M_{\rm gas}$ in the 
Galaxy, local spirals, local ULIRGs, and high-redshift SMGs by assuming for all 
a spectral slope $\beta = 1.8$ for the far-IR/sub-mm dust emission flux density 
as determined from the extensive {\it Planck} data throughout the Galaxy 
\citep{planck11a,planck11b} and a ``Galactic'' CO--H$_2$ conversion factor 
\footnote{\citet{scoville14} provide a solid justification of why using the 
``Galactic'' CO--H$_2$ conversion factor in local ULIRGs and high-redshift 
SMGs instead of the much smaller $\alpha = 1.1$ factor often proposed for 
these galaxies.}. 

\begin{figure}
\centering
\includegraphics[width=9cm,clip]{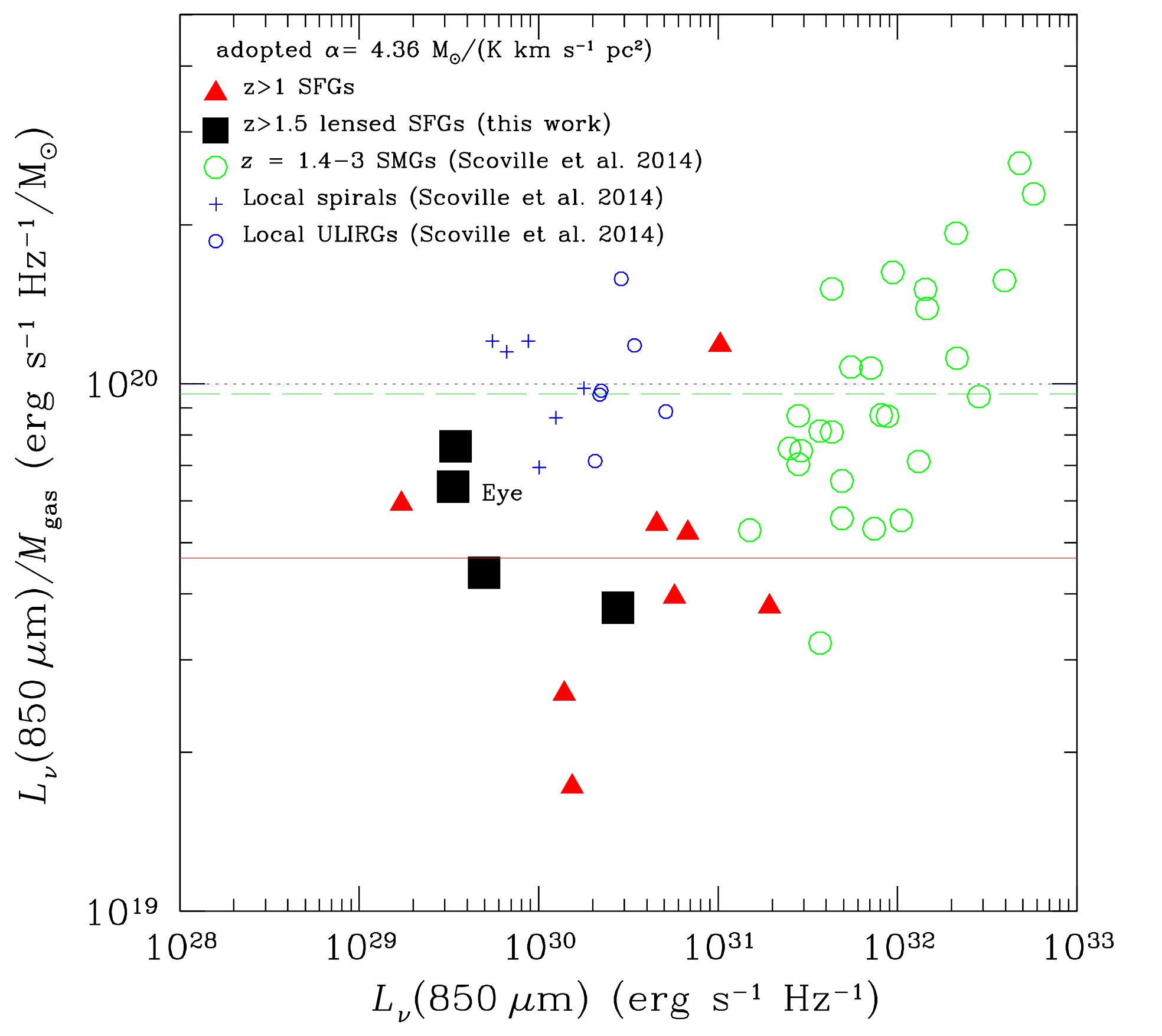} 
\caption{Dust emission at rest-frame 850~$\mu$m per unit mass of gas, 
$L_{\nu}(850\,\mu{\rm m})/M_{\rm gas}$, plotted for our low-$L_{\rm IR}$ 
selected SFGs (squares) and our comparison sample of $z>1$ SFGs with CO 
measurements from the literature and published far-IR/sub-mm photometry 
(triangles). Only SFGs 
with metallicities $Z/Z_{\sun} > 0.8$ are considered. These results are 
compared with the compilation from \citet{scoville14} for the SMGs at $z=1.4-3$ 
and local spirals and ULIRGs. We assume a ``Galactic'' CO--H$_2$ conversion 
factor for all objects and $M_{\rm gas} \simeq M_{\rm H_2}$ for high-redshift 
galaxies. All galaxies have a similar $L_{\nu}(850\,\mu{\rm m})/M_{\rm gas}$ 
within $1-2\,\sigma$, with a trend toward a lower dust-to-gas ratio in $z>1$ 
SFGs by $\sim 0.33~\rm dex$ at fixed near-solar metallicity, as shown by the 
lines corresponding, respectively, to the 
$L_{\nu}(850\,\mu{\rm m})/M_{\rm gas}$ means of local spirals and ULIRGs (blue 
dotted line), high-redshift SMGs (green dashed line), and $z>1$ SFGs (red 
solid line).}
\label{fig:L850Mgas-L850}
\end{figure}

Following the \citet{scoville14} prescriptions, 
we add to their sample our low-$L_{\rm IR}$ selected SFGs (see their rest-frame 
850~$\mu$m luminosities in Table~\ref{tab:SED-properties}) and the compilation 
of $z>1$ SFGs with CO measurements from the literature 
(Sect.~\ref{sect:literature}) for which published far-IR/sub-mm photometry is 
available \citep{magdis12b,saintonge13}. In addition, in order to bypass 
effects linked to the possible dust-to-gas ratio dependence on metallicity, we 
consider only the star-forming galaxies with near-solar metallicities 
$Z/Z_{\sun} > 0.8$. The metallicities are estimated using the mass--metallicity 
relation at $z\sim 2.2$ from \citet{erb06} recalibrated by \citet{maiolino08} 
and scaled to the \citet{chabrier03} IMF, when direct metallicity measurements 
from nebular emission lines are not available. 12 out of 73 CO-detected SFGs at 
$z>1$ (including our low-$L_{\rm IR}$ selected SFGs) satisfy the above 
prescriptions. Similarly to Scoville et~al., we adopt the ``Galactic'' 
CO--H$_2$ conversion factor, which in the case of these high-redshift SFGs is 
reasonable given their near-solar metallicities (and see also 
Sect.~\ref{sect:luminosity-corrections}).
The $L_{\nu}(850\,\mu{\rm m})/M_{\rm gas}$ results of these $z>1$ SFGs are 
shown in Fig.~\ref{fig:L850Mgas-L850} as a function of 
$L_{\nu}(850\,\mu{\rm m})$ and are compared to local spirals, local ULIRGs, and 
high-redshift SMGs from Scoville et~al. The respective means and dispersions 
are:
\begin{eqnarray*}
\langle L_{\nu}(850\,\mu{\rm m})/M_{\rm gas} \rangle_{\rm SFGs} &=& 
4.66^{+3.03}_{-1.84}\times 10^{19}~\rm erg~s^{-1}~Hz^{-1}~M_{\sun}^{-1}\\
\langle L_{\nu}(850\,\mu{\rm m})/M_{\rm gas} \rangle_{\rm SMGs} &=& 
9.54^{+6.00}_{-3.68}\times 10^{19}~\rm erg~s^{-1}~Hz^{-1}~M_{\sun}^{-1}\\
\langle L_{\nu}(850\,\mu{\rm m})/M_{\rm gas} \rangle_{\rm local} &=& 
1.01^{+0.27}_{-0.21}\times 10^{20}~\rm erg~s^{-1}~Hz^{-1}~M_{\sun}^{-1}\,.
\end{eqnarray*}

%

%

Measurements in local galaxies (spirals and ULIRGs), $z=1.4-3$ SMGs, and $z>1$ 
SFGs, at fixed near-solar metallicity ($Z/{\rm Z}_{\sun} > 0.8$), 
show similar $L_{\nu}(850\,\mu{\rm m})/M_{\rm gas}$ means 
within $1-2\,\sigma$ dispersion. While the means of local galaxies and 
high-redshift SMGs are nearly the same, $z>1$ SFGs sustain a trend for a shift 
toward a lower $L_{\nu}(850\,\mu{\rm m})/M_{\rm gas}$ mean by 
$\sim 0.33~\rm dex$ with a large dispersion of 0.22~dex. 
Such a shift 
in $z>1$ SFGs was already reported by \citet[][by about 0.23~dex]{saintonge13} 
when compared to Local Group galaxies \citep{leroy11}. 
If the dust-to-gas ratio truly is lower in high-redshift SFGs, this may 
suggest that a smaller fraction of dust grains is produced for the same 
metallicity under the specific conditions prevailing in the interstellar medium 
of these galaxies in comparison with local galaxies and high-redshift SMGs.

The data thus
support a non-universal $\delta_{\rm DGR}$.
Hence, deriving molecular gas masses from direct CO measurements remains 
highly recommended, as CO still appears a better tracer of the molecular gas 
mass, despite the uncertainty linked with the CO-to-H$_2$ conversion factor, 
than dust.


%

\section{Summary and conclusions}
\label{sect:summary}

We report new CO observations performed with the IRAM PdBI and 30~m telescope 
for five strongly-lensed star-forming galaxies at $z\sim 1.5-3$. These galaxies 
were selected from the {\it Herschel} Lensing Survey for their low intrinsic IR 
luminosities and high magnification factors. Four of them have IR luminosities 
as low as $L_{\rm IR} < 4\times 10^{11}~\rm L_{\sun}$ ($\rm SFR < 
40~M_{\sun}~yr^{-1}$) and reach the $L^{\star}$ to sub-$L^{\star}$ domain. 
While this regime is typical of SFGs, the molecular gas content of such 
galaxies has been poorly explored so far, because of mm/sub-mm instrumental 
sensitivity limitations. Thanks to the gravitational lensing, we achieve three 
CO emission detections in $L^{\star}$ to sub-$L^{\star}$ SFGs and a fourth in 
an SFG with a slightly higher $L_{\rm IR}$. These SFGs not only are among 
galaxies with the lower $L_{\rm IR}$ with accessible CO luminosities known, but 
also with the smaller stellar masses $M_* < 2.5\times 10^{10}~\rm M_{\sun}$. 
We add to this sample, cB58 and the Eye, two well-known strongly-lensed 
galaxies characterized by similar low $L_{\rm IR}$ and small $M_*$ properties, 
as derived by \citet{sklias14} from revised {\it Herschel} photometry and SED 
fitting. To put these SFGs in the general context of all CO-detected galaxies, 
we built up a large comparison sample of local and high-redshift galaxies with 
CO measurements reported in the literature that is exhaustive for high-redshift 
SFGs, complete for high-redshift SMGs, and only indicative for local spirals, 
LIRGs, and ULIRGs. 


This comparison sample allows us to obtain good estimates of the CO luminosity 
correction factors r$_{2,1}$ and r$_{3,1}$ for, respectively, the $J=2$ and 
$J=3$ CO rotational transitions for both $z=0$ and $z>1$ galaxies distributed 
over three $L_{\rm IR}$ intervals. 
Two main trends pop out in agreement with the models of \citet{lagos12}: 
(1)~shallower CO SLEDs for high-redshift galaxies compared to their $z=0$ 
counterparts at a fixed IR luminosity, and (2)~smaller differences in the CO 
SLEDs of faint- and bright-IR galaxies at $z>1$ than for $z=0$ galaxies. The 
inferred means of the observed CO luminosity correction factors valid for SFGs 
at $z>1$ are r$_{2,1} = 0.81\pm 0.20$ and r$_{3,1} = 0.57\pm 0.15$. 

The combination of the overall physical properties derived from the new CO 
measurements achieved in our SFGs covering a new dynamical range of 
$\rm SFR < 40~\rm M_{\sun}~yr^{-1}$ and 
$M_* < 2.5\times 10^{10}~\rm M_{\sun}$ with the comparison sample of 
CO-detected galaxies from the literature leads to the following main results.
\begin{enumerate}
\item 
A single linear relation within the $L_{\rm IR}$--$L'_{\rm CO(1-0)}$ log plane 
is observed, which best-fit gives a slope of 1.17. 
The current larger sample of $z>1$ SFGs shows an increased $1\,\sigma$ 
dispersion of 0.3~dex in the y-direction about the best-fit of their 
$L_{\rm IR}$--$L'_{\rm CO(1-0)}$ relation, which is sufficient to hide the 
reported 
bimodal behaviour between the `sequence of disks' and the `sequence of starbursts'.
\item Another way to represent the $L_{\rm IR}$--$L'_{\rm CO(1-0)}$ relation is 
through the star formation efficiency, defined as ${\rm SFE} = 
L_{\rm IR}/L'_{\rm CO(1-0)}\equiv {\rm SFR}/M_{\rm gas} = 1/t_{\rm depl}$. SFGs 
and SMGs at $z>1$ show very similar SFE distributions and large spreads; the 
SFE of $z>1$ SFGs are not confined to the low values of local spirals any more. 
The investigation of the SFE and $t_{\rm depl}$ dependence on several physical 
parameters leads us to conclude that it is the combination of the specific star
formation rate, stellar mass, and redshift that drives the large spread in SFE
of $z>1$ SFGs. The respective offset of SFGs from the main-sequence within the 
thickness of the SFR--$M_*$ relation, as well as the compactness of the 
starburst seem to play a minor role in the observed SFE spread.
\item The strongest dependence of $t_{\rm depl}$ is observed on the specific 
star formation rate both for $z>1$ SFGs and local galaxies. 
SFGs at $z>1$ show longer $t_{\rm depl}$ by about 0.75~dex than local galaxies 
with the same sSFR. This displacement of the $t_{\rm depl}$--sSFR relation with 
redshift 
is attributed to the larger molecular gas fractions found in $z>1$ SFGs that 
afford longer molecular gas depletion times at a given value of sSFR. 
In addition, the sSFR of $z=0$ galaxies are sealed on low values, because of 
the accumulation of more and more old stars in their bulge that have an 
important weight in the total stellar mass budget, but none on the present SFR.
\item A 
correlation between the molecular gas depletion timescale and stellar mass is 
observed thanks to the enlarged dynamical range of stellar masses down to 
$M_* = 10^{9.4}~\rm M_{\sun}$ sampled by our low-$L_{\rm IR}$ selected SFGs.
Although it needs to be confirmed with additional $z>1$ SFGs with small stellar 
masses, a $t_{\rm depl}$ increase with $M_*$ is seen in $z=0$ galaxies over the 
$M_*$ range from $10^{9}~\rm M_{\sun}$ to $10^{11.5}~\rm M_{\sun}$. 
If true, this $t_{\rm depl}$--$M_*$ correlation observed both in local galaxies 
and $z>1$ SFGs is opposed to the constant molecular gas depletion timescale 
assumed in the ``bathtub'' model and refutes the linearity of the 
Kennicutt-Schmidt relation.
%
\item We confirm the $(1+z)^{-1.5}$ evolution of the molecular gas depletion 
timescale for main-sequence galaxies as predicted by various models. The mean 
$t_{\rm depl}$ drops from 870~Myr at $\langle z_{0.2}\rangle = [0.055,0.4]$ 
down to 190~Myr at $\langle z_{3.0}\rangle = [2.7, 3.2]$ with a significant 
dispersion per redshift bin. This dispersion results from the 
$t_{\rm depl}$--sSFR and $t_{\rm depl}$--$M_*$ relations, which induce a 
gradient of $t_{\rm depl}$ per redshift bin, such that galaxies with the higher 
sSFR and smaller $M_*$ have the shorter $t_{\rm depl}$.
%
\item Observationally, it is now well established that high-redshift galaxies 
are gas-rich. What remains debated is the steady increase of the molecular gas 
reservoir with redshift predicted by models. While we do observe a net rise 
of the mean $f_{\rm gas}$ from $0.19\pm 0.07$ up to $0.46\pm 0.15$ between 
$\langle z_{0.2}\rangle = [0.055,0.4]$ and $\langle z_{1.2}\rangle = [1,1.6]$, 
it is followed by a very mild increase toward higher redshifts. 
A significant $f_{\rm gas}$ dispersion is seen per redshift bin, it is due to 
the strong dependence of the molecular gas fraction on the stellar mass, which 
we start testing observationally.
%
\item 
An $f_{\rm gas}$ drop with increasing $M_*$ because of the gas conversion into 
stars, as well as an $f_{\rm gas}$ upturn at the low-$M_*$ end whose strength is dependent on the outflow
are predicted. 
$z>1$ SFGs show a mild $f_{\rm gas}$ decrease 
between $10^{10.4} < M_*/{\rm M}_{\sun} < 10^{11.5}$. The large dispersion per 
stellar mass bin comes from the $f_{\rm gas}$ redshift evolution, as nicely 
illustrated by $z<0.4$ SFGs that have their $M_*$ dependence shifted toward 
lower $f_{\rm gas}$. We provide the first insights on $f_{\rm gas}$ of $z>1$ 
SFGs at the low-$M_*$ end $10^{9.4} < M_*/{\rm M}_{\sun} < 10^{9.9}$. The 
corresponding mean $\langle f_{\rm gas}\rangle = 0.69\pm 0.18$ value supports 
an $f_{\rm gas}$ upturn.
This shows that the average $f_{\rm gas}$ of SFGs not only increases with 
redshift, but this increase is even more substantial for low $M_*$ galaxies 
than for the high $M_*$ ones,
a behaviour 
resulting from the ``downsizing'' scenario.
\item To alleviate the uncertainties linked to the dust mass estimates, we 
consider the rest-frame 850~$\mu$m continuum as a dust mass tracer and compare 
the dust-to-gas mass ratios, $\delta_{\rm DGR} \approx 
L_{\nu}(850\,\mu{\rm m})/M_{\rm gas}$, inferred in a homogeneous way, of $z>1$ 
SFGs 
to those of local spirals, local ULIRGs, and high-redshift SMGs. 
While the $L_{\nu}(850\,\mu{\rm m})/M_{\rm gas}$ means of local galaxies and 
high-redshift SMGs are nearly the same, the one of high-redshift SFGs 
scatters systematically by a factor of two at fixed near-solar metallicity 
($Z/{\rm Z}_{\sun} > 0.8$). $z>1$ SFGs sustain a shift toward a lower 
$L_{\nu}(850\,\mu{\rm m})/M_{\rm gas}$ mean by 0.33~dex with a very large 
dispersion of 0.22~dex. 
This supports a non-universal $\delta_{\rm DGR}$. Thus, deriving molecular gas 
masses from direct CO measurements remains highly recommended.
\end{enumerate}

%

\begin{acknowledgements}
We are very grateful to M\'elanie Krips, Philippe Salom\'e, Claudia del P. 
Lagos, and Nicolas Bouch\'e for helpful discussions, advises, and adapted model 
predictions. 
We thank the IRAM staff of both the Plateau de Bure Interferometer and the 30~m 
telescope for the high-quality data acquired and for their support during 
observations and data reduction. This work was supported by the Swiss National 
Science Foundation. JPK acknowledges support from the ERC advanced grant LIDA 
and JR from the ERC starting grant CALENDS. 
\end{acknowledgements}

%

%

\appendix

\section{CO properties and kinematics of the individual low-$L_{\rm IR}$ selected galaxies}
\label{sect:appendix}


%

\subsection{A68-C0}

The CO(2--1) intensity peak is detected at $11\,\sigma$ in A68-C0, which makes 
it a very secure CO detection in this $z_{\rm H\alpha} = 1.5864$ galaxy, as 
shown in Figs.~\ref{fig:COmaps} and \ref{fig:COspectra}. The CO(2--1) emission 
is well centred on the HST 
images to within $0.5''$. The HST images clearly reveal the presence of two 
counterpart images of the same galaxy with a morphology typical of a spiral 
galaxy. The spiral structure is strongly stretched and extended by 
gravitational lensing, but the two counterpart images that are mirrored are 
easily identified by the two bright bulges observed north and south with the 
critical line passing in between. The two counterpart images of this galaxy are 
not resolved in CO, but they both contribute to the detected CO emission given 
the CO emission intensity precisely peaking on the critical line and the CO 
elongation in the direction of the stretch. The two counterpart images are 
amplified by the same magnification factor. 

The CO(2--1) spectrum reveals a double-peaked emission line profile with a 
total full width half maximum $\rm FWHM_{CO} = 334~km~s^{-1}$ and an observed 
integrated flux $F_{\rm CO} = 1.89\pm 0.28~\rm Jy~km~s^{-1}$. The measured CO 
redshift agrees very well with the H$\alpha$ redshift to within $\Delta z = 
(z_{\rm H\alpha}-z_{\rm CO}) = 0.001$. The derived lensing-corrected molecular 
gas fraction $f_{\rm gas} = 0.38$ is much higher than the molecular gas 
fractions of local spirals, but is comparable to the $f_{\rm gas}$ distribution 
of SFGs with CO measurements from the literature in the $\langle z_{1.2} 
\rangle = [1,1.6]$ interval which has a mean 
$\langle f_{\rm gas}^{\langle z_{1.2}\rangle} \rangle = 0.46\pm 0.15$ (see 
Sect.~\ref{sect:fgas-z}).

\begin{figure}
\centering
\includegraphics[width=9cm,clip]{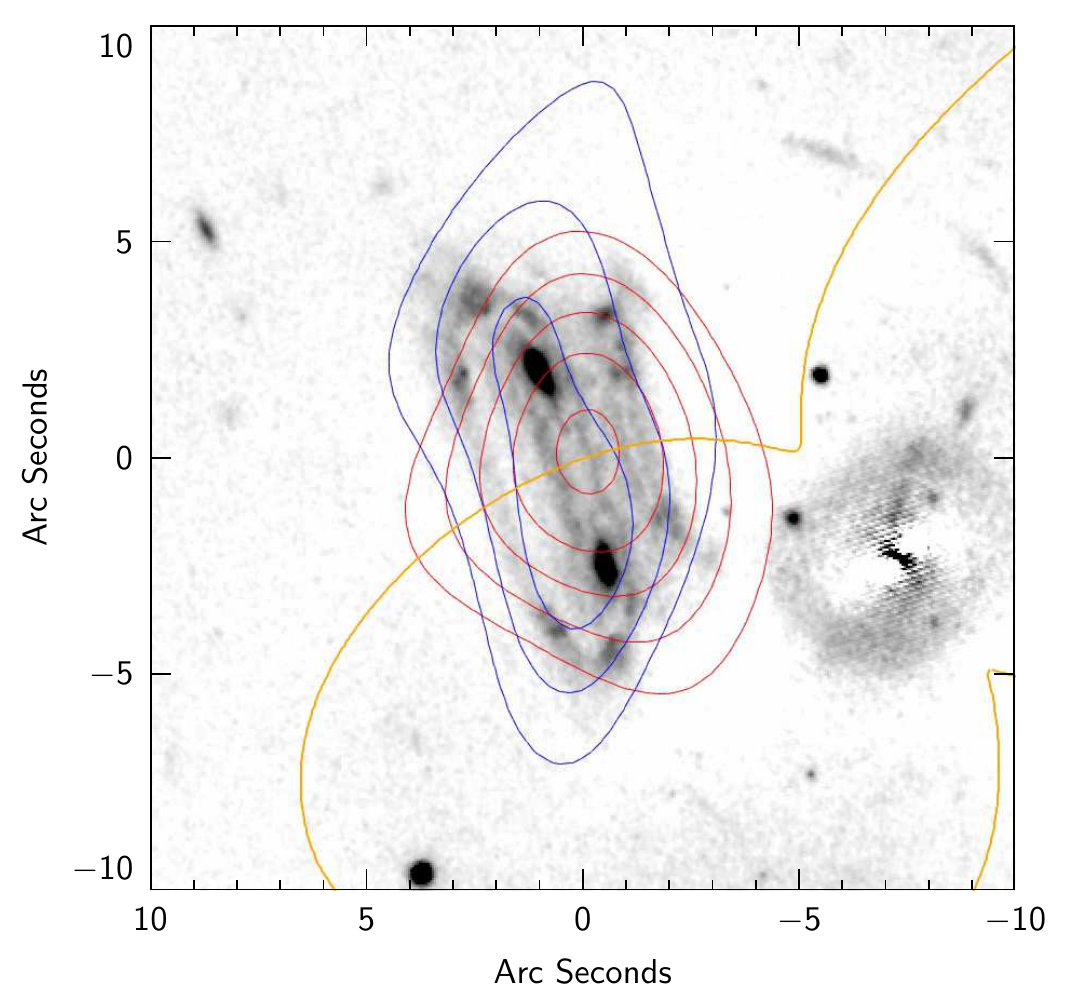} 
\caption{Velocity-integrated map of the CO(2--1) emission in A68-C0 overlaid on 
the HST image in band F110W. The extended blue contours are the contours 
integrated over the channels that define the blue-shifted component of the 
double-peaked CO emission line profile shown in Fig.~\ref{fig:COspectra} and 
the compact red contours are the contours integrated over the channels that 
define the red-shifted component of the CO profile. Contour levels start at 
$\pm 2\,\sigma$ and are in steps of $2\,\sigma$. The critical line (yellow 
line) is located right in the middle of the two multiple images easily 
identified by the two bright bulges resolved in the HST image.}
\label{fig:kinematic-A68C0}
\end{figure}

The observed double-peaked CO emission line profile is a first strong 
indication of rotation in A68-C0 \citep{daddi10a}. Two additional observational 
facts support rotation in this galaxy. First, its spiral morphology as seen in 
HST images (see Fig.~\ref{fig:COmaps}). Second, as shown in 
Fig.~\ref{fig:kinematic-A68C0} the CO contours integrated over the channels 
that define the blue-shifted component (blue contours) of the double-peaked CO 
emission line profile (Fig.~\ref{fig:COspectra}) appear to be much more 
extended in comparison to the contours integrated over the channels that define 
the red-shifted component (red contours) of the CO profile that are more 
compact. This is a direct consequence of rotation combined with gravitational 
lensing. Indeed, the critical line is located right in the middle of the two 
multiple images 
and is at an angle relative to the major axis of the spiral structure. Because 
the blue-shifted velocities relative to the observer are the farthest from the 
critical line and the red-shifted velocities the closest to the critical line, 
we see the blue-shifted CO component to be much more extended than the 
red-shifted component when plotting the CO blue-shifted and red-shifted 
velocity-integrated contours separately in Fig.~\ref{fig:kinematic-A68C0}.

%

%

\subsection{A68-HLS115}

The CO(2--1) line is also very well detected in A68-HLS115 as it can be seen in 
Figs.~\ref{fig:COmaps} and \ref{fig:COspectra}, with its intensity peaking at 
$14\,\sigma$. The CO(2--1) emission is nicely centred on the HST 
images to within $0.5''$. The HST images show that this $z_{\rm H\alpha} = 
1.5869$ galaxy has a very clumpy structure. The clumps are bright in the 
rest-frame UV, betraying intense star formation. Unfortunately they are 
unresolved in our CO observations, but would be of high interest for a deeper 
CO study at high spatial resolution. 

The CO(2--1) spectrum reveals a strong emission line very well fitted by a 
single Gaussian function with a full width half maximum $\rm FWHM_{CO} = 
267~km~s^{-1}$ and an observed integrated flux $F_{\rm CO} = 
2.00\pm 0.30~\rm Jy~km~s^{-1}$. The measured CO redshift perfectly agrees with 
the H$\alpha$ redshift to within $\Delta z = (z_{\rm H\alpha}-z_{\rm CO}) = 
0.001$. The derived lensing-corrected molecular gas fraction $f_{\rm gas} = 
0.75$ is very high; A68-HLS115 is among the gas-richer $z>1$ SFGs known (see 
Sect.~\ref{sect:fgas}). 
With a strong single-peaked CO emission line profile, we can hardly invoke 
evidence for rotation in this galaxy.

%

\subsection{MACS0451-arc}

The HST image reveals that this $z_{\rm H\alpha} = 2.013$ galaxy is strongly 
stretched by gravitational lensing as testified by the striking, elongated arc extending over $20''$ and composed of multiple images of the source. 
We detect the CO(3-2) emission at $4-5\,\sigma$. It does not cover the entire 
arc, but is confined to the southern part of the arc (see 
Fig.~\ref{fig:COmaps}). Such a confinement may either indicate that the lensing 
magnification is differential along the arc, 
and/or that the molecular gas is located 
in specific regions of the galaxy.
The spatially resolved kinematics of this arc has been studied in detail with 
H$\alpha$ observations by \citet{jones10}. This work allowed, in particular, to 
obtain a precise lens modelling and the reconstruction of 
the MACS0451-arc in the source plane. They show: 
(1)~the southern part of the arc is the image of the whole source and it 
also corresponds to the image in which the central red region of the source, 
the region where most of the CO emission is likely to originate, is most 
strongly magnified; and 
(2)~the northern part of the arc contains two images of the source very weakly 
magnified, in particular the central red region of the source is invisible in 
the HST images, the emission is dominated by one star-forming clump of the 
source close to the critical line which is popping up because of the increased 
magnification. Thus, the lack of CO detection in the northern part of the 
MACS0451-arc can be explained by the sensitivity limit of our PdBI data: if the 
total integrated CO flux of 
the southern part of the arc is detected at only $\rm S/N_{det} = 4-5\,\sigma$, 
the northern part of the arc 
is expected to have an integrated CO flux below $2\,\sigma$ of the noise. We 
may conclude with confidence that the detected CO emission 
provides a good estimate of the total molecular gas content of this 
strongly-lensed galaxy. Finally, the small offset of $1''$ observed between the HST 
images and the CO emission, as well as the elongated shape of the CO emission 
which does not follow the exact orientation of the arc in HST images (see 
Fig.~\ref{fig:COmaps}) are both most likely due to noise effects. Indeed, the 
low signal-to-noise ratio 
of the MACS0451-arc's CO detection combined with the low spatial resolution of 
D-configuration data imply that only the detection itself is reliable, while 
any spatial information remains uncertain.

There is evidence that the MACS0451-arc contains an AGN component, as 
determined from the analysis of its IR SED. In particular, the southern part of 
the arc, which contains the most magnified image of the center of the source 
(as said above), shows a much hotter IR SED than the northern part of the arc.
Indeed, while the SED of the northern part peaks at 250~$\mu$m, that of the 
southern part rises down to 100~$\mu$m (corresponding to a $\lambda_{\rm rest} 
< 33$~$\mu$m). This excess of flux at short wavelengths indicates the presence 
of very hot dust, which cannot be reproduced by a stellar component only. By 
modelling the IR emission of the southern part, Zamojski et~al.\ (in prep) find 
that roughly 55\% of the IR-flux originates from an AGN, while the remaining 
45\% is coming from star formation. This is taken into account when estimating 
the total IR luminosity of the MAS0451-arc corrected from the AGN component 
(Table~1).

The CO(3--2) spectrum indicates the presence of a double-peaked emission line 
profile (see Fig.~\ref{fig:COspectra}). The total full width half maximum is 
$\rm FWHM_{CO} = 261~km~s^{-1}$ and the observed integrated flux 
$F_{\rm CO} = 1.27\pm 0.32~\rm Jy~km~s^{-1}$. The measured CO redshift agrees 
very well with the H$\alpha$ redshift to within $\Delta z = 
(z_{\rm H\alpha}-z_{\rm CO}) = 0.0012$. The derived lensing-corrected molecular 
gas fraction $f_{\rm gas} = 0.62$ is high, 
it is the higher among SFGs with CO measurements from the literature in the 
$\langle z_{2.2} \rangle = [2,2.5]$ interval (see Sect.~\ref{sect:fgas-z}).

The observed double-peaked CO emission line profile in the MACS0451-arc may 
appear as tentative because of the low $\rm S/N_{det} = 4-5\,\sigma$, however 
the fit with two Gaussian functions is clearly favoured relative to the fit 
with a single Gaussian function. This prevents us from affirming conclusively 
whether the observed double-peaked CO emission line profile really originates 
from the presence of rotation \citep{daddi10a}, or is simply an artefact of a 
noisy line profile. \citet{jones10} showed that the kinematics of this galaxy 
as traced by the integrated H$\alpha$ emission line flux can be explained by 
rotation. Because, their data cover only part of the galaxy, either scenario, 
rotation versus merger, remains possible. The recent reconstructed morphology 
of the MACS0451-arc galaxy in the source plane found in \citet{sklias14} shows 
two tails, which can be most easily explained by a merger. Such a scenario 
would also naturally explain the presence of the AGN seen in the southern part 
of the arc.


%

\subsection{A2218-Mult}

This $z_{\rm H\beta} = 3.104$ galaxy is the highest redshift galaxy from our 
sample of low-$L_{\rm IR}$ selected galaxies. HST images reveal two counterpart 
images \citep{richard11}. We searched for CO in the western counterpart image. 
Comparatively we achieved similarly deep PdBI observations with an rms of 
0.77~mJy for this object, but the CO(3--2) emission remains undetected despite 
a high magnification factor $\mu = 14$. We provide $4\,\sigma$ upper limits on 
the CO(3--2) integrated line flux and its respective derivatives CO(1--0) 
luminosity and molecular gas mass 
in Table~\ref{tab:COresults}. 
This object was not included in the study of \citet{sklias14}, but we used the 
same tools and methods to derive its IR luminosity.

%

\subsection{A68-h7}

This $z_{\rm CO} = 2.1529$ galaxy is less magnified with $\mu = 3$ and 
characterized by a higher intrinsic IR luminosity $L_{\rm IR} = 18.3\times 
10^{11}~\rm L_{\sun}$ in comparison to the other galaxies from our sample. It 
has thus been observed at the IRAM 30~m telescope. The CO(3--2) spectrum shows 
a broad line detected at $3-4\,\sigma$ and well fitted by a single Gaussian 
function (see Fig.~\ref{fig:COspectra}). It has a full width half maximum 
$\rm FWHM_{CO} = 282~km~s^{-1}$ and an observed integrated flux 
$F_{\rm CO} = 1.12\pm 0.22~\rm Jy~km~s^{-1}$. A redshift offset is observed in 
A68-h7 between the CO redshift and the redshift determined in this case from 
weak absorption lines in the rest-frame UV, $\Delta z = 
(z_{\rm rest-UV}-z_{\rm CO})= -0.0029$. This offset is not surprising, 
because interstellar medium lines are expected to be blueshifted with respect 
to the systemic redshift of the galaxy as set here by the CO(3--2) line in 
presence of outflows \citep{quider09,dessauges10}. The derived 
lensing-corrected molecular gas fraction $f_{\rm gas} = 0.33$ is comparable to 
the $f_{\rm gas}$ distribution 
of SFGs with CO measurements from the literature in the 
$\langle z_{2.2} \rangle = [2,2.5]$ interval which has a mean 
$\langle f_{\rm gas}^{\langle z_{2.2} \rangle}\rangle = 0.45\pm 0.22$ (see 
Sect.~\ref{sect:fgas-z}).

The observed double-peaked CO emission line profile, which would require 
higher signal-to-noise ratio data to be confirmed (in particular for the second 
component sampled by two channels only given its small full width half maximum 
$\rm FWHM_{CO} = 90~km~s^{-1}$; see Fig.~\ref{fig:COspectra}), sustains evidence 
for rotation in A68-h7 \citep{daddi10a}. Its morphology in the HST images 
reveals several components, which probably all contribute to the CO emission 
signal and may be the analogues of giant star-forming clumps seen in disk-like 
simulated systems \citep{agertz09,genel10,bournaud11}.

%

\subsection{cB58 and Eye}

We adopt for these two strongly-lensed galaxies at $z_{\rm H\alpha} = 2.729$ 
and 3.0733, respectively, the CO(1--0) integrated fluxes obtained with the 
Expanded Very Large Array (EVLA) by \citet{riechers10}. The Eye is one of the 
rare $z>3$ star-forming galaxies with a CO detection. Detections of the 
CO(3--2) line with the PdBI and measurements of their respective integrated 
fluxes were also reported by \citet{baker04} in cB58 and \citet{coppin07} in 
the Eye. Those are used in Sect.~\ref{sect:luminosity-corrections} to study the 
CO luminosity correction factors of star-forming galaxies at $z>1$. Combined 
with our updated stellar masses from \citet{sklias14}, we infer 
lensing-corrected molecular gas fractions $f_{\rm gas} = 0.41$ in cB58 and 
$f_{\rm gas} = 0.18$ in the Eye. Although they are very similar to the 
molecular gas fractions reported by \citet{riechers10}, $f_{\rm gas} = 0.32$ 
in cB58 and $f_{\rm gas} = 0.13$ in the Eye, if we consider the stellar masses 
assumed by Riechers et~al.\ and our prescriptions for the CO luminosity 
correction factor and the CO--H$_2$ conversion factor, the corresponding 
molecular gas fractions get significantly higher by a factor of 2 and 2.5, 
respectively, compared to ours. The revised molecular gas fraction of the Eye 
is particularly low, it is the lowest among $z>1$ SFGs with CO measurements 
from the literature (see Sect.~\ref{sect:fgas}). This may partly be attributed 
to the ``post-starburst'' nature of the Eye as inferred by \citet{sklias14} 
from a detailed SED analysis.

\end{document}